\PassOptionsToPackage{hidelinks}{hyperref}
\documentclass[twocolumn]{aastex701}

\usepackage{amsmath}
\usepackage{array}
\usepackage{makecell}
\usepackage{cleveref}
\usepackage{orcidlink}

\begin{document}

\title{The stellar velocity anisotropy of strong lensing massive elliptical galaxies and its role in the inference of the Hubble parameter $H_0$ using spatially resolved kinematics}

\author{Vishal Verma~\orcidlink{0009-0003-9557-986X}}
\affiliation{Department of Astrophysics, American Museum of Natural History, Central Park West and 79th Street, NY 10024-5192, USA}
\affiliation{The Graduate Center of the City University of New York, 365 Fifth Avenue, New York, NY 10016, USA}
\affiliation{Department of Physics and Astronomy, Lehman College of the CUNY, Bronx, NY 10468, USA}
\email{vverma@gradcenter.cuny.edu}

\author{Quinn Minor~\orcidlink{0000-0002-4270-6453}}
\affiliation{Department of Astrophysics, American Museum of Natural History, Central Park West and 79th Street, NY 10024-5192, USA}
\affiliation{The Graduate Center of the City University of New York, 365 Fifth Avenue, New York, NY 10016, USA}
\affiliation{Department of Science, Borough of Manhattan Community College, City University of New York, New York, NY 10007, USA}
\email{qeminor@gmail.com}

%\maketitle
\submitjournal{ApJ}
\received{7 November 2025}
\revised{27 January 2026; revised manuscript currently under review}

\begin{abstract}

One of the biggest challenges in cosmology, the Hubble Tension, requires independent measurements of $H_0$, and strong lensing with time-delay cosmography is a promising avenue. The inclusion of spatially resolved kinematic data helps break the mass--sheet degeneracy, a key limitation in strong lensing. Kinematics, however, suffers from its own degeneracy due to unknown stellar velocity anisotropy, which can bias galaxy mass profile inferences. We investigate the bias in $H_0$ using a sample of ten massive elliptical galaxies at $z=0.2$ from the Illustris $TNG100$ simulations. We generate mock line-of-sight velocity-dispersion maps resembling JWST NIRSpec observations and test four anisotropy models: Osipkov--Merritt (OM), Mamon--Lokas (ML), constant $\beta$, and a generalized--OM (gOM) profile, under both kinematics-only and joint kinematics plus strong lensing analyses. We find a sub-percent average bias in $H_{0}$ across ten galaxies with joint modeling for three models: $+0.2 \pm 1.6\%$ (ML), $-0.9 \pm 1.9\%$ (constant) and $-0.9 \pm 1.6\%$ (gOM), with $\sim 5\%$
scatter. Joint modeling reduces bias, improves precision, and mitigates outlier results. Overall, the gOM model best recovers galaxy parameters and delivers the most accurate $H_{0}$ relative to posterior uncertainties considering both analyses. However, the single-parameter OM model produces large systematic biases: with kinematics only data, $H_{0}$ errors can exceed $20\%$, and even with joint modeling, produces an overall bias of $+11.5 \pm 1.3\%$ (OM). The higher bias in OM is unlikely to average out across an ensemble of galaxies. Our findings highlight the impact of anisotropy assumptions on $H_{0}$ inference and, more broadly, in galaxy dynamics.

\end{abstract}
\section{Introduction}
\label{sec:introduction}

Significant strides have been made in our understanding of the universe over the past couple decades owing to advances in precision cosmology enabled by missions like HST \citep{freedman_final_2001}, Planck~\citep{planck_collaboration_planck_2020}, WMAP \citep{bennett_nine-year_2013}, SDSS \citep{york_sloan_2000} and DESI \citep{desi_collaboration_desi_2016}, as well as by improvements in observational and data analysis techniques. One of the eminent results include the firm establishment of $\Lambda$CDM --- a theory that the Universe is dominated by Dark Energy ($\Lambda$) and Cold Dark Matter (CDM), with the reminder being normal baryonic matter --- as the standard model of cosmology. In addition, developments have been made in probing both large-scale cosmic structure and small-scale structures within galaxies, which are believed to be embedded in dark matter halos, with unprecedented detail.

Despite these advances, several fundamental challenges to $\Lambda$CDM remain unresolved, such as the core-cusp problem \citep{de_blok_corecusp_2010} and the missing satellites problem \citep{klypin_where_1999}.  In the last decade, these small-scale issues have been alleviated when baryonic physics such as stellar feedback and gas outflows are incorporated properly into hydrodynamic simulations \citep{brooks_how_2017}. But perhaps most critically, the so-called Hubble Tension, the discrepancy between the early and late universe measurements of the Hubble parameter ($H_0$), which measures the expansion rate of the universe has persisted. The Hubble Space Telescope Key project \citep{freedman_final_2001} was an important early effort that studied Cepheid variable stars as standard candles --- a late, nearby universe measurement --- and produced $H_0 = 72 \pm 8\ \mathrm{km\ s^{-1}\ Mpc^{-1}}$. In contrast, independent estimates based on the Cosmic Microwave Background (CMB) from the early universe provided a significantly lower value. The WMAP mission (2003–2013) initially estimated $H_{0} = 71 \pm 4~\mathrm{km~s^{-1}~Mpc^{-1}}$ but later converged to $H_{0} = 69.32 \pm 0.80~\mathrm{km~s^{-1}~Mpc^{-1}}$ \citep{spergel_firstyear_2003,hinshaw_nine-year_2013}. The Planck mission further refined this result, providing the most precise estimate based on the CMB of $H_0 = 67.4 \pm 0.5 \mathrm{km\ s^{-1}\ Mpc^{-1}}$, assuming the standard $\Lambda$CDM model.
Constraints from Baryonic Acoustic Oscillations (BAO) observable in surveys such as SDSS \citep{york_sloan_2000} estimated $H_0 = 68.18 \pm 0.79 \, \text{km s}^{-1} \text{Mpc}^{-1}$, a value close to the CMB measurements --- an expected agreement since both CMB and BAO measurements rely on the same physical calibration of the sound horizon. More recent analyses that looked at Cepheid stars such as the SH0ES project \citep{riess_expansion_2019} found $H_0 = 73.5 \pm 1.4 \mathrm{km\ s^{-1}\ Mpc^{-1}}$. In contrast, updates from the Chicago--Carnegie Hubble Program, \citep{freedman_status_2025} via JWST observations of the tip of the red giant branch (TRGB) and J-region asymptotic giant branch (JAGB) stars yield lower values of the Hubble parameter thereby bridging the gap, with estimates ranging from $H_0 = 70.39 \pm 1.22\;(\mathrm{stat}) \pm 1.33\;(\mathrm{sys}) \pm 0.70\;(\mathrm{\sigma_{SN}})~\mathrm{km\,s^{-1}\,Mpc^{-1}}$ (TRGB) to $H_0 = 67.80 \pm 2.17\;(\mathrm{stat}) \pm 1.64\;(\mathrm{sys})~\mathrm{km\,s^{-1}\,Mpc^{-1}}$ (JAGB).

Due to improvements in precision, the statistical difference of this tension has increased over time and has become significant, exceeding 5$\sigma$ in some cases. The resolution of this tension could be via the uncovering of systematics affecting the measurement, our incomplete understanding of the early or late universe, or perhaps more remarkably could point to the existence of unknown physics beyond the standard model. Thus, this challenge must be addressed, and one way to do this is to use independent methods that measure $H_0$. The use of strong lensing data to measure $H_0$ \citep{refsdal_possibility_1964} via time-delay cosmography is one such independent approach, and it becomes particularly robust when combined with spatially resolved kinematics. Strong lensing is a phenomenon arising out of predictions of General Relativity, where light rays from a distant object such as a quasar undergo significant deflection from their original path while traveling around a massive foreground object such as a massive elliptical galaxy ---  behaving effectively like a lens --- resulting in visibly distinct signatures like Einstein rings, arcs, and multiple images of the source. Time-delay cosmography involves measurement of time delays between the lensed images of intrinsically variable sources such as quasars (or transient sources like supernovae), and $H_0$ is inversely proportional to the difference in observed time delays. If the lens potential is known or can be accurately modeled, then in principle $H_0$ can be determined.

Early measurements of $H_0$ from time delay cosmography seem to align with the late universe measurements (for example, \cite{shajib_strides_2020}  found $ H_0 = 74.2^{+2.7}_{-3.0}\ \mathrm{km\ s^{-1}\ Mpc^{-1}}$). More recently \cite{tdcosmo_collaboration_tdcosmo_2025} estimated a value of $ H_0 = 71.6^{+3.9}_{-3.3}\ \mathrm{km\ s^{-1}\ Mpc^{-1}}$, providing a range that encompasses $H_0$ values from both Planck and SH0ES. One of the key milestones for time delay cosmography to be a competitive independent probe will be to constrain $H_0$ to one percent precision  \citep{treu_strong_2022}. Works by the TDCOSMO collaboration \citep{millon_tdcosmo_2020} have shown promise in this direction. \cite{shajib_strides_2020}  reported a 3.9 percent precision while  \cite{shajib_improving_2018}  found they could constrain $H_{0}$ to 1 percent precision when they used 40 gravitational lenses although they used simulated data. 

Importantly, however, while the measurements have become increasingly precise over time, the accuracy and reliability of these results will depend on several challenges and key systematics being investigated thoroughly. One major issue when using strong lensing data is the so called mass-sheet degeneracy (MSD) \cite{falco_model-dependent_1985}. In essence, different lens mass distributions and source positions can yield the same lensed image configurations. The MSD limits our ability to reconstruct the density profile of the deflector galaxy from lensing data alone. In contrast, time delays scale proportionally with any mass-sheet contribution to the lensing density, which makes this a key systematic in estimating  $H_0$ \citep{schneider_mass-sheet_2013-1}. Without additional information, it is not possible to break this degeneracy using strong lensing data alone. One of the most robust ways to address this is to include stellar kinematic data which encompasses information about the motion of stars in a galaxy.

The field of strong lensing, already experiencing rapid growth, is poised for a significant expansion with the advent of wide-field time-domain surveys such as Euclid and LSST, which are predicted to discover up to $100,000$ galaxy-galaxy strong lenses \citep{collett_population_2015}. What makes this even more exciting is that the large number of expected lenses will allow for an $H_0$ estimation across a wide range of redshifts. In principle, follow-up observations of the stellar kinematics can be made for systems of interest.  Early works involving kinematic data such as \cite{treuMassiveDarkMatter2004_svd}, \cite{koopmans_2009_svd} used a single velocity dispersion measurement of the central region of the galaxy to obtain the mass profile of a deflector. This additional independent measurement of the mass profile of the galaxy being observed, while useful, lacked the resolution to explore more detailed dynamics of stellar motion. The advent of Integral Field Spectroscopy (IFS) revolutionized the ability to study galaxies in greater detail, where the entire field of view can be divided into spatially resolved pixels. At the position of each pixel we obtain both the stellar spectrum and line-of-sight velocity information. 
\textbf{
}The SAURON \citep{bacon_sauron_2001} project followed by the ATLAS \citep{cappellari_atlas3d_2011} survey were landmark missions that used IFS to revolutionize our understanding of elliptical galaxies. The James Webb Space Telescope is actively obtaining high-quality IFS data via the Near Infrared Spectrograph (NIRSpec) instrument while other projects include OSIRIS on Keck \citep{larkin_osiris_2006} and the upcoming IRIS on Thirty Meter Telescope \citep{larkinIRIS}. The inclusion of IFS data while studying strong lensing systems has become increasingly common, as demonstrated by works such as \cite{tdcosmo_collaboration_tdcosmo_2025}, \cite{shajib_tdcosmo_2025}, \cite{shajib_tdcosmo_2023}, \cite{yildirim_tdcosmo_2023}, \cite{yildirim_time-delay_2020} and \cite{shajib_improving_2018}.  Combining spatially resolved kinematic IFS data allows us to mitigate the mass sheet degeneracy as it provides a detailed, independent mass profile of the deflector.  In addition, spatially resolved kinematic data offers other advantages.  For example, \cite{shajib_improving_2018} show that it can help add precision to the results of the galaxy parameters being studied  and measure cosmological parameters like the dark energy equation of state; and that it can help constrain parameters like the distance to the lens, which may not be determined uniquely by strong lensing data alone. It allows for a more detailed exploration of the inner dark matter distribution in spiral galaxies and thereby addresses the core-cusp debate.

However, one needs to take into account a key potential problem in kinematics itself --- the mass anisotropy degeneracy \citep{MADbinneyVelocityAnisotropyObservations1982}. This arises because apart from a few galaxies within the local group for which the proper motion of individual stars can be measured, the only velocity information we can measure for galaxies further away is the second moment of the velocity which lies along the line-of-sight. The velocity anisotropy parameter describes whether the motion of stars in a galaxy are preferentially oriented along the radial or tangential direction. The net effect of this degeneracy is that for the same observables one can infer the wrong galaxy mass profiles and orbital structure. Using higher-order velocity moments with anisotropic distribution-function–based models can mitigate, and in some cases, break the mass–anisotropy degeneracy \citep{cappellari_early-type_2025}. However, in practice their constraining power is quite sensitive to data quality, radial coverage, and the underlying assumptions about the adopted form of the distribution function.

Often, when only central velocity dispersion measurements or photometric data exist and having some radial variation in the anisotropy is preferred, the single-parameter Osipkov-Merritt \citep{osipkov_spherical_1979, merritt_spherical_1985, merritt_distribution_1985-1}  model has been the \textit{de facto choice}. The assumptions in the anisotropy model has wide ranging implications and extends to studies that involve galaxy dynamics of different morphologies, in galaxy formation, in exploring the nature of dark matter, and in cosmology. For instance, \cite{tan_project_2024} found that when using a single parameter Osipkov-Merritt there could be potential deviations in the expected power-law profile of massive ellipticals compared to the constant model.   \cite{el-badry_when_2017} tested different anisotropy models in dwarf galaxies and found the mass inference to be model dependent. \cite{evans_cores_2009} found that the identification of a core or cusp in dwarf spheroidal galaxies was heavily dependent on the assumptions of anisotropy and spherical symmetry rather than the data itself. Works such as \cite{gomer_tdcosmo_2022} and \cite{yildirim_time-delay_2020} have highlighted the importance of assumptions made in anisotropy, and \cite{birrer_mass-sheet_2016} found that the anisotropy model used had an effect on the value of $H_0$ inferred.

Spatially resolved kinematic data provide a more detailed analysis of stellar motion, and thereby improve the chances of breaking the mass-anisotropy degeneracy \citep{shajib_improving_2018}. However, even here the choice of anisotropy parameterization is a potential source of systematic. The primary focus of this work is the investigation of assumptions in the anisotropy model using spatially resolved kinematic data and the role they play in inferring galaxy parameters and thereby the measurement of $H_0$. We used a dataset of massive elliptical galaxies from the IllustrisTNG simulations and created IFS data to mimic a NIRSpec observation. We assume a value of $H_0 = 70 \mathrm{km\ s^{-1}\ Mpc^{-1}}$ wherever it is used and aim to find a relative bias that can be expected for a given set of time delays. We first model purely via the stellar dynamics and assume that the galaxy being modeled is a deflector in a strong lensing system. Next, we do a joint modeling that includes reliable information strong lensing provides into the stellar dynamics. The different anisotropy models that we test include Osipkov-Merritt, constant anisotropy, Mamon-Lokas anisotropy \citep{mamon_dark_2005}, and a more flexible anisotropy model, the generalized Osipkov-Merritt. 

This paper adopts the following layout. In \autoref{sec:background_theory}, we introduce the theory of strong lensing, stellar kinematics, and time-delay cosmography relevant to this study. In \autoref{sec:data_methods}  we introduce the galaxies dataset from the IllustrisTNG simulations and discuss the methods for obtaining the final simulated data to mimic a JWST observation of an elliptical galaxy at $z = 0.2$, via a Jeans analysis. In \autoref{sec:kinematic_modeling}, we talk about the parameters we use to represent a galaxy, the four different anisotropy models, and give the framework for Bayesian modeling using Nested Sampling. In \autoref{sec:modeling_results_kinematics} we look at a representative galaxy and present and discuss the results with kinematics only data for the different velocity anisotropy models and their effect on $H_0$ inference in detail. We also look at the results for all the remaining galaxies in the data set. We follow this with \autoref{sec:modeling_results_kin_plus_lensing}, in which we investigate the results of joint modeling with kinematics and strong lensing. In the subsequent \autoref{sec: additional_runs}, we show the results of kinematics only and joint modeling when the data quality is reduced while the methods and assumptions of the previous two sections remain the same. In \autoref{sec:discussions},  we discuss the performance of the different models,  the limitations and caveats of this study, make a comparison with other works, and discuss how this work can be improved. We end this study with \autoref{sec:summary_and_conclusions}, where we summarize the key results.

\section{Background and Theory}
\label{sec:background_theory}

\subsection{Strong Lensing and MSD}

If the light rays emerging from an astrophysical source at position $\beta$ undergo strong lensing by a deflector leading to formation of multiple images, then the positions $\theta$ of these images is given by the lens equation:
\begin{equation}
    {\beta} = {\theta} -  {\alpha}({\theta})
\end{equation}
Strong lensing leads to magnified images of the original source, thus acting like a giant optical telescope, and allows us to study the lens and the source at a much greater resolution than would be normally possible. The chief challenge in the reliability of results from strong lensing is due to the aforementioned problem called mass-sheet degeneracy. Consider a lens whose normalized surface mass density (convergence) $\kappa$ is given by:
\begin{equation}
    \kappa= \frac{\Sigma(\theta)}{\Sigma_{\text{crit}}}
    \label{eq: kappa}
\end{equation}
Here $\Sigma(\theta)$ is the two-dimensional projected surface mass density and $\Sigma_{\text{crit}}$ is the critical density for strong lensing defined by:
\begin{equation}
    \Sigma_{\text{crit}} = \frac{c^2}{4\pi G} \frac{D_s}{D_d D_{ds}}
    \label{eq:critical_dens}
\end{equation}
, with c being the speed of light and G the universal gravitational constant. The second term contains three key angular diameter distances: $D_s$ (between the observer and the source), $D_d$ (between the observer and the deflector), and $D_{ds}$ (between the defector and the source).
Suppose that the lens surface mass density $\Sigma(\theta)$ and the source position $\beta$ are rescaled a factor $\lambda$, then such a transformation of these quantities would leave the observables, the image positions and magnifications unchanged \citep{falco_model-dependent_1985}. For example, the presence of a line-of-sight structure like the extended outer halo of a nearby galaxy or large-scale structure can effectively mimic a uniform mass sheet and lead to incorrect inference of the lens surface mass density and the source position. This can result in biased galaxy inferences affecting $H_0$ measurement.

\subsection{Stellar Velocity Anisotropy $\beta$}

The velocity anisotropy parameter (denoted $\beta$) gives information about the velocity distribution of the stars in different directions. For spherical systems $\beta$  is given by: 
\begin{equation}
\beta(r) = 1 - \frac{(\sigma_\theta^2 + \sigma_\phi^2)}{2 \sigma_r^2} 
\label{eq:beta}
\end{equation}
Where $\sigma_r$, $\sigma_{\theta}$, and $\sigma_{\phi}$ are the standard deviations (velocity dispersions) in the velocities of the stars in the $r$, $\theta$, and $\phi$ directions. In general, $\beta$ can vary with distance from the center of the galaxy and lies in the range $(-\infty, 1)$.  A value of $\beta = 0$ denotes an isotropic system, where the velocity dispersion in all three directions is equal, meaning that there is no preferred direction of motion for the stars. For $\beta  > 0$ , the system is radially biased, so stars on average have motions in the radial direction. A value of $\beta  < 0$  denotes a tangentially biased system. For the limiting case of $\beta = -\infty$, the orbits are perfectly circular, and stars prefer to move in a plane.

\subsection{Jeans Modeling}

The governing set of equations that are fundamental to describing the motion of stars in a galaxy are the Jeans equations \cite{jeans_theory_1915}, which are derived from the collisionless Boltzmann equation that describes the dynamics of a collisionless system like the stars in a galaxy. The spherical Jeans equations \citep{Bovy2026} are given by:
\begin{equation}
    \frac{d (\nu \sigma_r^2)}{dr} + \frac{2\beta}{r} \nu \sigma_r^2 = -\nu \frac{d\Phi}{dr}
\end{equation}

Here, $\nu(r)$ is the normalized stellar mass density profile, $\sigma_r(r)$ are the velocity dispersions in the radial direction, and $\Phi(r)$ is the total gravitational potential due to stars and dark matter. From the Jeans equations we can arrive at an equation that connects the observable quantity, the velocity dispersions along the line-of-sight $\sigma_{\text{los}}(R)$ to other quantities like $\nu(r)$, the stellar velocity anisotropy profile $\beta (r) $, the normalized projected stellar mass density profile $\Sigma(R)$, and the velocity dispersions in the radial direction $ \sigma_r$:
\begin{equation}
\Sigma(R) \sigma^2_{\text{los}}(R) = 2 \int_{R}^{\infty} \mathrm{d}r \left(1 - \beta(r) \frac{R^2}{r^2}\right) \frac{\nu(r) \sigma_r^2 r}{\sqrt{r^2 - R^2}}
\label{eq:LOSVD}
\end{equation}
Here, the upper limit $\infty$ in the line-of-sight velocity dispersion integral corresponds to a distance far from the galaxy center, where the stellar mass density $\nu(r) $ becomes vanishingly small.

The projected stellar mass density $\Sigma(R)$ is related to the three-dimensional mass density $\nu(r)$ by:
\begin{equation}
\Sigma(R) = 2 \int_{0}^{\infty} \mathrm{d}z \, \nu(\sqrt{R^2 + z^2}), 
\label{eq:SIGMA_R}
\end{equation}
and the velocity dispersions in the radial direction $ \sigma_r$ are given by \citep{van_der_marel_velocity_1994}: 

\begin{equation}
\nu\,\sigma_r^2(r) =\int_r^{\infty} \nu(r') \left.\frac{d\Phi}{dr}\right|_{r'}
\exp\!\left[2 \int_r^{r'} \frac{\beta(r'')}{r''}\,dr''
\right]\,dr'
\label{eq:sigma_r}
\end{equation}

Here, $\Phi(r)$ is the total gravitational potential due to stars and dark matter. In reality, we aim to obtain the galaxy parameters --- the stellar and dark matter masses, and stellar anisotropy profiles --- from the observables, the line-of-sight velocity dispersions, and the two-dimensional light profile. The stellar light profile is chosen by template fitting. The projected stellar mass density $\Sigma(R)$ can be obtained from the stellar light profile if the mass-to-light ratio is known or assumed. The three-dimensional mass density $\nu(r)$ can then be found by inverting $\Sigma(R)$. For extra-galactic systems at high redshifts, we can only measure the velocity dispersions along the line-of-sight. This means that without additional information we cannot know the exact stellar velocity anisotropy profile $\beta(R)$.  Because the total gravitational potential of the galaxy  (Equation \ref{eq:sigma_r}) also requires assumptions, we end up with two unknowns in Equation \ref{eq:LOSVD}, the anisotropy $\beta(R)$ and the total galaxy mass profiles $M(r)$ and different combinations of either could still lead to the same observable, the line-of-sight velocity dispersions. This is the mass-anisotropy degeneracy, a chief problem in kinematic modeling.

\subsection{Dark Matter Model}

The modeling of dark matter is often done using an Navarro-Frenk-White (NFW) profile \citep{navarro_universal_1997}, which is a spherical density profile that describes the mass distribution of dark matter halos based on results from numerical simulations of structure formation. The dark matter density profile $\rho(r)$ in the NFW model is given by:
\begin{equation}
\rho(r) = \frac{\rho_0}{\frac{r}{r_s} \left(1 + \frac{r}{r_s}\right)^2}
\label{eq:dmrho}
\end{equation}
Here, $\rho_0$ is the characteristic inner density of the NFW halo, and is tied to the mass, concentration, and cosmic critical density of the halo at formation. $r_s$ is the scale radius, which defines the transition between the inner and outer regions of the halo. The NFW profile is widely used in galaxy formation and evolution studies, as it provides a good approximation to the results of cold dark matter simulations.  While this profile has its origin in N-body simulations, it may still be a good approximation to actual dark matter halos in galaxy-scale lenses. Whereas the presence of stars and gas leads to the density profile becoming steepened by adiabatic contraction due to baryon cooling, stellar and AGN feedback have the opposite effect of reducing the potential. Observationally, NFW appears to be a reasonable approximation at these scales for many lenses. \cite{shajib_dark_2021}  and \cite{sheu_project_2025} found that this profile seems to be a good fit for elliptical galaxies at $z = 0.2.$ This forms the motivation for using the NFW profiles to describe dark matter in this work.

\subsection{Bayesian Inference}

For kinematic modeling, we employ Bayesian analysis, in which the goal is to find a target posterior probability distribution $P(\Theta \mid D, M)$ of the parameters $\Theta$ given observed data $D$ under an assumed model $M$. The likelihood $P(D \mid \Theta, M) (\Theta)$, represents the probability of obtaining the data $D$ assuming a given set of parameters $\Theta$. The prior probability, $P(\Theta \mid M) $, encodes our initial knowledge or assumptions about the parameters before considering the data. The evidence, $P(D\mid M) $, acts as a normalization constant. These are related by Bayes' theorem, which is expressed as:
\begin{equation}
    P(\Theta \mid D, M) = \frac{P(D \mid \Theta, M) P(\Theta \mid M)}{P(D \mid M)} 
\end{equation}
To infer the galaxy parameters from the data, we utilize Dynesty \citep{speagle_dynesty_2020}, a Python-based Nested Sampling package that employs Bayesian inference to efficiently explore the parameter space and estimate the posterior distributions of the galaxy parameters. These parameters include the stellar mass, dark matter mass, dark matter scale radius, and anisotropy parameters, which describe the orbital distribution of stars. 

\subsection{Time Delay Cosmography and Hubble Parameter $H_0$ Estimation}

The Hubble parameter $H_0$, time delays and projected density of the lens are related by the following Equation \citep{kochanek_what_2002}:
\begin{equation}
    H_0 \propto \frac{1 - \kappa_E}{\Delta t}
    \label{eq:$H_0$}
\end{equation}
Where $\Delta t$ are the time delay differences  between images and $\kappa_E$ is the  normalized total projected mass density (dimensionless convergence) at the Einstein radius given by:
\begin{equation}
    \kappa_E = \frac{\Sigma(\theta_{E})}{\Sigma_{\text{crit}}}
    \label{eq:kappa_Ein}
\end{equation}
Here $\Sigma(\theta_{E})$ is the total projected surface mass density at the Einstein radius and $\Sigma_{crit}$ is the critical surface density for lensing described earlier. The crucial thing to note is that for an observed time delay measurement, the Hubble parameter $H_0$ depends on the total projected surface mass density at the location of the Einstein radii. 

\section{Data and Methods}
\label{sec:data_methods}
\subsection{Data}

We obtain our galaxies dataset from the state-of-the-art IllustrisTNG suite of simulations, a successor to the original Illustris Project. They are a suite of large scale cosmological magneto-hydrodynamical simulations within $\Lambda$CDM cosmology. They employ the moving-mesh code AREPO \citep{springel_e_2010}, which solves the equations of hydrodynamics alongside sub-grid models for star formation, stellar evolution and chemical enrichment, supernova feedback, black hole growth and AGN feedback.  The version of Illustris TNG simulations that we use is $TNG100$ \citep{nelson_illustristng_2019}, which corresponds to a cosmological box size of $100Mpc$. The final set of galaxies consists of a carefully chosen sample of $10$ massive elliptical galaxies at a redshift of $z = 0.2$. This redshift was selected keeping in mind its proximity with the redshift of several elliptical galaxy-scale lenses from the SLACS survey \cite{bolton_sloan_2006}. Our fiducial dataset consists of ten galaxies, with line-of-sight velocity dispersion maps with a peak signal-to-noise ratio of $60$ at the center. Unless otherwise indicated, all discussions refer to this fiducial dataset. 

\subsection{Sample Selection: Filtering Suitable Galaxies}

We identify galaxies from the simulation that can be potential deflectors in strong lensing systems and have properties suitable for spherical Jeans modeling. These are massive elliptical galaxies with low ellipticity that have most of their stellar mass in the bulge, and are slowly rotating. \cite{pulsoni_stellar_2020} discovered a lower than expected fraction of slow-rotating massive elliptical galaxies in the TNG50 simulations compared to the TNG100 simulations. This is the primary motivation for using TNG100 for this work. \cite{knabel_spatially_2024} found that a majority of their sample of strong lensing massive elliptical galaxies from the SLACS survey were slow rotators. In \cite{cappellari_structure_2016}, it is summarized that the first generation of IFS surveys found no spiral galaxies with stellar mass above $2\times10^{11}\,M_\odot$. The galaxies above this threshold are massive elliptical galaxies that are dominated by round or weakly triaxial slow rotators. They also note that the slower rotating massive ellipticals are found in the central regions of clusters or groups, with the slowest and most massive ones often being at the center of clusters or groups. Keeping all the above points in mind, for the first level of filtering, we select massive elliptical galaxies with stellar mass greater than this threshold of $2\times10^{11}\,M_\odot$. Next, we select only central or primary elliptical galaxies of the simulation. To ensure that the selected galaxies have low ellipticity, low angular momentum, and are bulge-dominated, we make use of the work done by \cite{genel_galactic_2015} in a TNG supplementary data catalog. They measure the ellipticity, stellar circularities (quantities which give the stellar mass fractions in the bulge and the disk), and the specific angular momentum (angular momentum per unit mass) of galaxies in the TNG simulations. The ellipticity is calculated for a radius which is twice the stellar half-mass radius, whereas the stellar circularities and the specific angular momentum are calculated within $10$ times the stellar half-mass radius. We remove more elongated galaxies by choosing galaxies that have ellipticity $\varepsilon$ lower than $0.3$. Next, we select galaxies that have a stellar mass bulge fraction greater than $0.9$ (excluding the contribution of disk stars). Finally we select slowly rotating ellipticals, by picking the $10$ galaxies with the least specific angular momentum (SAG) in the simulations after applying the above filters. We find that each of them has an SAG value less than $500$. 

\begin{figure}
    \centering
    \includegraphics[width=0.48\textwidth]{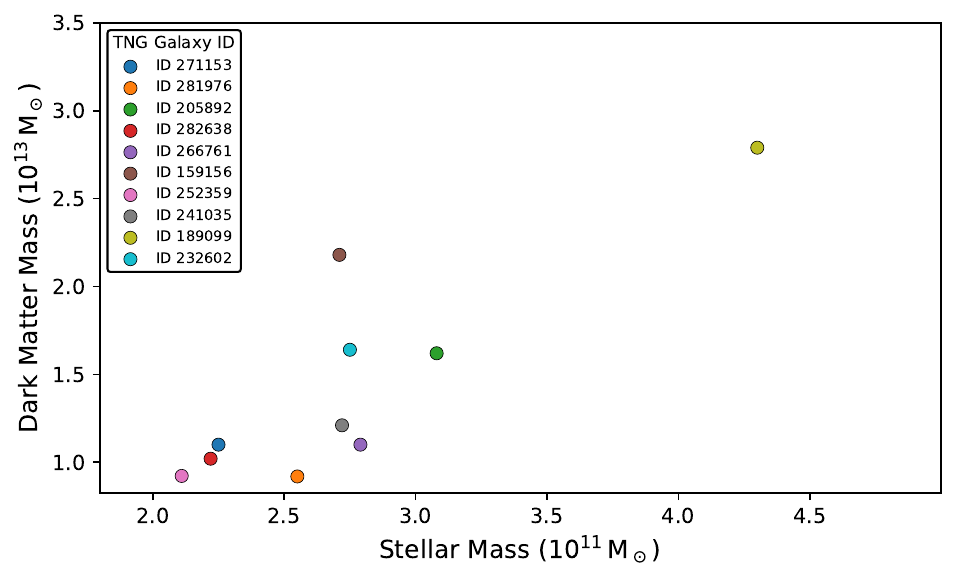}
    \caption{
    Stellar Mass versus Dark Matter Mass for the galaxies in the dataset. The galaxies are marked by different colors and the IDs of these galaxies in TNG100 simulations at $z = 0.2$ are as indicated. 
    }
    \label{fig:stellar_dm_mass}
\end{figure}

This leaves us with $10$ massive elliptical galaxies whose stellar and dark matter masses are shown in Figure \ref{fig:stellar_dm_mass}. In Table \ref{tab:galaxy_dataset_properties}, we summarize the dataset of ten galaxies and their properties. The first column represents the IDs of these galaxies at a redshift of $z = 0.2$. The remaining columns are their properties, and they are described in the table caption. In this work, we set a maximum radius of study for every galaxy --- considered by both the data and the model --- to $r^{\star}_{0.95}$, which is the spherical radius containing $95\%$ of the stars in the galaxy. This choice provides a realistic and computationally efficient upper limit. This is motivated by the fact that we found that in the simulations a small number of stars are present at extremely large distances, ex. several hundred kiloparsecs (kpc) from the galaxy center. We tested that accounting for every star is computationally slow and has a minimal effect on the data, since at these large distances $\nu(r) $ becomes vanishingly small and so does its contribution to the line-of-sight velocity dispersions via Equation \ref{eq:LOSVD}. The stars further away have more motion in the radial direction than along the line-of-sight. Put together, the contribution of these far away stars to the observables, the line-of-sight velocity dispersions is minimal. In general $r^{\star}_{0.95}$ varies for every galaxy.  The galaxies in the table have been sorted by their increasing specific angular momentum.

\begin{deluxetable*}{ccccccccc}
\tabletypesize{\footnotesize}
\decimals
\tablecaption{Properties of Galaxies Sorted by Increasing Specific Angular Momentum \label{tab:galaxy_properties}}
\tablecolumns{9}
\tablewidth{0.99\textwidth}
\tablehead{
\colhead{Galaxy ID} &
\colhead{$M_\star$} &
\colhead{$M_{\rm DM}$} &
\colhead{$R_{eff}$} &
\colhead{$r^{\star}_{0.95}$} &
\colhead{$e$} &
\colhead{BMF} &
\colhead{$j_\star$} &
\colhead{$M_{200}$} \\
\colhead{} &
\colhead{($10^{11}\,\text{M}_\odot$)} &
\colhead{($10^{13}\,\text{M}_\odot$)} &
\colhead{(kpc)} &
\colhead{(kpc)} &
\colhead{} &
\colhead{} &
\colhead{(kpc km s$^{-1}$)} &
\colhead{($10^{13}\,\text{M}_\odot$)}
}
\startdata
271153 & 2.25 & 1.10 & 4.61 & 84.85 & 0.20 & 0.978 & 74.86 & 1.09 \\
281976 & 2.55 & 0.92 & 11.53 & 122.86 & 0.25 & 0.967 & 138.22 & 1.03 \\
205892 & 3.08 & 1.60 & 9.92 & 151.69 & 0.26 & 0.974 & 204.54 & 2.03 \\
282638 & 2.22 & 1.02 & 9.37 & 127.23 & 0.20 & 0.914 & 217.12 & 1.03 \\
266761 & 2.79 & 1.10 & 8.14 & 102.84 & 0.29 & 0.967 & 249.57 & 1.25 \\
159156 & 2.71 & 2.18 & 9.59 & 169.16 & 0.19 & 0.930 & 271.46 & 2.92 \\
252359 & 2.11 & 0.92 & 7.95 & 107.44 & 0.21 & 0.993 & 299.19 & 1.08 \\
241035 & 2.72 & 1.21 & 8.44 & 125.84 & 0.27 & 0.965 & 326.36 & 1.23 \\
189099 & 4.30 & 2.79 & 14.87 & 214.97 & 0.22 & 0.967 & 334.10 & 3.40 \\
232602 & 2.75 & 1.64 & 9.06 & 143.95 & 0.24 & 0.977 & 371.56 & 1.97 \\
\enddata
\tablenotetext{}{\textbf{Note.} Total Stellar Mass $M_\star$ is in units of $10^{11}\,\text{M}_\odot$, Total Dark Matter Mass $M_{\rm DM}$ in $10^{13}\,\text{M}_\odot$, and Group Mass $M_{200}$ in $10^{13}\,\text{M}_\odot$. $R_{eff}$ is the projected (2D) stellar half-mass radius, while $r^{\star}_{0.95}$ is the 3D radius enclosing $95\%$ of the stars. $e$ is the ellipticity. BMF is the stellar bulge mass fraction. $j_\star$ denotes the specific angular momentum, and $M_{200}$ is the total mass within $r_{200}$, the radius enclosing an average density $200\,\rho_{\mathrm{crit}}$. The galaxies are sorted by increasing specific angular momentum.}
\label{tab:galaxy_dataset_properties}
\end{deluxetable*}

\subsection{\textbf{Constructing Mock Observations}}

We used the following procedure to obtain the data. For all the datasets and modeling in this study we always use the exact stellar mass and stellar velocity anisotropy parameter profiles of the galaxies in TNG. We get cutouts for the galaxies from the simulations at a redshift of $z = 0.2$. The stellar data includes the stellar mass, positions $x, y, z$, and velocities $v_x$, $v_y$ and $v_z$ of the individual stars. The dark matter portion includes the $x, y,z$ positions of the dark matter particles. The mass of each dark matter particle in the simulation is the same. We obtain this information for all the stellar and dark matter particles that lie within the radius $r^{\star}_{0.95}$.

\subsubsection{Obtaining the stellar velocity anisotropy parameter}

\begin{figure}[ht!]
    \centering
    \includegraphics[width=\columnwidth]{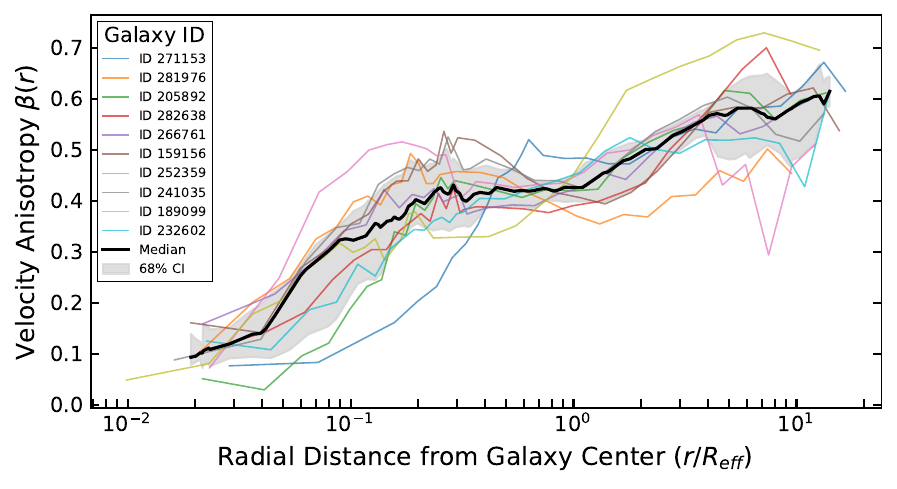}
    \caption{The stellar velocity anisotropy profiles of the galaxies in the dataset. The velocity anisotropy parameter $\beta(r)$ is plotted as a function of 3D distance r from galaxy center in units of the effective radius $R_{eff}$ of each galaxy. Individual galaxy curves are marked in color and the median anisotropy curve is in bold black. 
    \label{fig:anisotropy}}
\end{figure}

We transform the Cartesian velocities $v_x$, $v_y$ and $v_z$ of the stars obtained from the cut-outs to their spherical versions $v_r$, $v_\theta$, and $v_\phi$. We create concentric spherical 3D bins for the stars with the origin at the center of the galaxy, and subsequently find the velocity dispersions $\sigma_r$,  $\sigma_\theta$  and $\sigma_\phi$ for each bin. We then obtain the velocity anisotropy parameter $\beta(r)$ in each bin from the velocity dispersions using Equation \ref{eq:beta}. This lets us obtain the velocity anisotropy profile for the entire galaxy. The procedure is repeated for every galaxy in the dataset. Figure \ref{fig:anisotropy} shows the velocity dispersion profiles for all galaxies in the data set. We notice that in general,  the velocity anisotropy parameter starts off being fairly isotropic (or close to zero) in the center and begins to rise quite sharply with radial distance from the galaxy center. The anisotropy curves in Figure \ref{fig:anisotropy} show either a flattening or a slight dip in the approximately 2 - 10 kpc interval, likely due to the contribution from disk stars. Overall, we observe a significant radial variation of the anisotropy as a function of distance from the galaxy center, other examples of similar behavior include the the works of  \cite{dekel_lost_2005} and \cite{thomas_dynamical_2014} while studying massive ellipticals.
\subsection{Generating Dark Matter Profiles}

\begin{figure*}
    \centering
    \includegraphics[width=0.65\textwidth, height=0.6\textwidth]{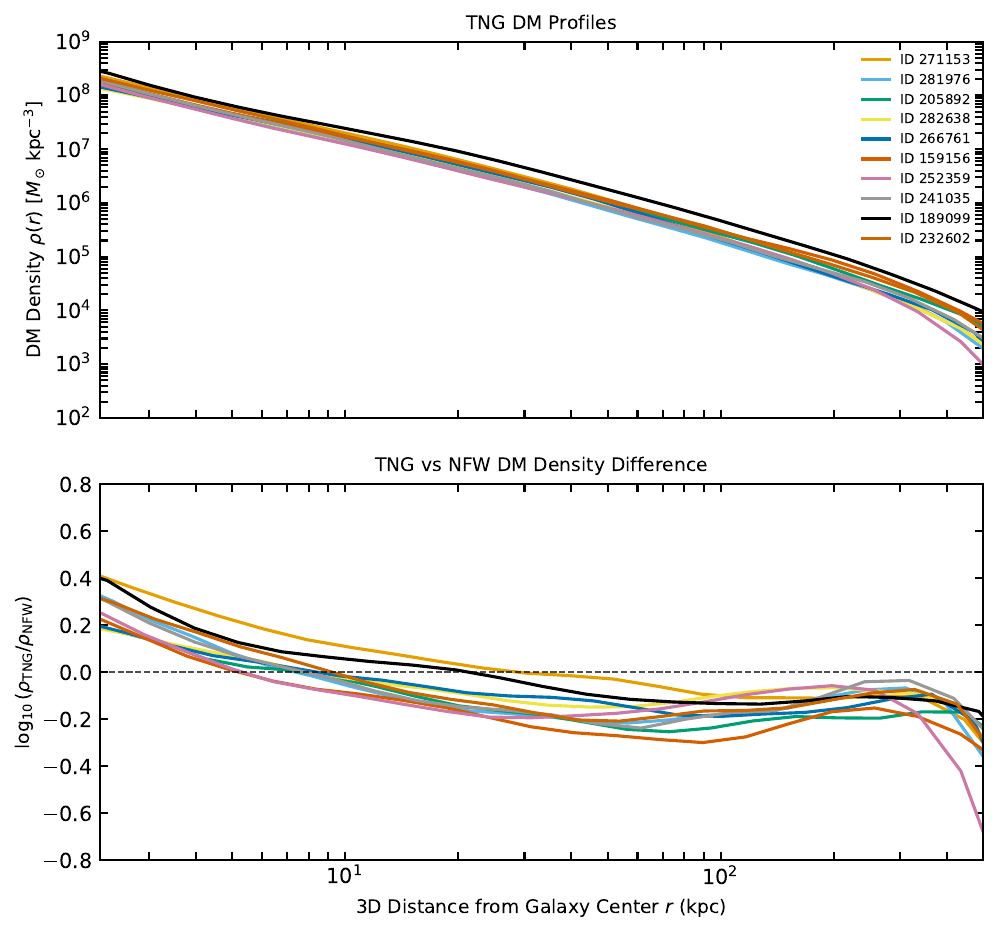}
    \caption{
    Comparison of Dark Matter Density Profiles $\rho(r)$ for the ten TNG galaxies in the Dataset. Both panels show $\rho(r)$ as a function of 3D radius $r$ from the galaxy center.  Also shown is the shared legend with the galaxies ID visible.
    \textbf{Top panel:} Dark matter density curves $\rho(r)$ of the galaxies from the native TNG simulation. 
    \textbf{Bottom panel:} Corresponding log ratio of TNG dark matter profiles of the galaxies  with their NFW counterparts derived via Colossus package using their redshift and M200 values.
}
    \label{fig:tng_vs_nfw_dm}
\end{figure*}

Using the mass profiles for dark matter from the cutouts, we can find the density profiles $\rho(r)$ of the galaxies in the dataset. We plot them in the upper panel of Figure \ref{fig:tng_vs_nfw_dm}. The left portion of the plot is cut-off at the convergence radius, below which numerical artifacts from the simulation become relevant. We estimate the convergence radius, by following \cite{power_inner_2003} and \cite{pillepich_simulating_2018} to be approximately \(2.8\,\epsilon_{\rm DM}\), where \(\epsilon_{\rm DM}=0.74\,\mathrm{kpc}\) is the gravitational softening length for dark matter in the TNG100 simulations. For radii less than the convergence radius, we found the dark matter density to flatten, indicating the presence of a small core (region of constant density). This core is potentially artificial, which arises due to resolution effects in the simulation \citep{pillepich_first_2018}. 
 
The primary goal of this study is to examine in detail the contribution of the velocity anisotropy parameter to galaxy modeling by isolating it as a systematic, quantifying any associated biases, and to examine ways to mitigate them. Because dark matter is another important systematic that is typically not known, we aim to avoid working with two unknowns simultaneously, which may make the results difficult to interpret. To achieve this while also getting rid of the central core, we transform the dark matter profiles of the TNG galaxies in the data set into exact NFW profiles while generating the simulated data. We also do our modeling runs with an NFW dark matter profile. The recipe for doing this is as follows.  We use the Python package Colossus \citep{diemer_colossus_2018} --- a framework for computing cosmological and halo-model quantities, including density profiles, mass–concentration relations, and cosmological distance functions. We selected the \cite{diemer_accurate_2019} model, which is known to provide accurate halo concentrations over a wide range of galaxy mass and redshifts. With the $M200$ values for each galaxy (Table \ref{tab:galaxy_dataset_properties}) and the galaxy redshift of $z = 0.2$ as input, we obtain the expected halo concentration for an NFW profile. From this, the characteristic density $\rho_0$ and the scale radius $r_s$ are found.  With the above information, we obtain the dark matter density profiles via Equation \ref{eq:dmrho}.

After doing so, the resulting dark matter profiles of the galaxies have the typical cusp associated with an NFW profile in the inner region, and we have thus removed the small artificial core. The difference (log ratio) of the dark matter density profiles of the TNG galaxies, with their NFW counterparts obtained using this prescription, is illustrated in the lower panel of Figure \ref{fig:tng_vs_nfw_dm}.  The profiles are truncated at \(r = 500\,\mathrm{kpc}\), which is approximately the virial radius \(r_{200}\) of these galaxies. Beyond this radius, the number of dark-matter particles decreases rapidly, resulting in a steep decline in the estimated density and increased susceptibility to numerical artifacts due to reduced particle statistics and simulation edge effects.

\subsection{Generating the line-of-sight Velocity Dispersion profiles}

We obtain the observable quantity, the 2D line-of-sight velocity dispersion profile $\sigma_{\text{los}}(R)$ for each galaxy from the stellar velocity anisotropy profile $\beta (R) $, the projected stellar number density profile $\Sigma(R)$, the 3D number density $\nu(r)$ and the velocity dispersions in the radial direction $ \sigma_r$ using Equation \ref{eq:LOSVD}, where we set $\infty$ to $r^{\star}_{0.95}$ for each galaxy (refer Table \ref{tab:galaxy_dataset_properties} for $r^{\star}_{0.95}$ values for each galaxy).

\subsection{ Details on obtaining data}

\subsubsection{Simulating a JWST NIRSpec Observation}

We generate mock spectroscopic data for each galaxy in the dataset by assuming a JWST observation using the NIRSpec instrument. The specifications of the NIRSpec are summarized in Table \ref{tab:JWST_IFU_specs}.

\begin{table}[h]
\centering
\caption{Specifications of \textit{JWST}'s NIRSpec instrument for mock IFU stellar kinematics.  We approximate the PSF as a Gaussian with $0.1''$ FWHM.}
\begin{tabular}{lc}
\hline\hline
 & \textit{JWST} \\
\hline
Instrument & NIRSpec \\
Pixel size & $0.1'' \times 0.1''$ \\
Field of view & $3'' \times 3''$ \\
PSF FWHM & $0.1''$\\
\hline
\end{tabular}
\label{tab:JWST_IFU_specs}
\end{table}

There are $900$ pixels in a $3''$ x $3''$ field of view, and the viewing window is set for a lens galaxy at redshift $z = 0.2$. To include the effects of realistic measurement error and to account for seeing, we follow a prescription similar to that in \cite{yildirim_time-delay_2020} and \cite{shajib_improving_2018}.  The effects of seeing are added by including the contribution of a point spread function (PSF) through Equation \ref{eq:conv} : 
\begin{equation}
\tilde{\sigma}^2_{\text{los}}(x, y) = \frac{I \sigma^2_{\text{los}} * g(x, y)}{I * g(x, y)}
\label{eq:conv}
\end{equation}
The surface brightness$(I)$ weighted line-of-sight velocity dispersions are convolved with a Gaussian PSF $g(x,y)$, after which they are normalized.  We approximated the PSF as a Gaussian with $0.1''$ (one pixel) FWHM. We set the signal-to-noise ratio (SNR) to be 60 in the central four pixels. This maximum Central SNR (CSNR) is the reference and the SNR for the remaining pixels $SNR_{\text{pixel}}(x, y)$ scales via the relation:
\begin{equation}
SNR_{\text{pixel}}(x, y) = CSNR \cdot \sqrt{\frac{I_{\text{conv}}(x, y)}{I_{\text{max}}}}
\label{eq:SNR}
\end{equation}
Here $I_{conv}(x, y)$ are the PSF convolved surface brightness values and ${I_{\text{max}}}$ is the maximum convolved surface brightness (i.e, at the four central pixels). As in \cite{yildirim_time-delay_2020}, we add a two percent measurement uncertainty to account for observational errors (e.g., due to stellar template mismatch). To add errors at each pixel, which are realistic, we draw a random number from a Gaussian distribution with mean $\mu$ zero and standard deviation $\sigma_{\text{stat}}$ given by the following equation:
\begin{equation}
\sigma_{\text{stat}} = \left( \overline{v^2_{\text{LOS}, l}} \right)^{1/2} \times \frac{1}{(SNR)_l}
\label{eq:measurement_error}
\end{equation}
Here, $\left( \overline{v^2_{\text{LOS}, l}} \right)^{1/2}$and $\frac{1}{(SNR)_l}$ are the velocity dispersions and signal-to-noise ratios at each pixel in the image, obtained using Equations \ref{eq:LOSVD} and \ref{eq:SNR}. We note that the resulting data obtained this way by using a prescription similar to \cite{yildirim_time-delay_2020} matched closely with the method used in \cite{shajib_improving_2018} who chose to employ a five percent measurement error.

\subsubsection{Exclusion of black holes and gas from data}

We found that the exclusion of black holes does not affect the line-of-sight velocity data. The sphere of influence for these blackholes is on the order of tens of parsecs, while even the innermost pixel is above two hundred parsecs, a distance within which the black hole mass is negligible compared to the stellar mass. The gas mass for some of the galaxies in our sample is higher than expected for massive elliptical galaxies which are often dry. However, the gas is highly diffuse and does not contribute meaningfully to the line-of-sight velocity dispersions --- a result also observed in \cite{mamon_dark_2005}. For an illustration of this, see the Appendix Section \ref{sec:appendix_gas}. For this reason, we exclude gas while obtaining data.

\section {Stellar Kinematics Modeling}
\label{sec:kinematic_modeling}
\subsection{Setting up the Model}

The goal of the galaxy model is to use the observable data --- the line-of-sight velocity dispersions and stellar light profiles --- to recover the stellar and dark-matter mass and stellar anisotropy components of the galaxy. In accordance with the earlier stated approach of minimizing modeling systematics, we assume that the stellar light profile is known and that the mass-light ratio is some constant but unknown quantity.

The galaxy model contains 4 parameters: The total stellar mass $M^*_{tot}$ within a sphere of radius $r^{\star}_{0.95}$, the dark matter mass $M_{DM<5kpc}$ which is the dark matter mass within a spherical radius of 5 kpc, the dark matter scale radius $r_{s}$ and the anisotropy radius $r_{ani}$. The radius $r^{\star}_{0.95}$ (which is different for each galaxy) is the threshold radius mentioned earlier that contains $95\%$ of the stars in the galaxy. The NIRSpec field of view for a galaxy at redshift $z = 0.2$ is approximately $5\ \mathrm{kpc}$ from the origin, and we set the dark matter mass parameter $M_{DM<5kpc}$ in line with this. The dark matter scale radius $r_s$ for the NFW profile is the second dark matter parameter. The velocity anisotropy $\beta(r)$ is modeled for 4 different cases. 

\subsection{Velocity Anisotropy Profiles}

We examine 4 different anisotropy profiles. These include the often used Osipkov-Merritt, the anisotropy profile used in \cite{mamon_dark_2005}, and the constant anisotropy profile. Lastly, we examine the results of generalizing the Osipkov-Merritt profiles. 

\subsubsection{Osipkov-Merritt (OM) Anisotropy}

A commonly used anisotropy profile for spherical Jeans modeling is the Osipkov-Merritt model (henceforth, OM anisotropy). It is often used in the study of elliptical galaxies, dwarf spheroidal galaxies, dark matter halos, and globular clusters. Its widespread use, e.g., in \citep{shajib_improving_2018, birrer_mass-sheet_2016} stems from the fact that it allows for radial variation of the anisotropy while being analytically tractable. The anisotropy has the form:
\begin{equation}
\beta(r) = \frac{r^2}{r^2 + r_{ani^2}}
\label{eq:om}
\end{equation}
At small radii r $\ll$ $r_{ani}$, the anisotropy $\beta(r)$ approaches $0$, which is the case for isotropic orbits. However, for very large r, r $\gg$ $r_{ani}$  $\beta(r)$ approaches $1$, i.e., highly radial orbits. Instead of having to calculate line-of-sight velocity dispersions using Equations  \ref{eq:sigma_r} and \ref{eq:LOSVD}, which is effectively a double integral, the advantage of this profile is that a single integral version exists, making it computationally simpler (refer Equations \ref{eq:LOSVD_Kernel_Eq} and \ref{eq:OsipkovMerrittKernel} in the Appendix).

\subsubsection{Mamon-Lokas (ML) Anisotropy}

Another anisotropy profile that allows for radial variation is the profile in \textbf{\cite{mamon_dark_2005}} (henceforth Mamon-Lokas or ML profile). Like the OM profile, its advantage lies in the fact that it allows for radial variation of the anisotropy. It has the form:
\begin{equation}
\beta(r) =\frac{1}{2} \frac{r}{(r + r_{ani})}
\label{eq:ml}
\end{equation}
At small radii, r$\ll$ $r_{ani}$,  $\beta(r)$ approaches $0$, as in the OM profile. However, for the limiting case of a very large r, r $\gg$ $r_{ani}$ , $\beta(r)$ approaches $0.5$, that is, radial orbits. Another benefit of this profile is that, like the OM anisotropy, a one-integral version for the LOS velocity dispersions also exists. (Refer Equations  \ref{eq:LOSVD_Kernel_Eq},  \ref{eq:AnisotropicKernel_ua_eq_1} and \ref{eq:AnisotropicKernel_ua_not_1} in the appendix). A glance at Figure \ref{fig:anisotropy} suggests that this profile might be more suitable for the stellar anisotropy in massive elliptical galaxies in TNG (\cite{mamon_dark_2005} also found this profile to be a better fit in N-body simulations than OM). 

\subsubsection{Constant Anisotropy}

Another commonly used anisotropy assumption is that the anisotropy is a global constant everywhere in space, that is, $\beta(r)$ = $\beta$ = constant for all r.  However, as Figure \ref{fig:anisotropy} suggests, the anisotropy is radially varying, and we shall examine the implications of this assumption. 

\subsubsection{Implementing a more flexible model: Generalized Anisotropy Profiles}

The aforementioned models have one thing in common, they are single-parameter anisotropy models with an anisotropy radius $r_{ani}$. What if we attempt to improve the model by making it more flexible? This can be done by \textit{generalizing} the Osipkov-Merritt or Mamon-Lokas model. The flexibility of the model is added by allowing the model to allow the anisotropy at the center of the galaxy where $r = 0$, (and $\beta  = \beta_{0}$) and at large radii (where $\beta = \beta_{\infty}$) to vary.  In theory, this flexibility should allow the models to better capture the inner and outer regions of the anisotropy. The generalized Osipkov-Merritt model (gOM) and the generalized Mamon-Lokas anisotropy models (gML) have the following form:
\begin{equation}
\beta_{gOM}(r) = \beta_0 + (\beta_\infty - \beta_0) \frac{r^2}{(r^2 + r_{ani}^2)}
\label{eq:gom_full}
\end{equation}
\begin{equation}
\beta_{gML}(r) = \beta_0 + (\beta_\infty - \beta_0) \frac{r}{(r + r_{ani})}
\label{eq:gml_full}
\end{equation}
Thus, these are 3 parameter models. Notably, for massive elliptical galaxies, $\beta_{0}$ usually has a very low value, nearly zero, for example, in the TNG simulations themselves as in Figure \ref{fig:anisotropy}, or as noted in \cite{mamon_dark_2005}. Here, we approximate $\beta_{0}$ to be zero and reduce the complexity of the model while preserving a realistic representation of the anisotropy. Consequently, we end up with two parameters for anisotropy: the anisotropy radius $r_{ani}$ and $\beta_{\infty}$. In our tests, we found both the generalized Osipkov-Merritt (gOM) and generalized Mamon-Lokas (gML) curves to fit the anisotropy curves of the galaxies in the dataset reasonably well. To avoid redundancy, we decided to pick the generalized Osipkov-Merritt model as the fourth model of investigation. Our choice is in part due to the popularity of the single-parameter Osipkov-Merritt model. The gOM model we used with $\beta_{0}$ set to zero thus has the form:
\begin{equation}
\beta_{gOM}(r) =   \beta_\infty \frac{r^2}{(r^2 + r_a^2)}
\label{eq:gom}
\end{equation}
\subsubsection{Mock Anisotropy Diagram}

\begin{figure}[ht!]
    \centering
    \includegraphics[width=\columnwidth]{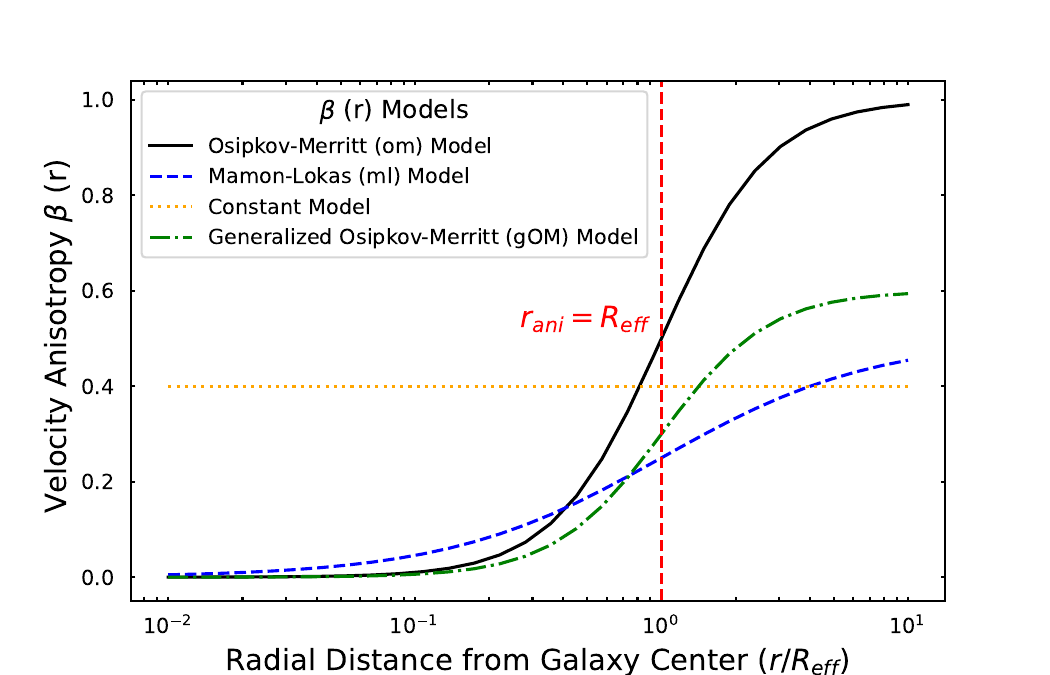}
    \caption{Mock example of the anisotropy profiles $\beta(r)$ of the four different models plotted as a function of the 3D radial distance r up to 10 effective radii $R_{eff}$. Here, the anisotropy radius $r_{ani}$ of the OM, ML and gOM models is assumed to equal the effective radius.  We assume that $\beta_{\infty}$ equals 0.6 for the gOM model and the constant model has $\beta$ equal to 0.4 everywhere in space.
    \label{fig:mock_beta_models}}
\end{figure}

To get a sense of the different anisotropy models, consider the following arbitrary example, a galaxy that has an effective radius $R_{eff}$ equal to 10 kpc. We plot the different anisotropy models in Figure \ref{fig:mock_beta_models} assuming that the OM, ML, and gOM have the anisotropy radius equal to the effective radius. For the constant model, we assume a global value of $\beta = 0.4$ at all radii, and for the gOM model, we assume that $\beta_{\infty}$ equals 0.6.

\subsection{Bayesian Inference Modeling using Nested Sampling}

We perform Bayesian inference via Dynesty \citep{speagle_dynesty_2020}, an open source Python package that makes use of nested sampling to obtain Bayesian posteriors and model evidence. The line-of-sight velocity dispersions for each galaxy are obtained via the procedure described earlier. The galaxy models incorporating the Osipkov-Merritt, Mamon-Lokas, and constant velocity anisotropies have 4 parameters in total, while the generalized Osipkov-Merritt model has an additional parameter $\beta_{\infty}$, making it 5 parameters in total.

\subsection{Details on Priors}

Given that the high quality mock data which is spatially resolved with high signal to noise ratio, we make very few assumptions about the priors. The stellar and dark matter masses are in units of solar masses ($M_\odot$), the dark matter scale radius and stellar anisotropy radius are in units of kiloparsecs (kpc), while $\beta $ for the constant model and $\beta_{\infty}$ for the gOM model are dimensionless. For the stellar mass, dark matter mass, dark matter scale radius, and anisotropy radius, we use uniform priors in log space. For constant anisotropy $\beta$ and for the parameter $\beta_{\infty}$ in the generalized Osipkov-Merritt profile, we use uniform priors.
Because the OM model reaches an unphysical asymptotic value of 1 away from the galaxy center (corresponding to purely radial stellar orbits which are unrealistic), the approach is often to set a lower limit on the anisotropy radius $r_{ani} $, for example, in \cite{sheu_project_2025}, to avoid arbitrarily small anisotropy radii which correspond to unstable stellar orbit configurations. We choose this lower limit to be equal to half the effective radius $R_{eff}$ for each galaxy. The ML and gOM models avoid unphysical orbits because the former approaches an asymptotic value of 0.5 at large radii, while the latter has freedom in the value that $\beta_{\infty}$ can take. We note that we also tried uniform priors on the anisotropy radius but the results did not change appreciably compared to the uniform priors in log space used here. In addition, we also tested informed priors on the dark matter scale radius, similar to those used in \citep{gavazzi_sloan_2007}, but the results did not change significantly. The priors used are summarized in Table \ref{table:priors}. 

\begin{deluxetable}{ll}
\tabletypesize{\small}
\tablewidth{\columnwidth}
\tablecaption{Priors for Model Parameters \label{tab:priors}}
\tablehead{
\colhead{Parameter} & \colhead{Prior}
}
\startdata
$\log(M_{\star}^{\mathrm{tot}}/M_\odot)$ & Uniform in log space over [10.5, 12] \\
$\log(M_{\mathrm{DM}<5\,\mathrm{kpc}}/M_\odot)$ & Uniform in log space over [9.5, 11] \\
$\log(r_s/\mathrm{kpc})$ & Uniform in log space over [1, 2.3] \\
$\log(r_{\mathrm{ani}}/\mathrm{kpc})$ & Uniform in log space over [$-0.5^*$, 2] \\
$\beta$ (constant model) & Uniform in [0.1, 1] \\
$\beta_\infty$ (generalized OM) & Uniform in [0.2, 1] \\
\enddata
\tablenotetext{}{\textbf{Note.} All logarithmic parameters use base-10. 
The $\beta$ prior applies only to the constant anisotropy model, 
while the $\beta_\infty$ prior applies only to the generalized Osipkov–Merritt model. 
\textbf{(*)} The lower limit of the prior for the anisotropy radius $r_{\mathrm{ani}}$ 
for the Osipkov–Merritt model is set to a value equal to half the effective radius $R_{eff}$ for each galaxy, 
whereas for the Mamon–Łokas and generalized Osipkov–Merritt models 
it is set to $10^{-0.5}$\,kpc.}
\label{table:priors}
\end{deluxetable}

\section{Modeling Results: Using Kinematics Only}
\label{sec:modeling_results_kinematics}

In this section, we summarize the results while modeling with only kinematic data, and in the subsequent section we present results from joint modeling.

\subsection{Modeling Overview}

We model the dark matter component with an NFW profile and test the data with the previously mentioned four different anisotropy profiles.  In this section, we model with the assumption that the elliptical galaxy at redshift $z_l = 0.2$ acts as a lens and there are sources at redshifts of $z_s$ equal to $0.6, 1, 2 \text{ and } 5$. We used purely kinematic data to estimate the bias in $H_0 $ that would result from combining kinematics with lensing time delays. We don't explicitly use time delay data, instead we assume that it is available and aim to find the relative bias in the Hubble parameter $H_0$ and assume its true value to be $H_0 = 70  \mathrm{km\ s^{-1}\ Mpc^{-1}}$.
The positions of the Einstein radii are obtained from the data for different source positions, by finding the radius where $\overline{\kappa}_{<E} = 1$ for each source-lens combination. 

The likelihood with kinematics only data is given by : 

\begin{equation}
\log \mathcal{L}_{\mathrm{kin}} =
-  \sum_{x,y}
\frac{\left[ \sigma_{\mathrm{los,model}}(x,y) -
\sigma_{\mathrm{los,data}}(x,y) \right]^2}
{2\sigma_{stat}^2(x,y)}  
\label{eq:logl_kin}
\end{equation}

Here $\sigma_{\mathrm{los,model}}(x,y)$ and $\sigma_{\mathrm{los,data}}(x,y)$ are the line-of-sight velocity dispersion values for the model and the data obtained after convolving with the PSF (Equation \ref{eq:conv}), and $\sigma_{stat}(x,y)$ is the measurement error for each pixel from kinematic modeling (Equation \ref{eq:measurement_error}) and the log refers to the natural logarithm (ln).\textbf{ }The summation is performed across the $900$ pixels in the NIRSpec field of view.

We apply a physical condition that $M^*_{tot} > M_{DM<5kpc}$, --- i.e., the total stellar mass of the galaxy be greater than the mass of dark matter within $5kpc$. This is motivated by the fact that massive elliptical galaxies have a significant stellar component ($> 2 \times 10^{11}  M_{\odot}$) as noted in \cite{cappellari_structure_2016}. We don't anticipate that the dark matter contribution within $5kpc$ --- a radius less than the stellar half-mass radius for the galaxies in the dataset --- would be greater than the total galaxy stellar mass. Works by \cite{gerhard_dynamical_2001, barnabe_two-dimensional_2011, oguri_stellar_2014} add confirmation that within the effective radius the stellar mass dominates the dark matter mass for massive elliptical galaxies. 

\subsection{One representative case: TNG Galaxy with ID 271153}

\begin{figure*}
    \centering
    \includegraphics[width=0.75\textwidth, height=0.7\textwidth]{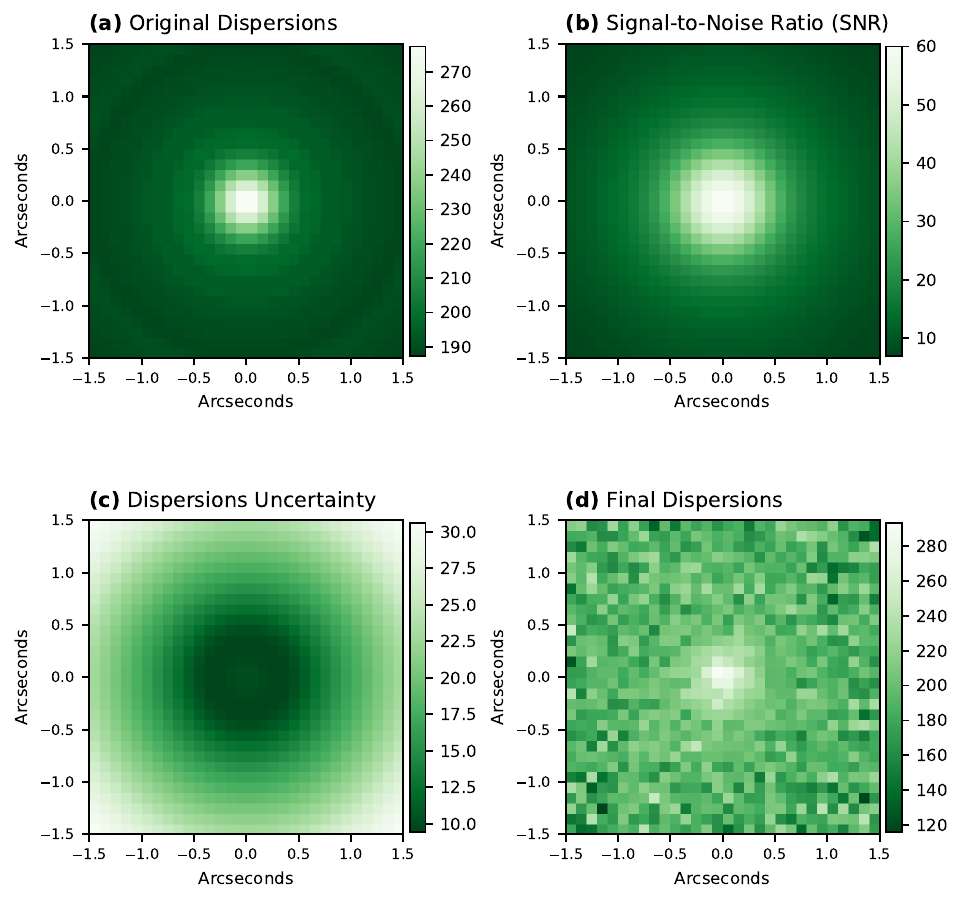}
    \caption{
   Simulated JWST NIRSpec maps for the reference galaxy at a redshift of $z = 0.2$. The dispersions refer to the line-of-sight velocity dispersions. The sub plots included are
    \textbf{(a)} Original dispersions, \textbf{(b)} Signal-to-noise ratio (SNR), \textbf{(c)} Uncertainty in the Dispersions, and \textbf{(d)} Final dispersions which include PSF effect and noise. 
    All panels share the same spatial scale in arc-seconds and correspond to a $3'' \times 3''$
 field of view.
}
    \label{fig:Four_in_One}
\end{figure*}

We continue the discussion by examining a specific galaxy, the galaxy with ID 271153 (corresponding to the first row of Table \ref{tab:galaxy_properties}). In Figure \ref{fig:Four_in_One} we illustrate the different components of the mock data. The top left plot shows the original line-of-sight velocity dispersions generated via Equation \ref{eq:LOSVD} in NIRSpec's 3" x 3" FOV. The top right shows the signal to noise ratio profile that peaks at $60$ at the center and falls according to Equation \ref{eq:SNR}.  In the bottom left is the measurement error obtained using Equation \ref{eq:measurement_error}. Lastly, the bottom right plot contains data that include effects of incorporating the PSF and measurement error. This is the final data seen by the model. The true values of the galaxy parameters for the reference galaxy are: $log_{10}M^*_{tot} = 11.32$, $log_{10}M_{DM<5kpc} = 10.48$, $log_{10} r_{s} = 1.77$. The stellar and dark matter masses have units of solar masses ($M_{\odot}$) while the dark matter scale radius has units of kiloparsecs (kpc).

\subsubsection{Results with Osipkov-Merritt (OM) Profile}

\begin{figure*}[ht!]
\gridline{
  \fig{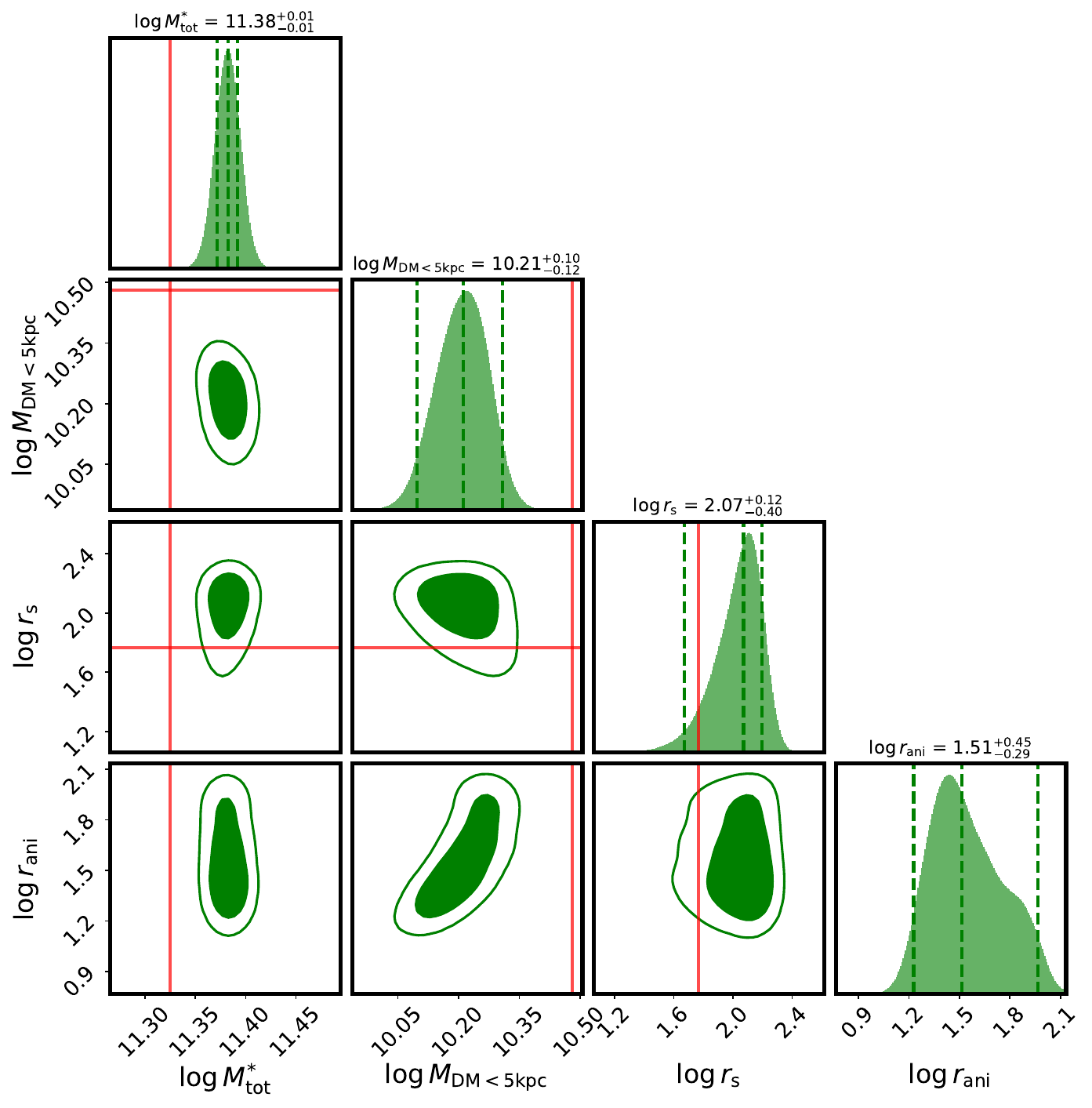}{0.45\textwidth}{\textbf{(a)} Osipkov-Merritt (OM)}
  \fig{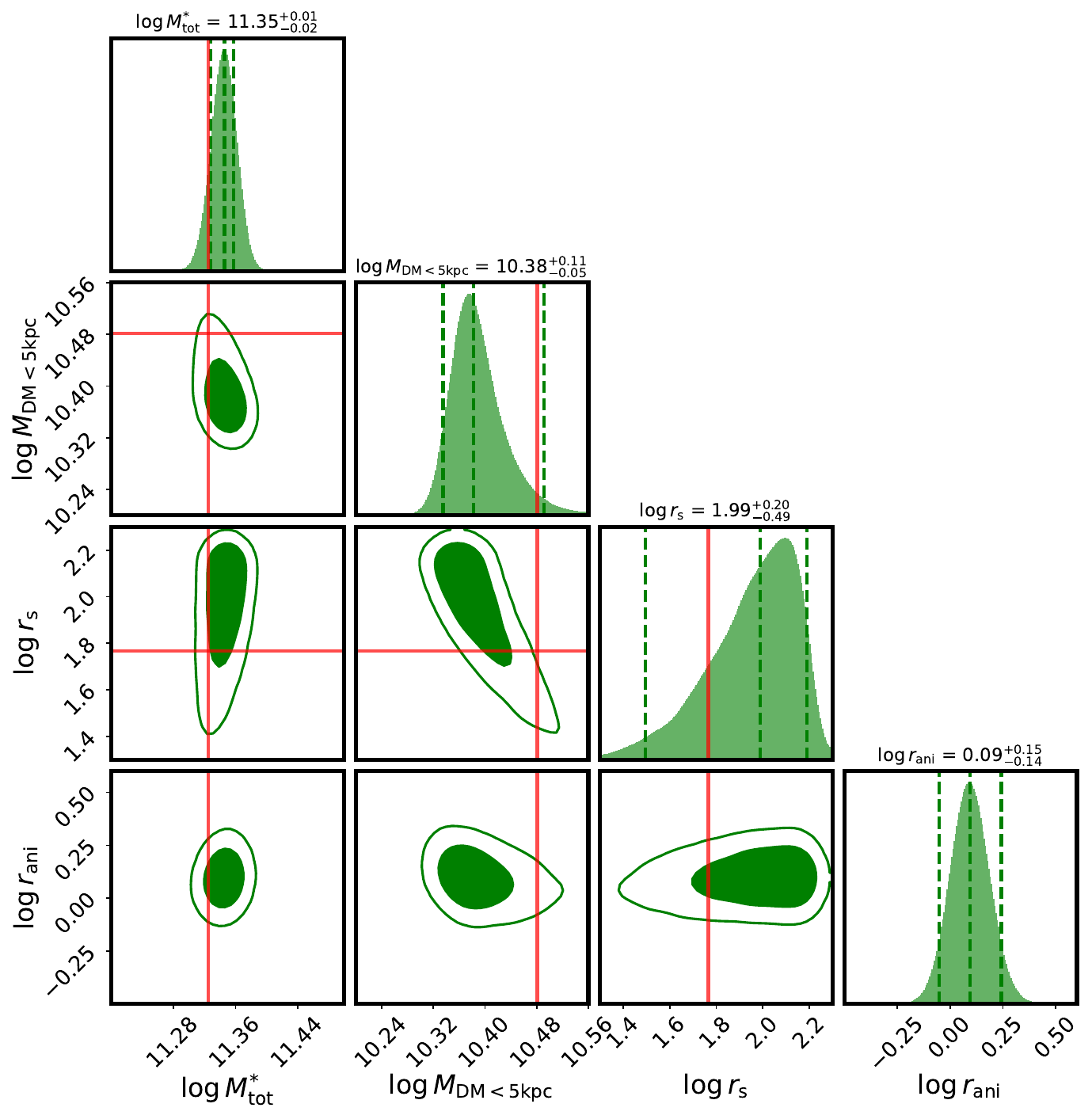}{0.45\textwidth}{\textbf{(b}) Mamon-Łokas (ML)}
}
\gridline{
  \fig{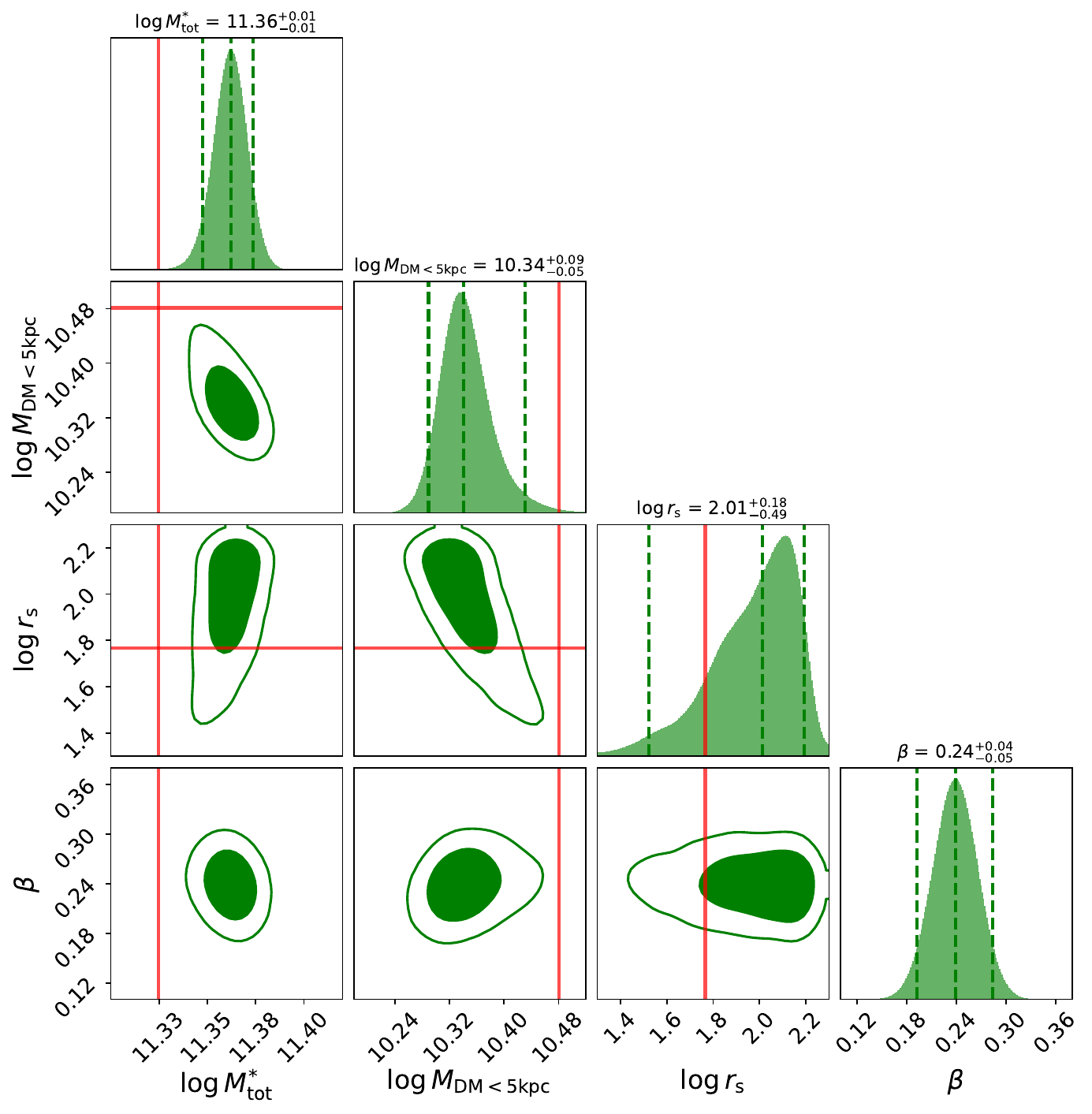}{0.45\textwidth}{\textbf{(c)} Constant}
  \fig{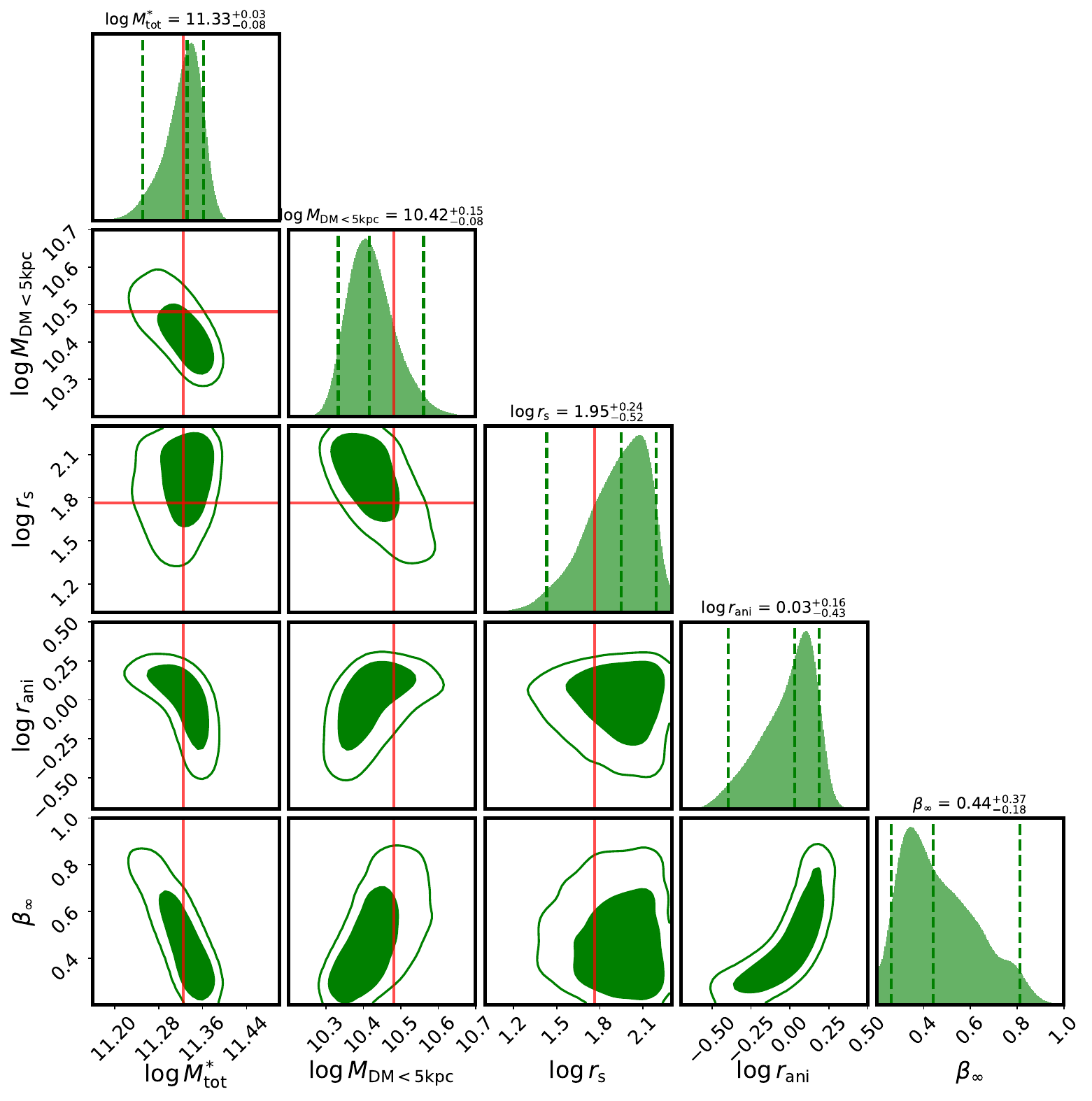}{0.45\textwidth}{\textbf{(d)} Generalized OM (gOM)}
}
\caption{Posterior distributions of the four anisotropy models for the reference galaxy. Red lines indicate true parameter values. The galaxy parameters with their units enclosed in brackets are: total stellar mass $M^*_{\mathrm{tot}}$ ($M_\odot$) within the radius $r^*_{0.95}$, dark matter mass within five kpc $M_{\mathrm{DM}<5\,\mathrm{kpc}}$ ($M_\odot$), dark matter scale radius $r_s $ (kpc).  The anisotropy parameters for the model are: anisotropy radius $r_{ani}$ (kpc) for the OM and ML models, and $\beta$ for the constant anisotropy. The gOM model has an additional anisotropy parameter $\beta_\infty$.  The 1D posteriors span the 95$\%$ credible interval. The inner and outer contours of the joint posteriors correspond to the 68$\%$ and 95$\%$ credible intervals.}
\label{fig:all_posteriors_four_in_one}
\end{figure*}

\begin{figure*}[ht!]
\centering
\includegraphics[width=\textwidth]{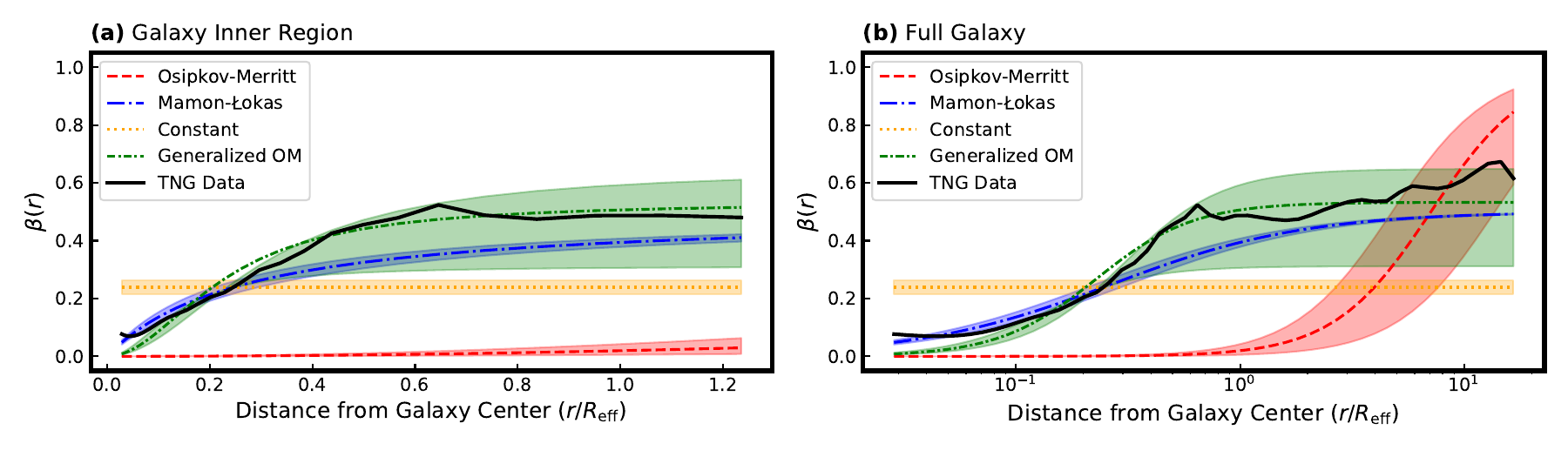}
\caption{
Comparison of Anisotropy profiles plotted as a function of the 3D distance r from the galaxy center, for the four different models (using best-fit parameter values) with the TNG data, for the reference galaxy with TNG ID 271153. The x-axis is scaled in units of the galaxy effective radius $R_{eff}$ which has a value of 4.61 kpc. 
\textbf{Left Panel:} $\beta(r)$ profile in the inner regions of the galaxy around the effective radius.
\textbf{Right Panel:}  $\beta(r)$ profile across the full galaxy.
Model lines represent Osipkov-Merritt (red-dashed), Mamon-Łokas (blue-dashed),
Constant (orange-dot), and Generalized Osipkov-Merritt (green-dashed).
The black solid line denotes the TNG data. 
}
\label{fig:combined_beta_profiles}
\end{figure*}

In the upper left panel of Figure \ref{fig:all_posteriors_four_in_one} we plot the posteriors of the nested sampling run. The results of the sampling indicate that the model  overestimates the stellar mass (posterior median $log_{10}M^*_{tot} = 11.38$) and the true value of the stellar mass lies well outside the 95$\%$ credible interval of the posteriors. At the same time, the model underestimates the dark matter mass (posterior median $log_{10} M_{DM<5kpc}$ = 10.21) whose true value also lies outside the $95\%$ credible interval of the model. Let us examine the anisotropy inferred by the model. Figure \ref{fig:combined_beta_profiles} compares the data and the best-fit anisotropy (with the $68\%$ error bars). The left sub-plot shows the inner region of the galaxy close to the effective radius, while the right one shows the entire range of the galaxy. Clearly, the Osipkov-Merritt model does a poor job of reproducing the true anisotropy curve in the interior regions of the galaxy, where it rises far too slowly, i.e., it prefers to be more isotropic. At large distances from the galaxy center (beyond ten effective radii), the model becomes extremely radial, although the contribution at these large radii to the dynamics in the inner regions is minimal.  What about the observables themselves, the line-of-sight velocity dispersions? The top panel of Figure \ref{fig:combined_los_dispersions_and_residuals} in the appendix shows the line-of-sight velocity dispersions of the best-fit model and the normalized residuals. The model fits the observables, the LOS velocity dispersions well overall (\textbf{$\chi^2_\nu = 1.04$}), within the limits of the measurement error. This highlights the fundamental mass-anisotropy degeneracy problem --- one can recover the  observable data, while still inferring incorrect masses (stellar, dark matter or both) and anisotropy profiles. 

\subsubsection{Results with Mamon-Lokas (ML) Profile}

The nested sampling posteriors for the ML profile are shown in the upper right of Figure \ref{fig:all_posteriors_four_in_one}. While the model is still struggling to capture the true values of the stellar and dark matter parameter masses, it seems to fare better than the OM model. The model stellar and dark matter masses lie close to the edge of the $95\%$ credible interval. As Figure \ref{fig:combined_beta_profiles} suggests, the model has a better fit for the anisotropy profile, particularly in the inner regions where it increases rapidly. The 68$\%$ credible interval suggests that the anisotropy radius is tightly constrained.  Like in the OM case, the model fits the observables,  the line-of-sight velocity dispersions well ($\chi^2_{\nu}$ = $0.94$) as in Figure \ref{fig:combined_los_dispersions_and_residuals}. 

\subsubsection{Results with Constant Profile}

For the constant model, the anisotropy $\beta$ is assumed to be the same everywhere in space. The posteriors for in the bottom left of Figure \ref{fig:all_posteriors_four_in_one} indicate that the true masses lie outside the $95\%$ credible interval. However, the mass biases are reduced compared to the OM model. The model's line-of-sight velocity dispersions ($\chi^2_{\nu}$ = $0.95$) and the residuals are as shown in the third panels of Figures \ref{fig:combined_los_dispersions_and_residuals}. Although, as indicated in Figure \ref{fig:combined_beta_profiles}, the assumption of constant anisotropy in the model is a poor fit to the true anisotropy curve at different radii from the galaxy center, the model performs better than a radially varying OM. 

\subsubsection{Results with Generalized Anisotropy Profile} 

Lastly, let us look now at the results of modeling with the generalized Osipkov-Merritt model for the representative TNG galaxy. In the posteriors plotted for this case in the bottom right of Figure \ref{fig:all_posteriors_four_in_one}, we see a marked improvement over the previous models, as the model captures the true stellar mass and dark matter parameters within the $95\%$ credible interval. As Figure \ref{fig:combined_beta_profiles} indicates, this model captures the shape of the anisotropy both in the inner and outer regions of the galaxy. The model's line-of-sight velocity dispersions ($\chi^2_{\nu}$ = $0.95$) and the residuals are shown in the last panel of Figure \ref{fig:combined_los_dispersions_and_residuals}.

\subsection{$H_0$ Inference Method }

We now examine what these results mean for $H_0$ inference when the elliptical galaxy being modeled acts as a lens in a strong lensing system. A key quantity in time delay cosmography is the total projected density at the Einstein radius $\kappa_{E}$.
As described in \cite{kochanek_over-constrained_2020}, the Hubble parameter $H_0$ and the time delays between the images $\Delta t$ are related by the equation:

\begin{equation}
H_0 \propto \frac{1 - \kappa_{\text{E}}}{\Delta t}
\label{eq:$H_0$}
\end{equation}

Where $\kappa_{E}$ is the dimensionless convergence given by $\kappa_{E}$ = $\Sigma (\theta_{E})/\Sigma_{crit}$, $\Sigma (\theta_{E})$ is the total projected surface mass density at the Einstein radius and $\Sigma_{crit}$ is the previously mentioned critical surface density. More precisely $H_0$ is proportional to the average projected density within the annulus between images, which is well approximated by $\kappa_{E}$ \citep{kochanek_what_2002}. This implies that, given a particular source-lens configuration, if the time delay differences between the images are measured, one can infer $H_0$ if the total projected surface mass density at the Einstein radius is calculated.  If the total projected density at the Einstein radius as inferred by the model differs from the true value, a similar bias can be expected in the inference of $H_0$. This is addressed further in Section \ref{sec:discussions}. We assume that sources exist at different redshifts of $z_s = 0.6,\ 1,\ 2, \text{and}\ 5$, while the elliptical galaxy acting as a lens is at $z _l= 0.2$. We examine and plot the projected density profiles of the data and model at the Einstein radii for these different configurations. As mentioned earlier, we do not use time delay data directly; instead, we assume the true value of $H_{0} = 70 \mathrm{km\ s^{-1}\ Mpc^{-1}}$ and aim to find the relative bias in $H_0$ according to:
\begin{equation}
\frac{\delta H_0}{H_{0,\mathrm{true}}} = \frac{\kappa_{E,\mathrm{true}} - \kappa_{E, model}}{1 - \kappa_{E,\mathrm{true}}}.
\label{eq:H0_bias}
\end{equation}
Lastly, we note that if $H_0$ is varied during fitting, the length scales corresponding to the pixel positions and the Einstein radius will change, as will the projected densities; however, in Appendix \ref{sec:varying_H0} we show that the estimated convergence $\kappa_{E,model}$ remains invariant as $H_0$ is varied. Hence, equation \ref{eq:H0_bias} for the bias in $H_0$ holds even though we do not explicitly vary $H_0$ during the kinematics analysis.

\subsubsection{$H_0$ Inference for the Representative Galaxy}

\begin{figure*}[ht!]
\gridline{
  \fig{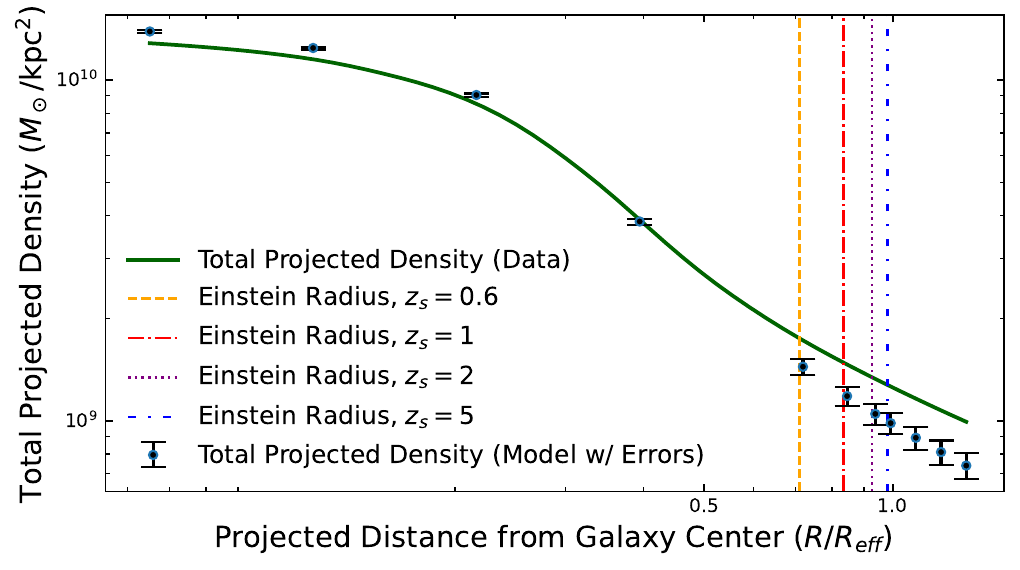}{0.45\textwidth}{\textbf{(a)} Osipkov-Merritt (OM)}
  \fig{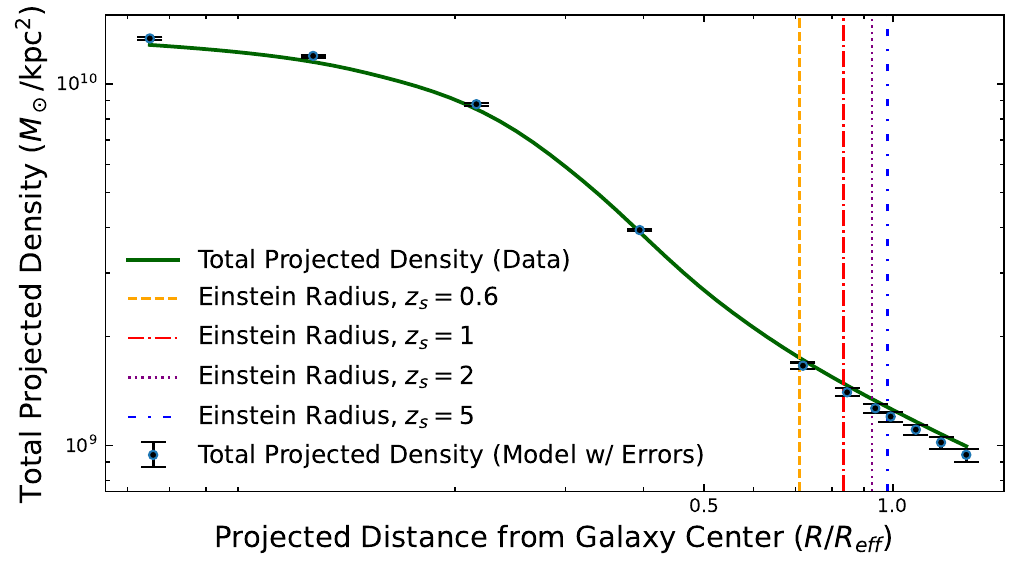}{0.45\textwidth}{\textbf{(b)} Mamon-Łokas (ML)}
}
\gridline{
  \fig{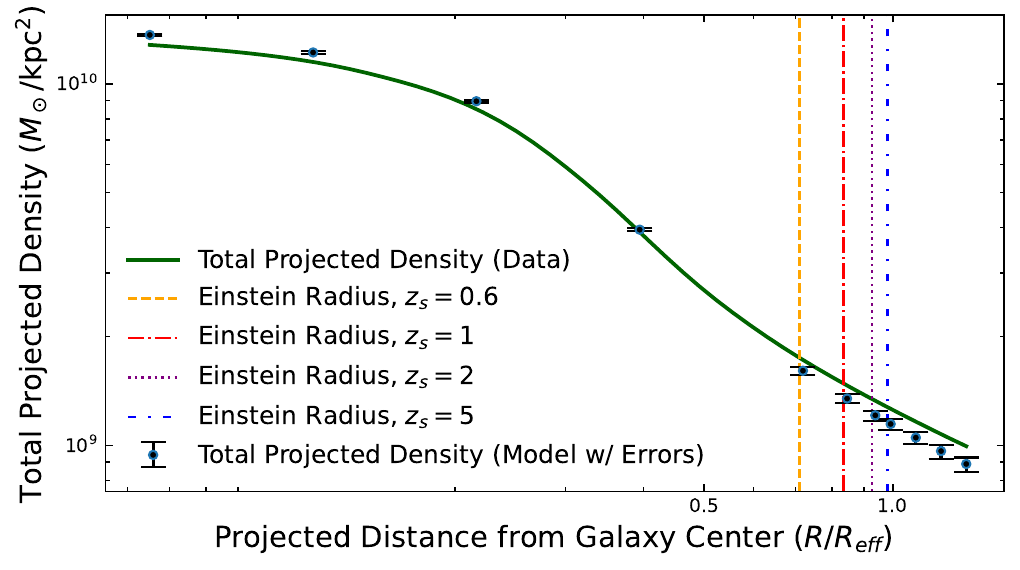}{0.45\textwidth}{\textbf{(c)} Constant}
  \fig{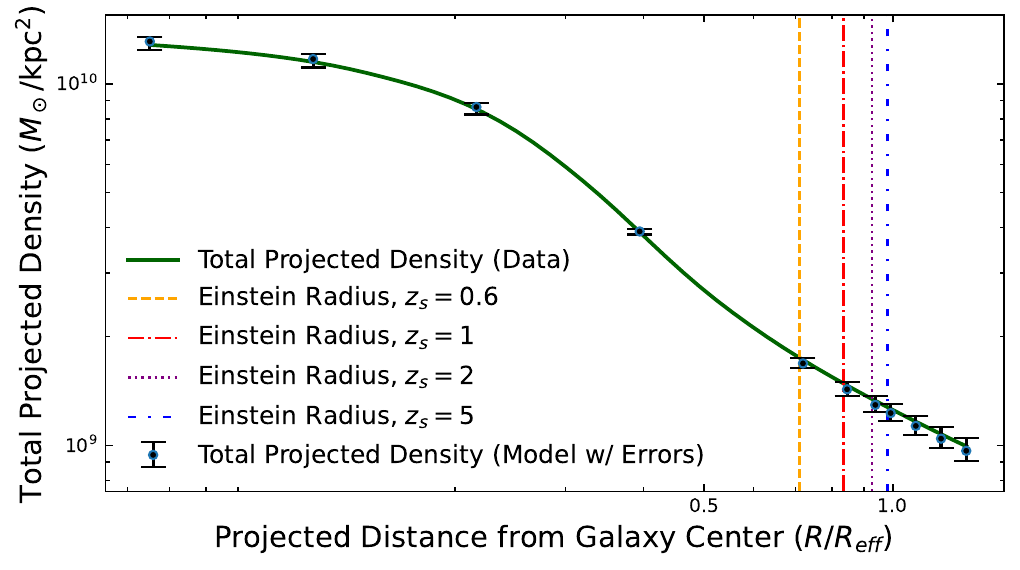}{0.45\textwidth}{\textbf{(d)} Generalized OM (gOM)}
}
\caption{Total projected density profile comparisons between the TNG data and the four different anisotropy models for the reference galaxy, plotted as a function of the 2D distance R. The x-axis is scaled in units of the effective radius $R_{eff}$ which has a value of 4.61 kpc for this galaxy.
The solid green line shows the data, and the black points with error bars represent the models with the 68\% credible interval. 
Vertical lines indicate the Einstein radius positions for sources at redshifts $z_s = 0.6, 1, 2$, and $5$, with different line styles for each. 
The reference galaxy is at $z = 0.2$.}
\label{fig:all_proj_densities_four_in_one}
\end{figure*}

The value of the Hubble parameter $H_0$ inferred for the reference galaxy can be found using the projected density profiles for each model, illustrated in Figure \ref{fig:all_proj_densities_four_in_one}. The positions of the Einstein radii for the sources at different redshifts $z_s = (0.6,\ 1,\ 2,\ 5)$  are indicated by the vertical lines, which in the units of the effective radius $R_{eff}$ are $(0.72, 0.85, 0.94, 0.99)$ respectively. 
We find that the Osipkov-Merritt model performs poorly, the true projected density at the Einstein radius lies well outside the 68$\%$ confidence region for the model, greater than $2\sigma$ away for the closest source at $z = 0.6$. This effect seems pronounced for sources further away whose Einstein radii are larger and the model extrapolates worse. For the farthest source at $z=5$, the true projected density is more than $3.5\sigma$ away.

The Mamon-Lokas  and constant models are less biased, with the true projected density at Einstein radius being $1.2\sigma$ and $1.8\sigma$ away respectively for the closest source, compared to the model. The change in the bias for sources across different redshifts is much smaller for these two models.
The generalized Osipkov-Merritt model shows the least bias, and the true total projected density at the Einstein radii is captured within the model's $68\%$ credible interval for sources at all redshifts, is within about $0.5\sigma$ and does not vary appreciably for the sources at different redshifts.

To quantify the effect on $H_0$ inference, we can use the aforementioned Equation \ref{eq:H0_bias}, to find the percentage bias in $H_0$, using the median values of the total projected density of the model and the true projected density at the Einstein radius. For this galaxy, the trend is that any $H_0$ bias gets worse for sources further away across all models. The OM model shows a significant bias of  $13.8\%$ for the closest source at a redshift of $z_s = 0.6$ which increases to  $19.6\%$ for the farthest galaxy at a redshift $z_s = 5$.
The ML model performs better with bias ranging from $3.6\%$ to about $5\%$. For the constant model, the bias lies between $6\%$ and about $8.5\%$. The generalized Osipkov-Merrit model shows the least bias of $2.4\%$ for the closest source, which increases modestly to a maximum of $3.1\%$ for the most distant source.
Furthermore, this model captures the true value of $H_0$ within the $68\%$ credible interval for all source redshifts.

\begin{figure*}
    \centering
    \includegraphics[width=0.75\textwidth]{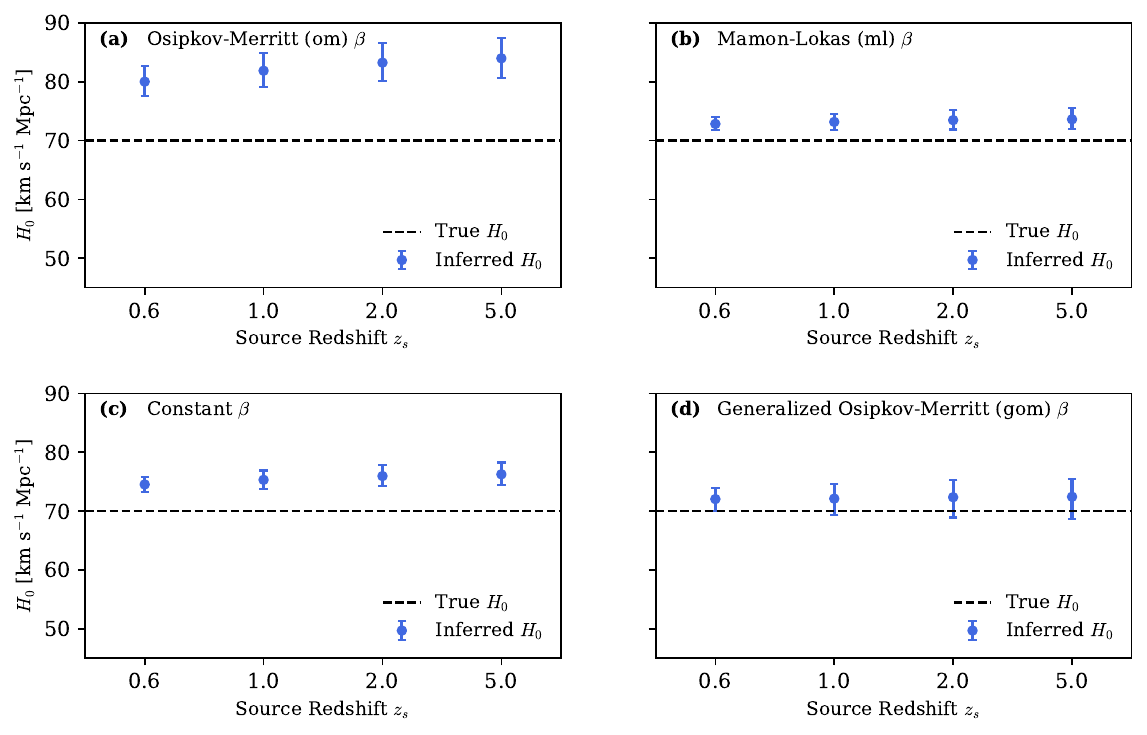}
    \caption{Hubble inference results for the four different anisotropy $\beta$ models for the reference galaxy at redshift $z_l = 0.2$, which acts as a gravitational lens. Each panel corresponds to a distinct anisotropy $\beta$ model: \textbf{(a)} Osipkov-Merritt (om),\textbf{ (b)} Mamon-Łokas (ml), \textbf{(c)} Constant, and \textbf{(d)} Generalized Osipkov-Merritt (gom).  The dashed line indicates an assumed true value of $H_0 = 70~\mathrm{km\,s^{-1}\,Mpc^{-1}}.$ Points show the median inferred $H_0$ values with 68\% credible intervals for each of the four different source redshifts $z_s = (0.6,\ 1,\ 2,\ 5)$ considered. Differences between the models reflect the impact of  anisotropy assumptions on $H_0$. The modeling in this figure uses only kinematic data while assuming that time delay information is available.
 }
    \label{fig:$H_0$_beta_models}
\end{figure*}

\subsubsection{$H_0$ Inference for all Galaxies in the Dataset}
 
\begin{figure*}[ht!]
\centering
\includegraphics[width=0.75\textwidth]{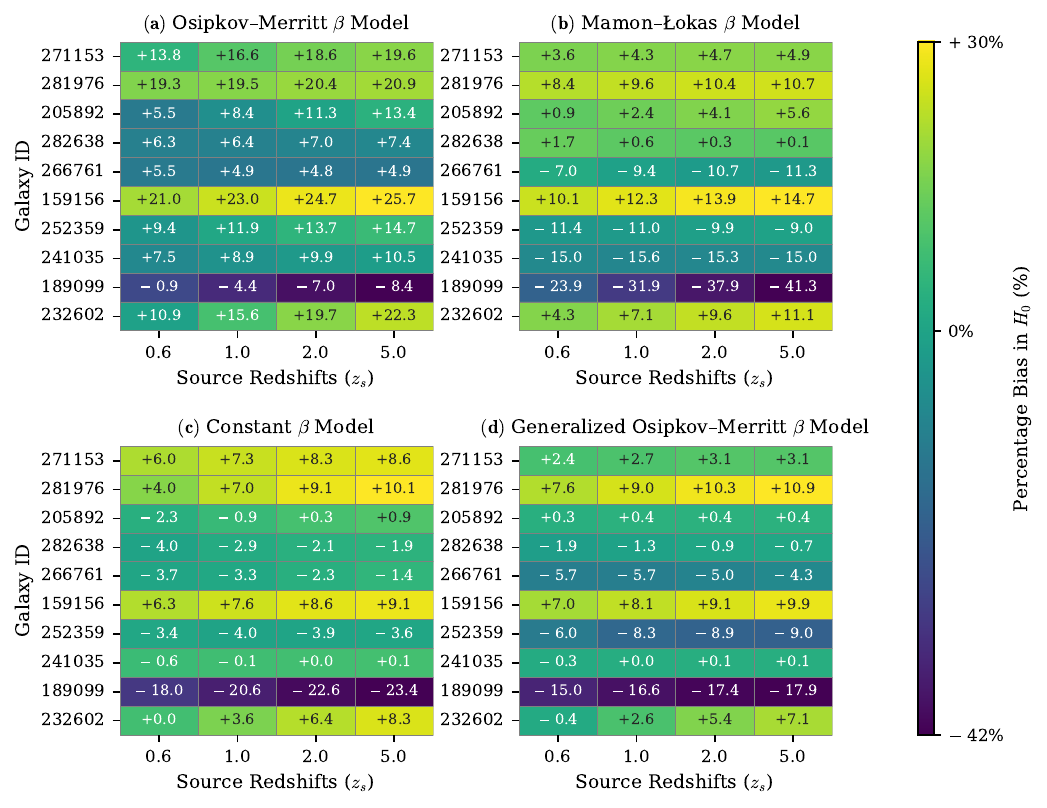}
\caption{
Results of  $H_0$ inference for all the galaxies in the dataset. 
Each of the 4 subplots correspond to a different anisotropy model indicated by the label at the top \textbf{(a)}~Osipkov-Merritt, \textbf{(b)}~Mamon-Łokas, \textbf{(c)}~Constant, and \textbf{(d)}~Generalized Osipkov-Merritt. Rows correspond to galaxy IDs, and columns span different source redshifts $z_s = (0.6,\ 1,\ 2,\ 5)$. Each pixel corresponds to percentage bias in the $H_0$ estimate for a given galaxy using an anisotropy model and the source at the indicated redshift.
The results highlight the mass anisotropy degeneracy and its effect on the Hubble inference. The assumed true value of the Hubble parameter is $H_0 = 70~\mathrm{km\,s^{-1}\,Mpc^{-1}}$. The modeling in this figure uses only kinematic data while assuming that time delay information is available.
}
\label{fig:hubble_kin_only_abs_vals_map_all_galaxies}
\end{figure*}

\begin{figure*}[ht!]
\centering
\includegraphics[width=0.75\textwidth]{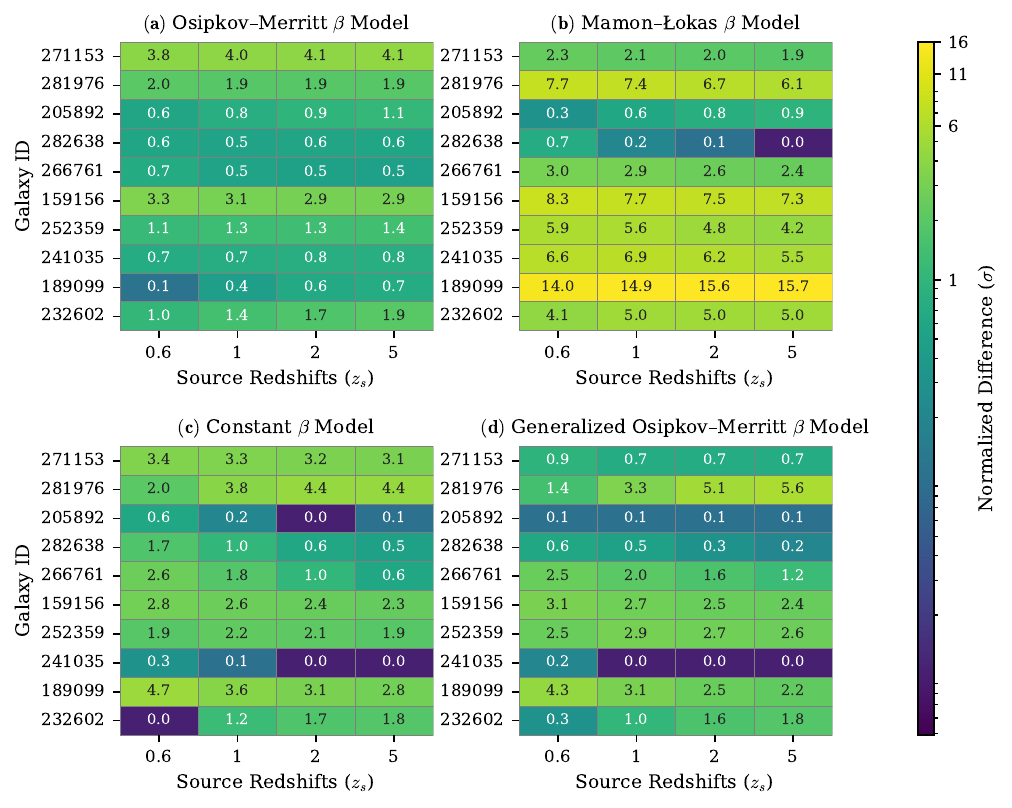}
\caption{
Normalized bias in inferred $H_0$ across galaxies and source redshifts for the four anisotropy models:
\textbf{(a)}~Osipkov-Merritt, \textbf{(b)}~Mamon-Łokas, \textbf{(c)}~Constant, and \textbf{(d)}~Generalized Osipkov-Merritt. Rows correspond to galaxy IDs, and columns span different source redshifts $z_s = (0.6,\ 1,\ 2,\ 5)$.
Each pixel shows the absolute value of $(H_{0,\ \text{model}} - H_{0,\ \text{data}})/\sigma_{H_0}$, where $\sigma_{H_0}$ is the 68$\%$ posterior uncertainty on the model-inferred $H_0$.
The modeling in this figure uses only kinematic data while assuming that time delay information is available.
}
\label{fig:hubble_kin_only_error_bars_map_all_galaxies}
\end{figure*}

We now summarize the results of the all the galaxies in the data set across the different anisotropy models and calculate the Hubble bias for each run in the same manner as the representative galaxy. In Figure \ref{fig:hubble_kin_only_abs_vals_map_all_galaxies} we plot the absolute value of $H_0$ inferred for each of the anisotropy models, and in Figure \ref{fig:hubble_kin_only_error_bars_map_all_galaxies} we show the normalized deviations between the model and the data using four different heatmaps. From these two plots, we can make the following observations. The Osipkov-Merritt model shows a high bias for nearly every run.  The tendency of this model is to systematically overestimate $H_0$ well above $70  \mathrm{km\ s^{-1}\ Mpc^{-1}}$ for most cases, while severity varies. In some cases, the bias is extremely high, greater than $20\%$. Such large biases born from the mass anisotropy degeneracy are problematic for addressing both the mass sheet degeneracy and the Hubble tension. In general, the results show a high bias in both inferred absolute values and in the number of normalized deviations between the model and the data. For some cases, the true $H_0$ lies beyond $3\sigma$.  The primary reason for this is that the model's anisotropy profile differs significantly from the true anisotropy profile, and the model generally prefers a large anisotropy radius, corresponding to more isotropic orbits in the central regions. We also observe significant scatter in the inference across different source redshifts, where the change in bias in the Hubble inference between the closest redshift source ($z_s =0.6$) and the farthest one ($z_s = 5$) can be quite high.

The Mamon-Lokas anisotropy performed better, and we see a general reduction in the biases. However, there are still several cases with biases greater than $10\%$. This model faces a similar problem to Osipkov-Merritt, that it is highly constrained, and we observe that the true value of $H_0$ lies beyond 3$\sigma$ for several cases. There is comparatively less scatter in the bias for sources at different redshifts.

Despite the strongly radial nature of the actual anisotropy curves (Figure \ref{fig:anisotropy}), the constant anisotropy yields a good fit to the data. For a few cases, the biases are under $5\%$ while for nearly all the others (except the outlier galaxy with ID $189099$), the bias lies between $5-10\%$. Like in the Mamon-Lokas case, the biases in general are worse for sources farther away. The model exhibits modest biases, generally under $3\%$ between sources at different redshifts. For most cases, the true value of $H_{0}$ captured by the model lies within $3\sigma$ of the true value. 

Lastly, we find that the gOM model recovers the true galaxy parameters better than any other model for the ten galaxies. Most results can be found with biases less than $5\%$, and almost all lie within the $10\%$ range, except for the galaxy with ID $189099$. Being a two-parameter model, it is less constrained and the size of credible intervals for model parameters and $H_0$ is larger. The true $H_0$ lies within 1$\sigma$ for several cases and within $3\sigma$ for nearly all cases. The scatter in the results between sources at redshifts was minimal, and on average the change in the bias between the closest and farthest sources was about $2\%$.

The three galaxies with IDs 281976, 159156, and 189099 exhibit systematically larger biases. This behavior is likely driven by a combination of their anisotropy structure, the location of their Einstein radii, and their large projected stellar half-mass radii $R_{\mathrm{eff}}$ (Table~\ref{tab:galaxy_properties}). The inferred results are most sensitive to the dynamical region around $R_{\mathrm{eff}}$, which contributes significantly to the line-of-sight velocity dispersion measurements. In the range $0.2$--$1.5\,R_{\mathrm{eff}}$ (Figure~\ref{fig:anisotropy}), the $\beta(r)$ curves for galaxies $281976$ and $159156$ drop sharply and becoming less radial, while for galaxy $189099$ it increases precipitously. In contrast, the anisotropy profiles assumed by the models are smooth and gradually increasing, preventing them from capturing either the dip in $\beta(r)$ for the first two galaxies or the sharp rise for the third. As a result, the models compensate for this mismatch through adjustments to the mass profile, leading to biased inferences which we believe is the likely explanation for their large biases. In general, biases in the inferred mass profile and projected density grow with radius. Galaxies $159156$ and $189099$ have the largest Einstein radii owing to their higher dark-matter fractions (Table~\ref{tab:galaxy_properties}). Because the inferred Hubble parameter depends on the total projected density at the Einstein radius (Equation \ref{eq:$H_0$}), this helps explains their larger $H_0$ biases.

For the closest source, when considering the ten galaxies, the average biases were: $+9.8 \pm 2.1\%$ (OM), $-2.8  \pm 3.5\%$ (ML), $-1.6\pm 2.2\%$ (constant) and $-1.2  \pm 2.1\%$ (gOM), thus highlighting the high systematic bias in the OM model. The median deviations between true $H_0$ and the model inferred $H_0$ using the posterior uncertainties were: $0.9\sigma$ (OM), $5.0\sigma$ (ML), $2.0\sigma$ (constant) and $1.2\sigma$ (gOM). 
Among the four models, gOM provides the most accurate recovery of $H_{0}$ and exhibits the lowest bias relative to the inferred uncertainties.

We plot the inferred stellar and dark matter masses for the four models and all the galaxies in Figures \ref{fig:M_star_model_vals} and \ref{fig:M_dm_model_vals}. The remaining parameters are tabulated in the Tables \ref{table: om_new}, \ref{table: ml_new}, \ref{table: constant_new} and \ref{table: gom_new}  with relevant columns for the pure kinematics modeling as indicated.


\section{Joint Modeling Results: With Kinematics and Strong Lensing}
\label{sec:modeling_results_kin_plus_lensing}

In this section, we combine lensing information into modeling. In the case of a fully formed ring, the Einstein radius $\theta_E$ is directly measurable with high-resolution imaging. The most reliable constraint we can obtain from strong lensing is the total projected 2D mass within the Einstein radius \citep{schneider_gravitational_1992}. This can be obtained by enforcing the condition for the formation of an Einstein ring, that the average convergence within the Einstein radius $\overline{\kappa}_{<E} =1$.\footnote{Another lensing quantity that is independent of the mass-sheet degeneracy is the quantity $R_E\alpha_E''/(1-\kappa_E)$, where $\alpha_E$ is the deflection at the Einstein radius \citep{kochanek_over-constrained_2020}. While this has been combined with kinematic constraints in prior work to provide constraints on the density slope \citep{shajib_dark_2021,treu_strong_2023}, it is unclear how robust this quantity is in the presence of ellipticity and external shear for real lenses as it is only strictly invariant for axisymmetric lenses. Additional unmodeled angular structure may also render the inference less robust \citep{koch2021}. Thus, to be conservative, we use only the Einstein radius constraint in this work.}  While this is a robust constraint and reliably gives the total projected mass within the Einstein radius, this includes both the projected mass of the lens galaxy and potential external contributions from structures outside the lensing galaxy, thereby making it susceptible to the mass-sheet degeneracy. For joint modeling, we only consider the case of the closest source, i.e., the configuration in which the lens is at $z_l = 0.2$ and the source at $z_s =0.6$.

\subsection{Modeling Overview}

We continue with the modeling assumptions and methods of the previous section, where we used purely kinematic data. In addition, we include the constraints from strong lensing. The contribution from strong lensing is incorporated by including an additional term in the log-likelihood function of the modeling, leading to its general form:
\begin{equation}
\log \mathcal{L} = \log \mathcal{L}_{\mathrm{kin}} + \log\mathcal{L}_{\mathrm{lensing}} + C
\end{equation}
Here C is some constant, the kinematic contribution term to the likelihood $\log \mathcal{L}_{\mathrm{kin}}$ is given by equation \ref{eq:logl_kin} and the contribution of the lensing term is given by: 

\begin{equation}
\log \mathcal{L}_{\mathrm{lensing}} =
- \frac{\left[ \overline{\kappa}_{E,model} - 1 \right]^2}
{2\sigma_{m}^2}
\label{eq:lc_term_likelihood}
\end{equation}

Here $\overline{\kappa}_{E}$ is the model's normalized average convergence within the Einstein radius given by:

\begin{equation}
\overline{\kappa}_{E}
= \frac{\overline{\Sigma}_{E, model}}
{\Sigma_{\mathrm{crit}}}
\end{equation}

where $\overline{\Sigma}_{E, model}$ is the average projected density of the model within the Einstein radius, ${\Sigma}_{crit}$ is the critical density (Equation \ref{eq:critical_dens}).

The measurement error from lensing $\sigma_{m}$ of Equation \ref{eq:lc_term_likelihood} is given by: 
\begin{equation}
\sigma_{m}^{2} =
 \sigma_{E}^{2}  + \sigma_{\kappa_{\mathrm{ext}}}^{2}
\label{eq:sigma_m_def}
\end{equation}
where $\sigma_E$ is the fractional uncertainty in the measurement of the Einstein Radius and $\sigma_{\kappa_{ext}}$ is the fractional uncertainty due to the line-of-sight contributions to the projected mass within the Einstein radius from structures external to the galaxy, i.e., the external convergence. We include this term because, while the external convergence contributes to the measured Einstein radius, it does not contribute to the galaxy mass itself; thus our inferred galaxy mass within the Einstein radius depends on our prior estimate of the external convergence. 

While the lensing contribution to the likelihood is from a single term, we found it to be significant. When $\overline{\kappa}_{E, model}$  moves away from unity, the contribution of the lensing term can rise rapidly from a few to tens of percent. When $\overline{\kappa}_{<(R_{\mathrm{E, model}})} = 1$, i.e., it exactly matches the data, the contribution of the lensing term in Equation \ref{eq:lc_term_likelihood} to the overall likelihood is exactly zero. We estimate the uncertainty in the measurement of the Einstein radius by referring to \cite{shajib_dark_2021} who looked at a sample of elliptical galaxies at $z=0.2$ from the SLACS survey, and set $\sigma_{E} = 0.005$. \cite{wells_tdcosmo_2024} modeled the $\kappa_{ext}$ contribution for 25 SL2S galaxy-galaxy lenses as a function of deflector and source redshifts, and \cite{birrer_tdcosmo_2020} estimated $\kappa_{ext}$ for a sample of TDCOSMO lenses. Using these two references, we approximate the uncertainty in $\kappa_{ext}$ to be
$\sigma_{\kappa_{ext}} = 0.02$ for our galaxy lenses at $z=0.2$.

We continue to apply the realistic condition that the total galaxy stellar mass be greater than the mass of dark matter within $5kpc$, i.e., $M^*_{tot} > M_{DM < 5kpc} $ and assume that the true value of the Hubble parameter is $H_0 = 70 \mathrm{km\ s^{-1}\ Mpc^{-1}}$. 

\subsection{Results from Joint Modeling}

\begin{figure*}[ht!]
\centering
\includegraphics[width=0.8\textwidth]{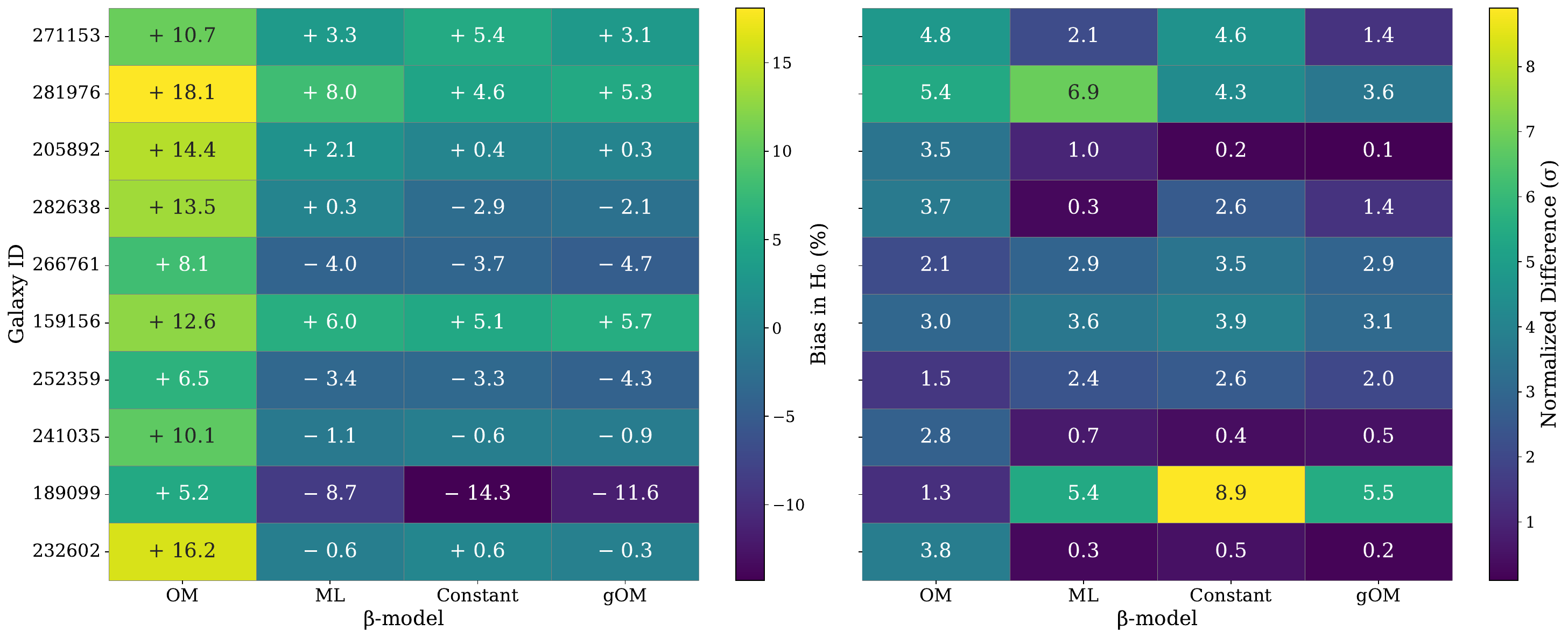}
\caption{Results from Joint Modeling with Kinematics and Strong Lensing. 
The lens is at redshift $z_l = 0.2$ and source at redshift $z_s = 0.6$. 
\textbf{Left panel:} Percentage Bias in $H_{0}$ inferred by the model. 
\textbf{Right panel:} Absolute values of the normalized residuals of the model versus data in units of $\sigma$. The assumed true value of the Hubble parameter is $H_0 = 70~\mathrm{km\,s^{-1}\,Mpc^{-1}}$.}
\label{fig:abs_vals_and_errors_lc}
\end{figure*}

\begin{figure*}[ht!]
\centering
\includegraphics[height=0.95\textheight,width=0.65\textwidth]{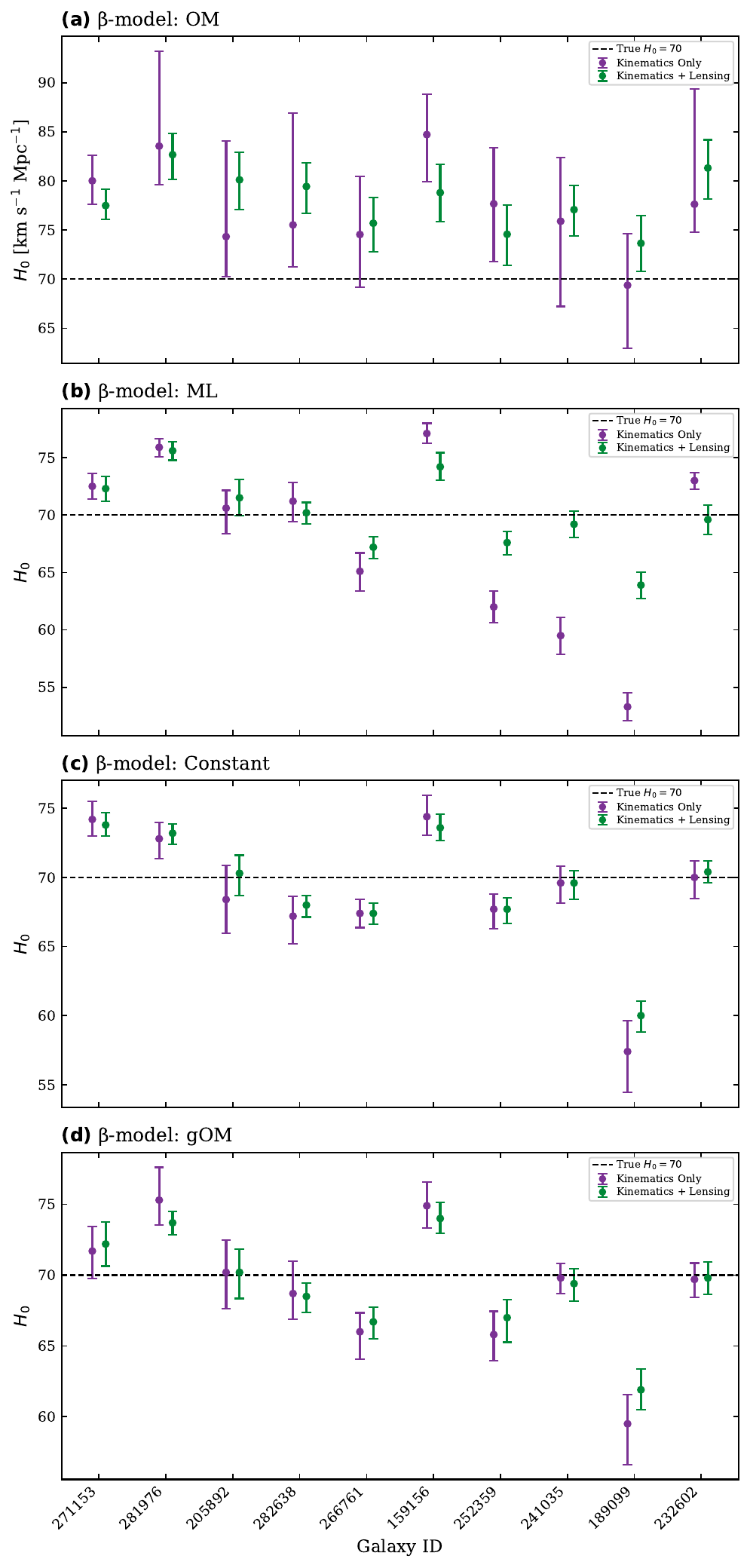}
\caption{Results from Kinematics Only, and Joint Modeling with Kinematics and Strong Lensing for the ten galaxies. The lens is at  redshift $z_l = 0.2$ and source at redshift $z_s =0.6$. The sub-plots from top to bottom show different anisotropy models \textbf{(a)} Osipkov-Merritt,\textbf{ (b)} Mamon-Lokas, \textbf{(c)} Constant, and\textbf{ (d)} Generalized Osipkov Merritt. The colored circles indicate the inferred $H_0$ values with the error bars indicating the $68\%$ credible interval. The assumed true value of the Hubble parameter is $H_0 = 70~\mathrm{km\,s^{-1}\,Mpc^{-1}}$.
}
\label{fig:lc_constraints_all_in_one}
\end{figure*}

In Figure \ref{fig:abs_vals_and_errors_lc}, we plot the inferred values of $H_0$, and the normalized residuals in $H_{0}$ --- the difference between the true value and the model's median prediction, divided by the measurement uncertainty --- expressed in units of $\sigma$. In Figure \ref{fig:lc_constraints_all_in_one}, we plot the results for both the kinematics only and the joint modeling cases. For direct comparison of the joint modeling with the kinematics only version, the relevant information for the latter is in the first columns of each of the subplots in Figures \ref{fig:hubble_kin_only_abs_vals_map_all_galaxies} and \ref{fig:hubble_kin_only_error_bars_map_all_galaxies}.

The effect of adding lensing constraints is to shift the model parameter values so that $\overline{\kappa}_{E}$ approaches unity. In theory, the combined constraints from lensing and kinematics should help the inferred galaxy parameters and $H_{0}$ be closer to the truth.  Overall, we find that the general effect of including strong lensing constraints is to increase the accuracy and precision compared to the kinematics only case for the ML, Constant and gOM models. We find that the benefit of adding lensing information is particularly powerful for the most biased results. The results are model dependent, and sometimes if there exists a tension between the kinematics and lensing constraints, the bias increases slightly, although it is minimal for these three models.

When considering the ten galaxies, the average biases were: $+11.5 \pm 1.3\%$ (OM), $+0.2  \pm 1.6\%$ (ML), $-0.9\pm 1.9\%$ (constant) and $-0.9 \pm1.6\%$ (gOM).  The median deviations between true $H_0$ and the model inferred $H_0$ using the posterior uncertainties were: $3.25\sigma$ (OM), $2.25\sigma$ (ML), $3.05\sigma$ (constant) and $1.7\sigma$ (gOM).

For the OM model, adding lensing constraints improved the $H_{0}$ estimates for the most biased results, but there was a slight increase in bias for a few other galaxies. The lensing constraints were at times in tension with the kinematic constraints for this model, and there was a degradation in the quality of fit (see Table \ref{table: om_new} for the $\chi^2_\nu$ values). Overall, the inclusion of lensing constraints did not improve the results and, in fact, led to an increase in average bias.

The Mamon-Lokas model benefited enormously from lensing constraints, with nine galaxies seeing an improvement in $H_0$ inference. Like OM, the cases with the most biased measurements of $H_0$ observed the maximum improvements. With joint modeling, seven out of ten galaxies ended up with a bias less than $5\%$, while all remaining ones had a bias less than $10 \%$. There was also a reduction in the size of the error bars. Seven of ten galaxies captured $H_0$ within $3\sigma$. 

The constant model also improved generally with the outlier galaxies receiving maximum help, for example the galaxy with ID $189099$ saw a reduction in the bias from $-18\%$ to $-14.3\%$. \textbf{ }Five out of ten galaxies captured true $H_0$ within $3\sigma$. Barring the outlier galaxy, all galaxies had close to or less than $5\%$ error in the $H_0$ measurement.

The gOM model also improved after including lensing information. All galaxies that had a bias greater than $5\%$ saw an improvement in the $H_0$ measurement. Any worsening of $H_0$ measurement due to lensing constraints was minimal, only about $0.5\%$. The outlier galaxy with ID $189099$ saw an improvement in $H_0$ with bias reducing from $-15\%$ to $-11.6\%$. Except for the outlier galaxy, where the inferred $H_0$ was $5.5 \sigma $ away, five of the remaining galaxies captured $H_0$ within $2\sigma$, while the remaining captured $H_0$ within or close to $3\sigma$.

We see in particular that the ML model shows drastic improvements, likely due to the fact that the anisotropy curve from this model mimics the behavior of massive elliptical galaxies in the TNG simulations. This means that if the anisotropy model resembles the true anisotropy curves, the inclusion of lensing constraints significantly alleviates the mass-anisotropy degeneracy.  The gOM model is the most accurate of the four models with the least bias in $H_0$ relative to the posterior uncertainties. We plot the inferred stellar and dark matter masses for the four models and all the galaxies in Figures \ref{fig:M_star_model_vals} and \ref{fig:M_dm_model_vals}. The remaining parameters are tabulated in Tables \ref{table: om_new}, \ref{table: ml_new}, \ref{table: constant_new} and \ref{table: gom_new} with relevant columns for joint modeling as indicated. 

\section{Results with data having a reduced Signal-to-Noise Ratio}
\label{sec: additional_runs}

In this section, we examine the impact of reduced data quality on the results. We lower the maximum central signal-to-noise ratio of the line-of-sight velocity dispersion data from $60$ in the fiducial dataset to $20$. Apart from the change in signal to noise, all the other assumptions and methods of the previous two sections remain exactly the same. We consider here the case of the closest source at redshift $z_s=0.6$. We find that the results exhibit similar trends even with lowered signal-noise for both kinematic and joint modeling. The results obtained are summarized via Figure \ref{fig:lc_constraints_all_in_one_snr_20} where for brevity only the final inferred $H_0$ values are shown. In Table 4, we summarize the average biases for the ten galaxies for both the kinematics-only and joint modeling runs, for the two data qualities considered. 

The Osipkov-Merritt model shows a slight increase in bias with the reduction in data quality. All the other three models remain consistent with sub-percent bias. There is generally a small increase in the scatter.  There is an increase in the uncertainties associated with the inference of the individual galaxy parameters and this is reflected in the Hubble parameter. 

\begin{deluxetable*}{lcccc}
\tablecaption{Mean bias in inferred $H_0$ over ten galaxies for kinematics-only and joint modeling, evaluated for the two data sets with different peak signal-to-noise ratios. Values are reported as average bias $\pm$ standard error (SE).\label{tab:h0_bias_summary}}
\label{tab:SNR_Comparison}
\tablehead{
\colhead{Modeling Setup} & \colhead{OM} & \colhead{ML} & \colhead{Constant $\beta$} & \colhead{gOM}
}
\startdata
Kinematics Only (SNR 60)   & $9.82 \pm 2.11$  & $-2.26 \pm 3.47$ & $-1.52 \pm 2.16$ & $-1.23 \pm 2.10$ \\
Joint Modeling (SNR 60)    & $11.54 \pm 1.32$ & $0.77 \pm 1.53$  & $-0.86 \pm 1.85$ & $-1.09 \pm 1.58$ \\
Kinematics Only (SNR 20)   & $13.46 \pm 3.29$ & $2.12 \pm 2.03$  & $0.17 \pm 2.31$  & $0.57 \pm 2.26$ \\
Joint Modeling (SNR 20)    & $12.38 \pm 1.73$ & $1.07 \pm 1.62$  & $-0.13 \pm 2.00$ & $0.35 \pm 1.85$ \\
\enddata
\end{deluxetable*}

\section{Discussions}
\label{sec:discussions}

\subsection{Impact of assumptions in the stellar anisotropy}

As we have seen, the choice of the anisotropy plays a major role in galaxy dynamics in general, and in $H_0$ inference. The results from Osipkov-Merrit modeling suggest that while having a radially varying anisotropy is desirable to model real galaxies, the form of the radial variation employed is equally important and can be problematic if it deviates significantly compared to the ground truth. Single-parameter models like Osipkov-Merritt can be highly constraining. A very small anisotropy radius means that the model becomes highly radial far too quickly corresponding to unstable orbits, and a larger anisotropy radius means that it tends to be too isotropic in the central regions (Figure \ref{fig:combined_beta_profiles}). In addition, $\beta$ asymptotically reaches an unphysical value of $1$ (purely radial orbits).  Thus, it can struggle to capture radial variation like those of massive elliptical galaxies in $TNG100$ (Figure \ref{fig:anisotropy}), while the Mamon-Lokas model can capture such behavior at small radii better, without being too radial at large radii, where it asymptotically reaches a value of $0.5$. 

Despite the significant radial variation in anisotropy in our $TNG100$ sample of galaxies (as Figure~\ref{fig:anisotropy} shows), the spatially constant anisotropy model performs reasonably well, presumably by globally averaging the anisotropy across all radii to fit the data. This makes the constant model a useful baseline for comparison with more complex models. Since the true nature of anisotropy in real systems is not known \emph{a priori} and may have irregularities, the constant model is a practical tool, it can fit the observable data well, with the caveat that it may not reproduce the true anisotropy profile at all radii. If real elliptical galaxies exhibit fewer variations in radial anisotropy compared to our dataset, the performance of this model should, in principle, improve. It is being used increasingly in the context of joint analysis \citep{tdcosmo_collaboration_tdcosmo_2025, shajib_tdcosmo_2023}, and its utility can extend to dynamical  modeling of galaxies, in general.

We note that imposing the physical condition that $M^*_{tot} > M_{DM<5kpc}$ --- the total stellar mass of the galaxy be greater than the mass of dark matter within $5kpc$ --- was crucial. Without imposing the condition, there was a lot of degeneracy in the stellar and dark matter components that satisfied the condition $\overline{\kappa_E} = 1$  leading to unpredictable results, sometimes an unrealistically high dark matter mass within $5kpc$ and a tiny scale radius. We have opted for an approach with least assumptions; another way would be to have a restrictive prior on the scale radius to prevent over-concentrated halos, provided the constraints on the scale radius can be estimated beforehand. 

\subsection{Comparison with JAM}

Another approach to Jeans modeling is $JAM$ --- Jeans Anisotropic Modeling, \cite{cappellari_5_2008} --- which has two standard flavors: an axisymmetric, cylindrically aligned version ($JAM_{cyl}$) and a spherically aligned version ($JAM_{\mathrm{sph}}$) both of which make use of Multi-Gaussian Expansion ($MGE$) to model stellar light. In $JAM_{cyl}$,  the assumption is that the mass distribution is assumed axisymmetric and the anisotropy parameter is defined by $\beta_z = 1 - \frac{\sigma_z^2}{\sigma_R^2}$. Here, $\sigma_z$ and $\sigma_R$ are the velocity dispersions in the z and R directions. A common assumption made here is that the ratio of the two velocity dispersions $\sigma_z/\sigma_R$ is constant, so that $\beta_z$ = constant, often to be inferred by modeling. It is possible that using a single degree of freedom via $\beta_z$ could be too restrictive and lead to biases in masses due to mass anisotropy degeneracy, as we saw in the single-parameter Osipkov-Merritt model. More recently, \cite{cappellari_6_2020} presented a computationally efficient version of $JAM_{sph}$ where each Gaussian component has its own constant anisotropy. In principle, the combination of several such Gaussian can lead to approximating an arbitrary profile, which can resemble, for example, the gOM anisotropy. The work of \cite{shajib_tdcosmo_2023} is an example of a step in this direction in which they model with two different Gaussian components, each allowed to have a different anisotropy profile. In \cite{simon_supermassive_2023}, they allow each Gaussian component of the MGE to have anisotropy $\beta(r) = \frac{\beta_{0} + \beta_{\infty}\,(r/r_{a})^{2\delta}}{1 + (r/r_{a})^{2\delta}}$, where $\delta > 0$ controls the sharpness of the transition, an example of a highly flexible anisotropy profile introduced in \cite{baes_dynamical_2007}. \cite{cappellari_6_2020} caution that it will be helpful to model with both flavors of JAM ($JAM_{sph}$ and $JAM_{cyl}$) to test the assumptions of either method. Because massive ellipticals are generally round and weakly triaxial \citep{cappellari_structure_2016}, $JAM_{sph}$ is likely a better choice for such galaxies. A study exploring the anisotropy and its interpretation in the two JAM variations and the spherical Jeans framework from this work might be useful. 

\subsection{Limitations and Caveats}

We do not claim that the anisotropy profiles of real massive ellipticals necessarily follow the behavior of those in TNG. \cite{mamon_dark_2005} found the anisotropy in the inner regions to rise rapidly  in \textbf{$\Lambda CDM$} simulations which were fit well by the ML profile, similar to what we observe for massive ellipticals in TNG. \cite{thomas_dynamical_2014} found that within the cores of massive ellipticals, the anisotropy was tangential but outside the core it grew rapidly in the radial direction, similar to the dataset we use. However, \cite{cappellari_early-type_2025} observe that, while applying Schwarzschild modeling to a subset of  real massive ellipticals, the anisotropy begins tangentially at the center of the galaxy and becomes only mildly radial within the effective radius.  \cite{sheu_project_2025} find that the OM model with a large anisotropy radius was mostly consistent with the constant model, indicating that the anisotropy prefers to be more isotropic in the inner regions. Thus, there may be potential discrepancies in the anisotropy between different simulated and observational works, and the radial nature of the anisotropy in the inner regions of massive elliptical galaxies remains an open question.

In the absence of proper motion data, we cannot \textit{beforehand} know the true nature of the anisotropy. As long as the anisotropy profiles are approximately increasing with distance from the galaxy center, whether they prefer to be more isotropic in the inner region or more radial like in this dataset, in either case, the behavior can be accounted for by using a model with more degrees of freedom, for example via the gOM profile, particularly in the range of the effective radius, the primary contributing region to the velocity dispersions. 

For dark matter, we have been self-consistent and mocked data and modeled with an NFW profile, which \cite{shajib_dark_2021} and \cite{sheu_project_2025} find to be the case for redshift of $z=0.2$. However, if the expected dark matter profile is different, e.g., \cite{sheu_project_2025} find the inner slope to be shallower at high redshifts, then dark matter could be modeled with a generalized NFW profile (gNFW) that accounts for a variable inner slope.

Another systematic not included here is the triaxiality of massive ellipticals. We created spherically averaged profiles and modeled them via spherical Jeans equations, thereby taking the ellipticity out of the picture. While this is a good approximation and massive elliptical galaxies are generally round and weakly triaxial \cite{cappellari_structure_2016}, for a more complete description or if the ellipticity is significant, it would have to be taken into account. \cite{huang_tdcosmo_2025} find that the use of $JAM_{sph}$ could introduce a slight bias in $H_0$ which was corrected by using $JAM_{cyl}$.

\subsection{Improving this work}

One way we can improve the accuracy of $H_0$ inference further could be by allowing the anisotropy to vary at the center of the galaxy, i.e., to have $\beta_0$ as an additional parameter. While $\beta_0 =0$ has been a good approximation, we see from Figure \ref{fig:anisotropy} that $\beta_0$ actually varies from about $0.05 - 0.15$. This addition may increase accuracy, as well as the $1\sigma$ credible interval of  inferred $H_0$. This flexibility can be incorporated into gOM and also be useful when the anisotropy in the inner regions of the galaxy is slightly tangential, for example, within the sphere of influence of a central black hole \citep{simon_supermassive_2023} or as in the elliptical galaxies obtained via Schwarzchild modeling in \cite{cappellari_early-type_2025}. In case some tangential anisotropy is anticipated, the priors of the constant model should also be adjusted to account for this. Similarly, a generalized extension of the Mamon-Lokas profile (gML) with $3$ degrees of freedom can be explored, and in principle this model can model can handle high radial gradients in the anisotropy profile at small radii better than gOM. If even more flexibility is sought, the previously mentioned anisotropy model in \cite{baes_dynamical_2007} can be explored.

The value of the uncertainty $\sigma_{\kappa_{ext}}$ we used was conservative for the lens and source at redshifts of 0.2 and 0.6 respectively. If the actual value is lesser (e.g., a closer source and lens configuration), or if there lie lesser line-of-sight structures, or a better estimate of $\kappa_{ext}$ exists, e.g., from weak lensing or galaxy number counts, then this should lower the value of the uncertainty on $\kappa_{ext}$, which should in principle lead to better $H_0$ inference.
As mentioned earlier, using the JAM method to explore arbitrary anisotropy profiles could be helpful.

If the coverage of the spatially resolved data is extended so that the data go well beyond the half-light radius, this will improve the analysis, as \cite{shajib_improving_2018} point out. In principle, this should also help us to get better constraints on the dark matter scale radius $r_s$.  Another approach to breaking the mass–anisotropy degeneracy is to include higher-order moments of the Jeans equations into the modeling, examples of which include \cite{lokas_dark_2003,richardson_analytical_2013,read_how_2017}, where the fourth-order moment kurtosis was explicitly modeled alongside the velocity dispersion. These studies showed that the inclusion of the fourth-order velocity moment can significantly reduce the degeneracy, but their effectiveness is strongly dependent on the availability of high-quality kinematic data. \cite{kaplinghat_too_2019}, while studying dwarf spheroidal galaxies of the Milky Way, found that the mass–anisotropy degeneracy persisted, even after supplementing the Jeans analysis with fourth-order velocity moment constraints. Third-order moments are sometimes included, but they are generally more sensitive to galaxy rotation than to random orbital anisotropy, and thus less useful for breaking the mass–anisotropy degeneracy.

\section{Summary and Conclusions}
\label{sec:summary_and_conclusions}

In this work we use spatially resolved kinematic data and simulate JWST NIRSpec observations of massive elliptical galaxies from the IllustrisTNG$100$ simulations to investigate the mass-anisotropy degeneracy, and quantify the effects that assumptions in stellar anisotropy play in the inference of galaxy parameters and the Hubble parameter $H_0$ via time-delay cosmography. The different anisotropy models we tested include the often used radially varying single parameter Osipkov-Merrit (OM), the Mamon-Lokas anisotropy (ML), a spatially constant anisotropy $\beta$, and a generalized Osipkov-Merritt model (gOM). First, we used purely kinematic data to estimate the bias in $H_0 $ that would result from using only kinematic data assuming that time delay-data is available (see Figures \ref{fig:$H_0$_beta_models} , \ref{fig:hubble_kin_only_abs_vals_map_all_galaxies}, and \ref{fig:hubble_kin_only_error_bars_map_all_galaxies} for the main results). In the second case, we performed joint modeling by including constraints from strong lensing in the kinematic modeling (Figures \ref{fig:abs_vals_and_errors_lc} and  \ref{fig:lc_constraints_all_in_one}). The primary results of this work are as follows:

\begin{itemize}
    \item For joint modeling with the closest source at redshift $z_s = 0.6$, the average bias across the ten galaxies in the Hubble parameter was: $+0.2 \pm 1.6\%$ for ML, $-0.9 \pm 1.9\%$ for constant, and $-0.9 \pm 1.6\%$ for gOM, i.e., less than one percent for these three models. However, for the OM model it was $+11.5 \pm 1.3\%$. This suggests that over an ensemble of galaxies, the ML, constant, and gOM models can recover the Hubble parameter with sub-percent bias, whereas the higher bias in OM is unlikely to average out. While the mean bias was close to zero for the first three models, the individual galaxies showed varying degrees of bias and the scatter for these models was $\sim 5\%$, approximately half the Hubble tension. The median deviations between true $H_0$ and the model inferred $H_0$ using the posterior uncertainties were: $3.25\sigma$ (OM), $2.25\sigma$ (ML), $3.05\sigma$ (constant) and $1.7\sigma$ (gOM). The gOM model was the best performer across the ten galaxies considering both the kinematics and joint modeling approaches. Overall, it most accurately recovered the true galaxy parameters, produced lesser outliers, and yielded the lowest $H_{0}$ bias relative to the inferred posterior uncertainties.\textbf{ }In general, for ML, constant and gOM models, the effect of joint kinematics and strong lensing analysis was to improve both accuracy and precision, where the galaxies with the most biased results showed the largest improvements. For the analysis with reduced data quality where the maximum signal-to-noise ratio is reduced from $60$ in the fiducial dataset to $20$, the OM model exhibits a slight increase in bias, whereas the other three models remain consistent with sub-percent average bias with a slight increase in the scatter (Figure \ref{fig:lc_constraints_all_in_one_snr_20})\textbf{.} 
    \item The single-parameter Osipkov-Merritt model shows a very high bias with a tendency to systematically overestimate $H_0$. For the kinematics data only case, more than half of the results show an absolute bias greater than $10\%$,and for the most biased results, the true $H_0$ is biased by more than $20\%$.  Inclusion of lens constraints added precision to the results, but due to tension with kinematic constraints for this model, the average bias in $H_0$ increased from $+9.8  \pm 2.1\%$ to $+11.5 \pm 1.3\%$.
    \item The Mamon-Lokas anisotropy model fared better than the OM anisotropy and showed a  lesser bias for $7$ of the $10$ galaxies in the dataset with kinematics only data and the results saw a significant improvement under joint modeling where the bias in all galaxies improved, with most galaxies showing a less than $ 5\%$ bias.  Given its advantages of being a radially varying anisotropy, being analytically tractable, and at the same time not getting too radial at large radii like OM, this model seems to be underused in the current literature, and similar models or extensions may be explored.
\item The constant anisotropy model also performed better than the OM model. For individual galaxies, with the kinematics-only case, the absolute bias in the Hubble parameter inference was generally between $5\text{-}10\%$, which dropped to below $5\%$ under joint modeling for most galaxies. The quality of fit remained good in either scenario. Since the true nature of the anisotropy cannot be known apriori without proper motion data, the constant model can be a good baseline model and serve as a reference against complex models.
    \item The generalized Osipkov-Merritt anisotropy model often captured the true galaxy parameters within the $1\sigma$ credible interval, and the Hubble biases were within $5\%$ for the kinematics only data. Joint modeling with lensing information increased the accuracy and precision of the results. The quality of fit was preserved in both versions. Thus, adding flexibility by introducing even a single degree of freedom to the OM model had significant implications for recovering the galaxy parameters and $H_0$ inference.

\end{itemize}

In summary, our work highlights that the mass-anisotropy degeneracy can be mitigated by using spatially resolved kinematic data and choosing a suitable stellar anisotropy model. Joint modeling with strong lensing constraints improves this further  and helps to achieve a more robust and reliable inference of the Hubble parameter $H_0$. More broadly, our results may be relevant to studies that explore galaxy dynamics, galaxy formation, the nature of dark matter, and cosmology. 

\begin{acknowledgments}

The authors are extremely grateful to the anonymous referee for a thoughtful and constructive review. Their insightful comments and suggestions greatly improved the scope, clarity, and overall presentation of this work. We deeply appreciate Tommaso Treu for his valuable and detailed feedback. We acknowledge Michele Cappellari, Simon Birrer, and Shawn Knabel for their useful inputs. We thank Claudia Pulsoni for helpful discussions on slow-rotating elliptical galaxies, and Sophia Nasdr for introducing us to the IllustrisTNG simulations. We acknowledge the use of data from the IllustrisTNG Project \citep{nelson_illustristng_2019}, made available at \url{https://www.tng-project.org}. 

This research received support through Schmidt Sciences, LLC. for V.V., and Q.M.

\vspace{5mm}
Database: IllustrisTNG \citep{pillepich_first_2018, springel_first_2018, nelson_first_2018, naiman_first_2018, marinacci_first_2018}

\software{Dynesty \citep{speagle_dynesty_2020}, 
          Colossus \citep{diemer_colossus_2018},
          NumPy \citep{harris_array_2020}, 
          SciPy \citep{virtanen_scipy_2020}, 
          Matplotlib \citep{hunter_matplotlib_2007},  
          Numba \citep{lam_numba_2015}
          }

\end{acknowledgments}
          
\clearpage

\appendix

\section{Line of Sight Velocity Dispersion Residual Maps for the Reference Galaxy}

\begin{figure}[ht]
\centering
\includegraphics[width=0.8\columnwidth,height=0.985\columnwidth]{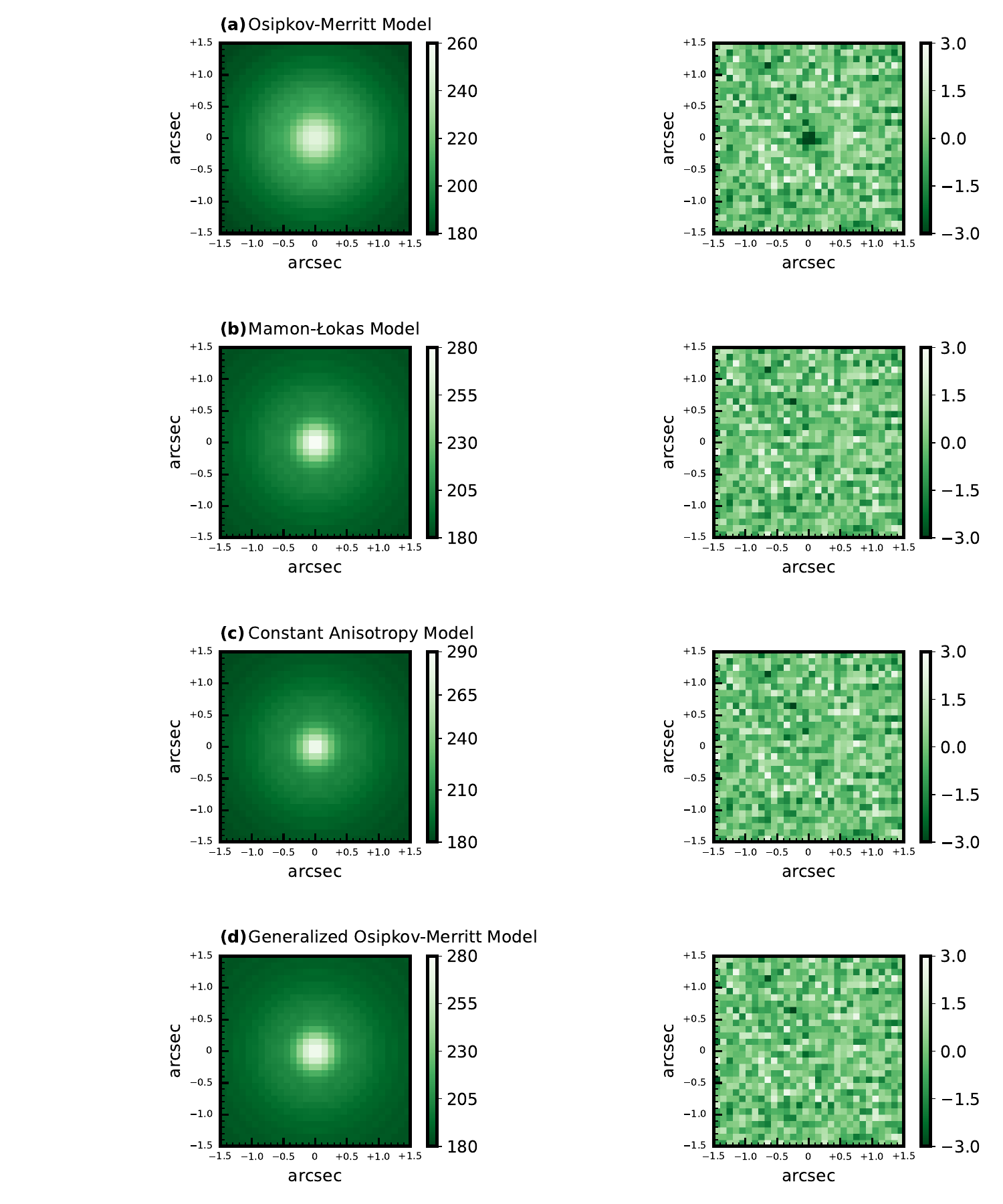}
\caption{
Line-of-sight (LOS) velocity dispersion maps and normalized residuals for the reference Galaxy with ID 271153
modeled under the four anisotropy models. Each row corresponds to a different model:
\textbf{(a)}~Osipkov-Merritt, \textbf{(b)}~Mamon-Łokas, \textbf{(c)}~Constant, and \textbf{(d)}~Generalized Osipkov-Merritt.
Left panels show the best-fit model LOS velocity dispersions, and right panels show residuals (model - data) velocity dispersions
normalized by the measurement error. The reduced $\chi^2_\nu$ values for the different models are: a. $1.04$ (OM), b. $0.95$ (ML), c. $0.95$ (Constant) and d. $0.95$ (gOM) indicating a good fit to the data by the four models.}
\label{fig:combined_los_dispersions_and_residuals}
\end{figure}

\clearpage

\section{Effect of including Gas}
\label{sec:appendix_gas}

\begin{figure}[ht!]
\centering
\includegraphics[width=0.5\textwidth,height=0.25\textwidth]{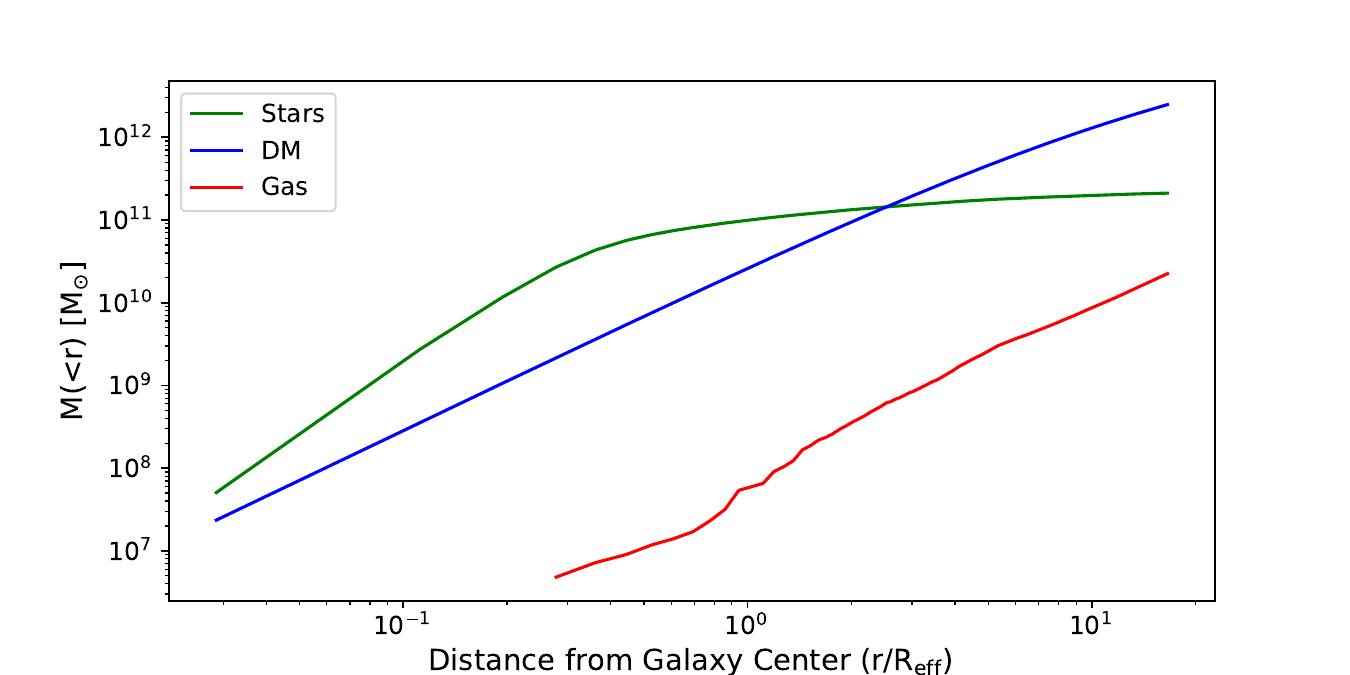}
\caption{Mass profiles of the stellar, dark matter and gas components within a 3D radius of $r^*_{0.95}$ for the reference galaxy. The x axis is scaled by the galaxy effective radius $R_{eff}$ which has a value of 4.61 kpc.}
\label{fig:mass_profiles_r_threshold}
\end{figure}

In Figure \ref{fig:mass_profiles_r_threshold}, we plot the 3D mass profiles of the stellar, dark matter, and gas components for the reference galaxy with ID 271153. The stellar matter dominates within the effective radius (4.61 kpc), and a little beyond 2 effective radii, the dark matter mass exceeds the stellar mass. The gas mass at these radii is insignificant, and indeed even at the maximum cutoff radius of study for this galaxy, $r_{0.95}^* \sim 84 \,\mathrm{kpc}$ (about 18 effective radii), the total stellar mass is still about an order of magnitude higher than the total gas mass, which are both dwarfed by the dark matter mass.

\begin{figure}[ht!]
\centering
\includegraphics[width=\columnwidth]{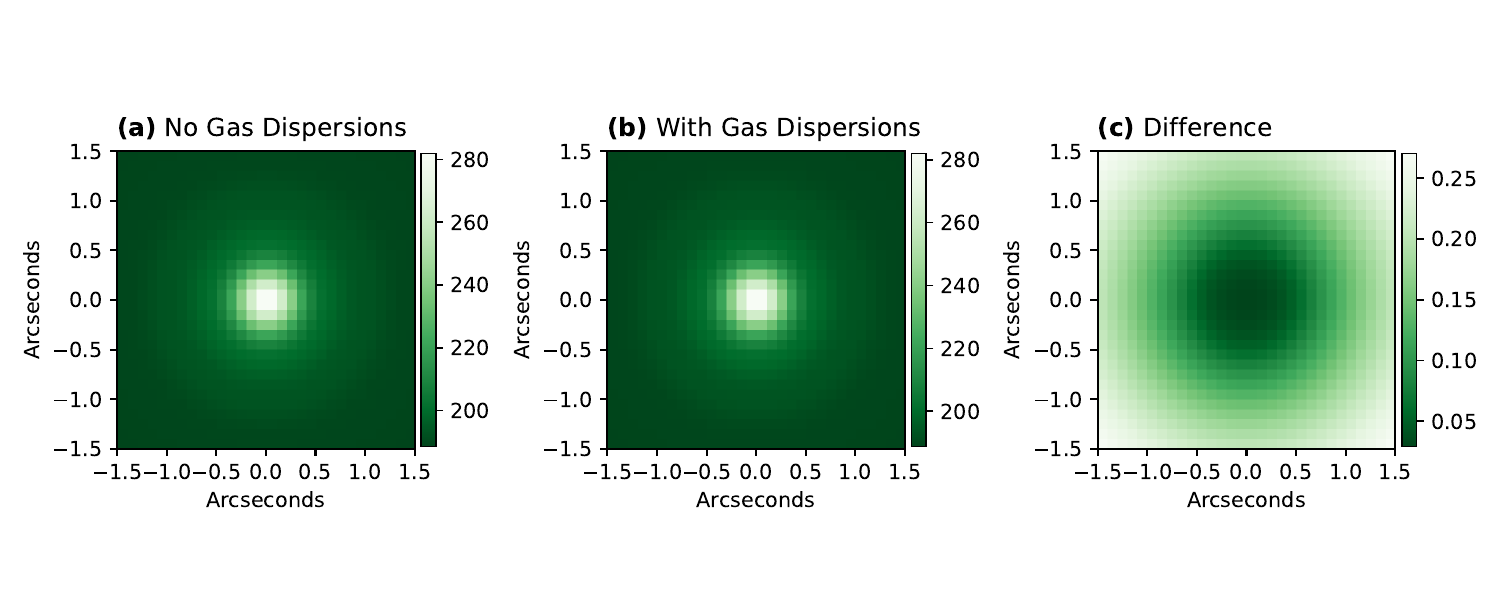}
\caption{Line-of-sight velocity dispersion profiles for the reference galaxy for different cases. \textbf{(a) }Excluding gas from the data \textbf{(b) }Gas is included \textbf{(c)} Absolute pixel based differences in the dispersions obtained by subtracting the data in (a) from (b).}
\label{fig:LOSVD_profiles_gas}
\end{figure}

In Figure \ref{fig:LOSVD_profiles_gas}, we plot the line-of-sight velocity dispersion data for the reference galaxy. The first two plots show the line-of-sight velocity dispersions without and with gas included. The third shows the difference between the two. We observe that the inclusion of gas in the dynamics increases the velocity dispersions by a maximum of $\sim 0.25 \,\mathrm{km}\,\mathrm{s}^{-1}$ to the velocity dispersions and therefore its exclusion can be justified.

The galaxy total masses for the different components are $M_{\mathrm{stars}} = 2.25 \times 10^{11}\,M_\odot$, $M_{\mathrm{gas}} = 8.72 \times 10^{11}\,M_\odot$, and $M_{\mathrm{DM}} = 1.10 \times 10^{13}\,M_\odot$, thus the total gas mass is greater than the stellar mass. If we went to radii well beyond $r_{0.95}^*$ to about fifty effective radii, the gas mass would then exceed the stellar mass, but it would still be negligible compared to the dark matter mass (\cite{mamon_dark_2005} also observed this). In addition, at such a large radius the stellar mass density $\nu(r)$ would drop significantly, so that the line-of-sight velocity dispersions obtained using Equation \ref{eq:LOSVD} would not change significantly. Overall, this is reassuring because, in general, massive ellipticals tend to be dry and have a lower gas to stellar mass fraction than the reference galaxy here, and if the gas contribution here to the velocity dispersions is negligible, one can expect that to be true for real massive ellipticals as well.   

\section{Equations for the LOS VD}
\label{sec:eqs_losvd}
For the Osipkov-Merritt and Mamon-Lokas profiles, instead of using Equations \ref{eq:sigma_r} and \ref{eq:LOSVD}, which involves two nested integrals, we used computationally simpler one-integral versions (adopted from the appendix of \cite{mamon_dark_2005}). The line-of-sight velocity dispersions can be put in the form: 

\begin{equation}
I(R)\, \sigma_{\text{los}}^2(R) = 2G \int_R^\infty K\left(\frac{r}{R}, \frac{r_{ani}}{R}\right) \ell(r)\, M(r)\, \frac{\mathrm{d}r}{r}
\label{eq:LOSVD_Kernel_Eq}
\end{equation}

The kernel $K(\frac{r}{R},\frac{r_{ani}}{R})$ can be rewritten in the form $K(u,u_a)$ with $u=r/R$ and $u_a=r_{ani}/R$ and its expansion for the different profiles is given below.

\subsection{Osipkov-Merritt Anisotropy}
\begin{equation}
K(u, u_a) = \frac{u_a^2 + \tfrac{1}{2}}{(u_a^2 + 1)^{3/2}} \left( \frac{u^2 + u_a^2}{u} \right) \tan^{-1} \left( \sqrt{ \frac{u^2 - 1}{u_a^2 + 1} } \right) 
- \frac{1}{2(u_a^2 + 1)} \sqrt{1 - \frac{1}{u^2}}
\label{eq:OsipkovMerrittKernel}
\end{equation}
\subsection{Mamon-Lokas Anisotropy}

\begin{align}
K(u, u_a) &= \frac{1}{2(u_a^2 - 1)} \sqrt{1 - \frac{1}{u^2}} 
+ \left(1 + \frac{u_a}{u} \right) \cosh^{-1} u \notag \\
&\quad - \operatorname{sgn}(u_a - 1)\, u_a \, \frac{u_a^2 - \tfrac{1}{2}}{|u_a^2 - 1|^{3/2}} 
\left(1 + \frac{u_a}{u} \right) \mathcal{C}^{-1} \left( \frac{u_a u + 1}{u + u_a} \right) 
\hspace{1cm} (u_a \ne 1)
\label{eq:AnisotropicKernel_ua_not_1}
\end{align}

\begin{equation}
K(u, u_a) = \left(1 + \frac{1}{u} \right) \cosh^{-1} u 
- \frac{1}{6} \left( \frac{8}{u} + 7 \right) \sqrt{ \frac{u - 1}{u + 1} }
\hspace{3cm} (u_a = 1)
\label{eq:AnisotropicKernel_ua_eq_1}
\end{equation}
, where the sign function $\operatorname{sgn}(x)$ is given by $\operatorname{sgn}(x) = -1$ if $x < 0$, $0$ if $x = 0$, and $1$ if $x > 0$, and $\mathcal{C}^{-1}(x) = \cosh^{-1} x$ for $u_a > 1$ \quad\text{and}\quad $\mathcal{C}^{-1}(x) = \cos^{-1} x$ for $u_a < 1$.

\section{Effect of varying Hubble constant during fitting}
\label{sec:varying_H0}

In the kinematic analysis in this paper, the equations are in terms of lengths $r$, and we calculate dispersions with an assumed $H_0 = 70\ \mathrm{km\ s^{-1}\ Mpc^{-1}}$. However, if one varies the Hubble constant during fitting, the length scales corresponding to a particular redshift can be expected to change. Here, we show that the $H_0$ scalings do not affect our results, in particular the convergence $\kappa_E$ at the Einstein radius estimated from kinematics, upon which our estimated bias in $H_0$ depends.

Although kinematics equations are usually expressed in terms of distances, the actual observables are the line-of-sight dispersions $\sigma_{i,j}$ observed at a pixel with an angular position $\theta_{i,j}$, and the Einstein radius $\theta_E$ (in angular units) from lensing. For a given angle, the corresponding lengths are given by $R = D_L\theta$, where $D_L$ is the angular diameter distance to the lens; since $D_L \propto 1/H_0$, we can relate the length scale $R_{7}$ under the assumption $h=0.7$ (which we denote as $h_7$) to the length scale $R$ for a different Hubble constant by $R = R_7\frac{h_7}{h}$. The task now is to derive how the projected density of our model (when fit to the kinematics data) must scale with $h$.

The dispersion in the radial direction $\sigma$ at a particular radius can be calculated from Jeans modeling according to:
\begin{equation}
\sigma_r^2 = c^2 r^{-2\beta}\frac{1}{\nu(r)}\int_r^\infty r'^{2\beta}\nu(r')\frac{d(\Phi/c^2)}{dr'} dr'
\end{equation}
where $\nu$ is the stellar density, and we are assuming a constant $\beta$ model, although the results can be easily generalized to more complicated anisotropies $\beta(r)$. One can see above that if one substitutes $r_7 = r\frac{h}{h_7}$, the length factors cancel, so that the exact same dispersions will result (and hence, will fit the data) provided that the potential $\Phi$ is kept the same. However, assuming that $\Phi$ is kept fixed, the mass enclosed within a particular radius is given by $M(r) = \frac{c^2r^2}{G}\frac{d(\Phi/c^2)}{dr} \propto D_L\theta^2\frac{d(\Phi(\theta)/c^2)}{d\theta} \propto 1/h$; thus we can say that $M(r) = \frac{h_7}{h}M_7(r_7)$. Thus, to fit the dispersions as $H_0$ is varied, the mass normalization must be scaled by $1/h$. The mass density can then be calculated via $\rho_{\mathrm{tot}} = \frac{1}{4\pi r^2}\frac{dM}{dr}$; plugging in for $M(r)$, after a bit of algebra, we find, 
\begin{equation}
\rho_{\mathrm{tot}}(r) = \left(\frac{h}{h_7}\right)^2\rho_{\mathrm{tot},7}\left(r\frac{h}{h_7}\right).
\end{equation}

With this in hand, the total projected density at a radius $R$ is given by
\begin{eqnarray}
\Sigma_{\mathrm{tot}}(R) & = & \int_{-\infty}^{\infty} \rho_{\mathrm{tot}}(\sqrt{R^2+z^2})dz \\
& =& \left(\frac{h}{h_7}\right)^2 \int_{-\infty}^{\infty} \rho_{\mathrm{tot},7}\left(\frac{h}{h_7}\sqrt{R^2+z^2}\right)dz \\
& = & \left(\frac{h}{h_7}\right) \int_{-\infty}^{\infty} \rho_{\mathrm{tot},7}\left(\sqrt{R_7^2+z_7^2}\right)dz_7 \\ 
& = & \left(\frac{h}{h_7}\right) \Sigma_{\mathrm{tot},7}(R_7), \\
\end{eqnarray}
where in the third line we have made the substitution $z = z_7\frac{h_7}{h}$ and similarly for $R_7$.

Now, to calculate the convergence at the Einstein radius for a general Hubble parameter $h$, we use the fact that the critical density for lensing $\Sigma_{\mathrm{crit}} \propto \frac{D_L D_S}{D_{LS}} \propto 1/h$. Hence, we have
\begin{eqnarray}
\kappa_E & = &\frac{\Sigma_{\mathrm{tot}}(R_E)}{\Sigma_{\mathrm{crit}}} \\
& = & \frac{\left(\frac{h}{h_7}\right)\Sigma_{\mathrm{tot},7}(R_{E,7})}{\left(\frac{h}{h_7}\right)\Sigma_{\mathrm{crit},7}} \\
& = & \kappa_{E,7} \\
\end{eqnarray}
since the $h$ factors exactly cancel. Thus, while the Einstein radius in distance units changes as $H_0$ is varied, the value of the convergence at this radius $\kappa_E$ estimated from kinematics is invariant regardless of the assumed Hubble constant during the analysis. Thus, our estimated bias in $H_0$ (which depends on $\kappa_E$ to good approximation) remains the same even though $H_0$ is not explicitly varied during kinematics analysis.

\newpage

\section{Additional Figures}

\subsection{Inferred Stellar and Dark Matter Masses}

\begin{figure*}[ht!]
\centering
\includegraphics[width=0.95\textwidth]{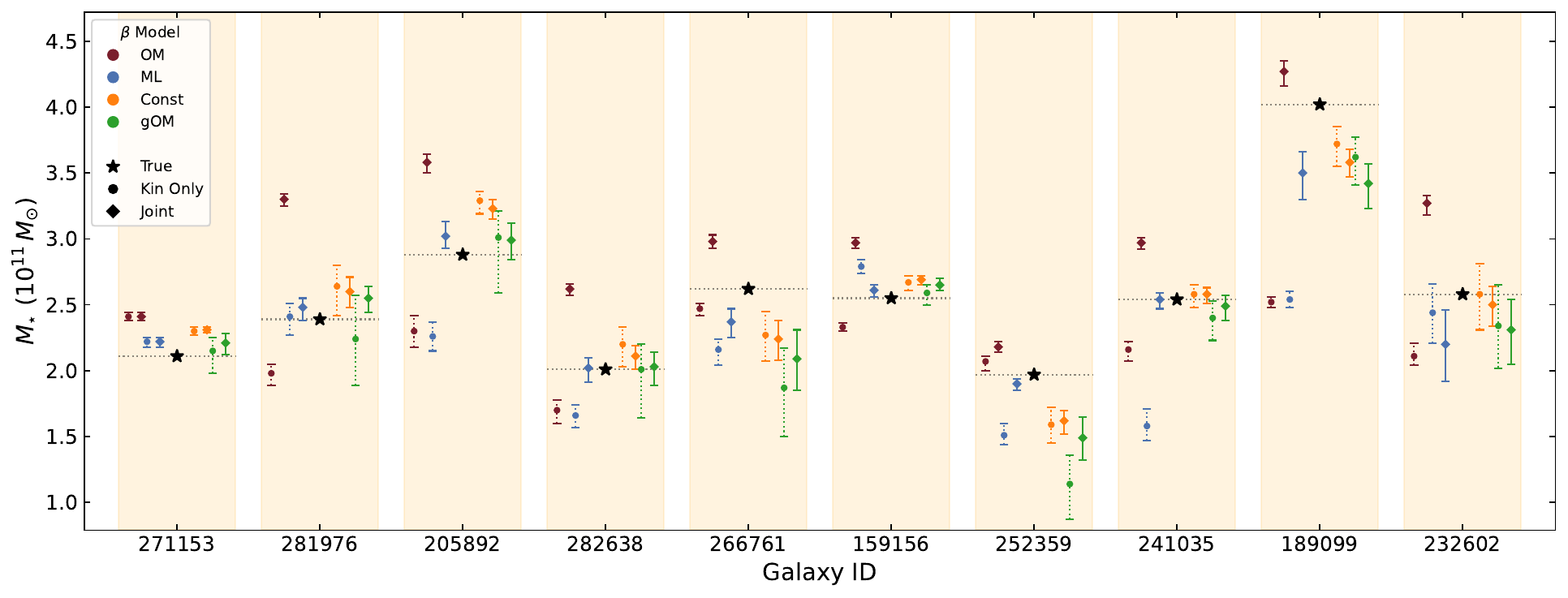}
\caption{
Comparison of true and inferred stellar masses for the ten galaxies. The shaded vertical bands mark individual galaxies.
For each galaxy, the true total galaxy stellar mass $M_\star^{\mathrm{true}}$ is shown by a black star,
with a horizontal dotted line indicating the same value within the shaded vertical band.
Colored symbols show marginalized stellar-mass constraints from dynamical modeling using the
four anisotropy models: Osipkov--Merritt (OM), Mamon--Łokas (ML), Constant $\beta$, and
generalized Osipkov--Merritt (gOM).
Circles (with dotted lines) and diamonds (with solid lines) denote kinematics-only and joint kinematics plus lensing models
respectively, with error bars indicating $68\%$ credible intervals.}
\label{fig:M_star_model_vals}
\end{figure*}

\begin{figure*}[ht!]
\centering
\includegraphics[width=0.95\textwidth]{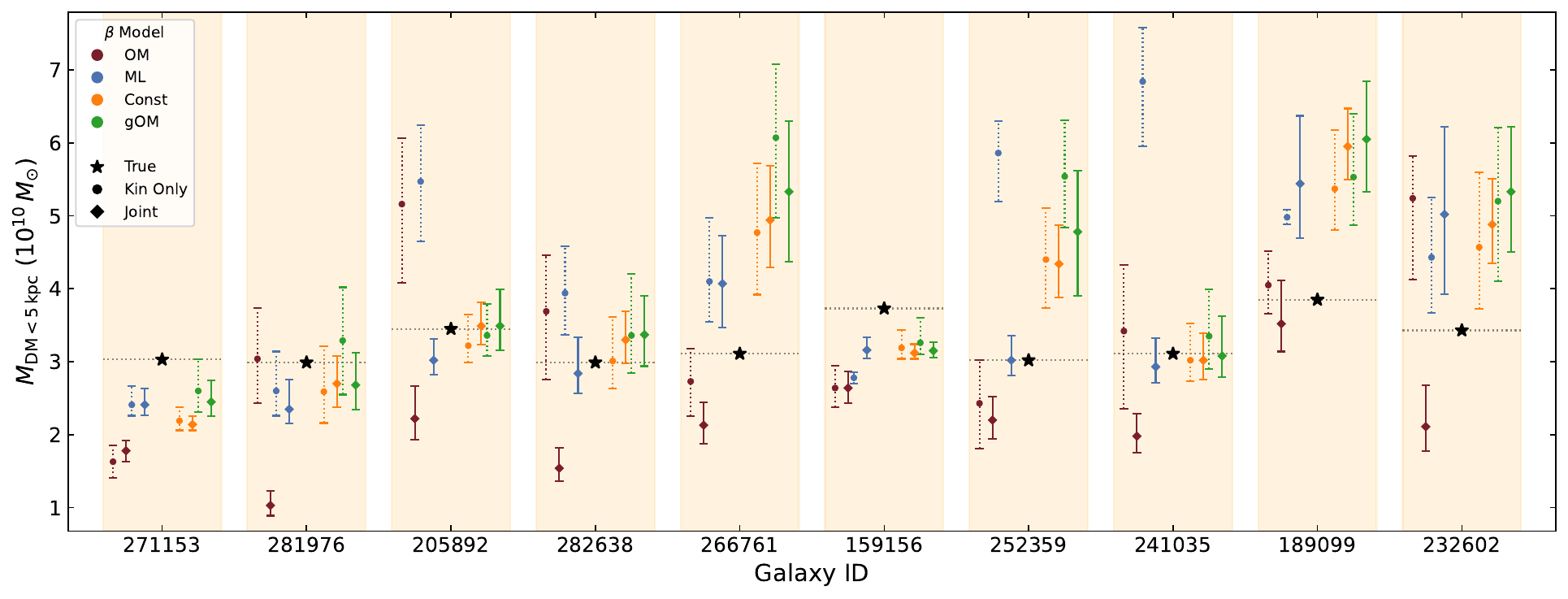}
\caption{
Comparison of true and inferred dark matter  masses within 5 kpc for the ten galaxies. The shaded vertical bands mark individual galaxies.
For each galaxy, the true dark matter mass within 5kpc, $M_{DM<5kpc}^{\mathrm{true}}$ is shown by a black star,
with a horizontal dotted line indicating the same value within the shaded vertical band.
Colored symbols show marginalized dark-mass constraints from dynamical modeling using the
four anisotropy models: Osipkov--Merritt (OM), Mamon--Łokas (ML), Constant $\beta$, and
generalized Osipkov--Merritt (gOM).
Circles (with dotted lines) and diamonds (with solid lines) denote kinematics-only and joint kinematics plus lensing models,
respectively, with error bars indicating $68\%$ credible intervals.}
\label{fig:M_dm_model_vals}
\end{figure*}
\newpage

\subsection{Results with Lower Signal to Noise Ratio}

\begin{figure*}[h]
\centering
\includegraphics[height=0.85\textheight,width=0.6\textwidth]{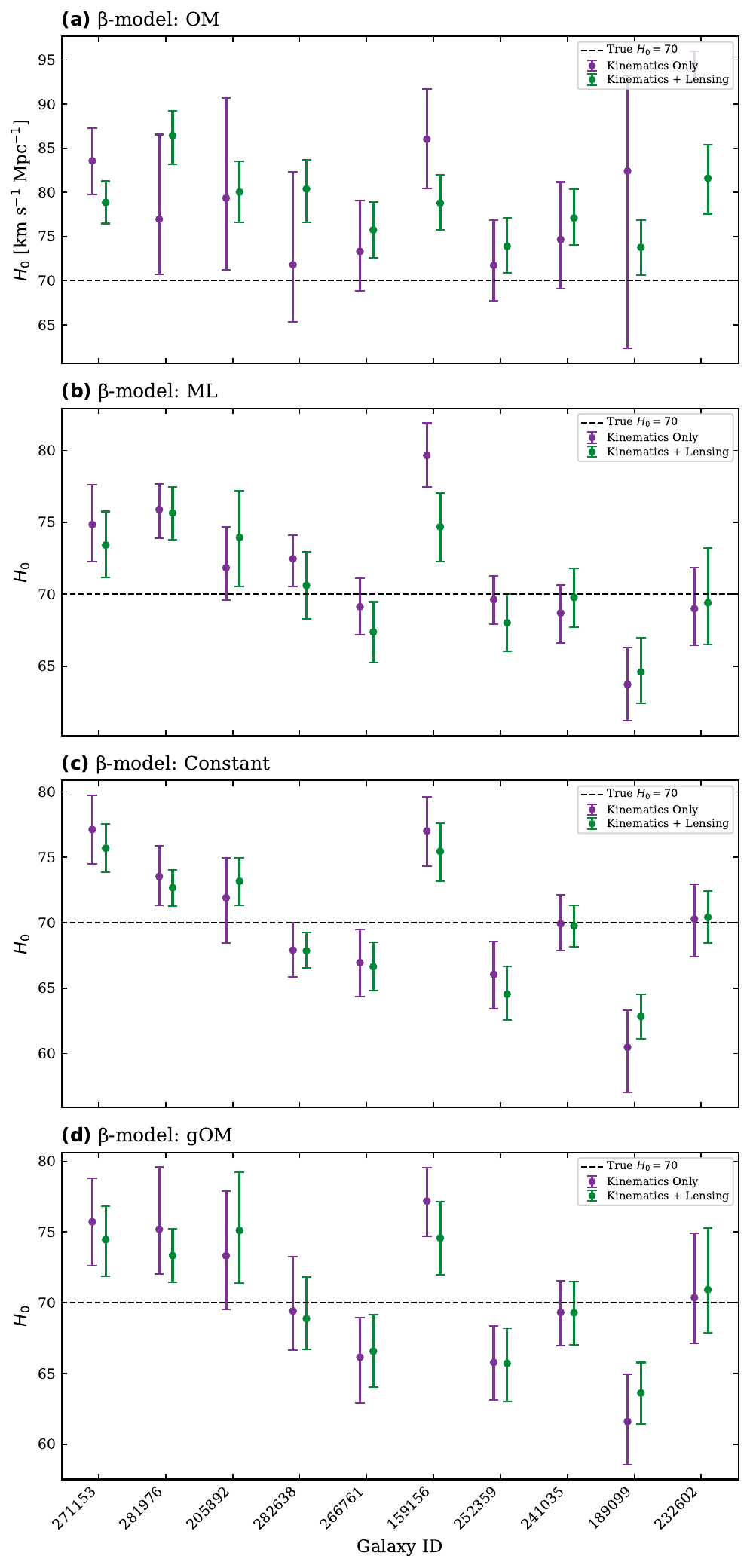}
\caption{Results from Kinematics Only, and Joint Modeling with Kinematics and Strong Lensing for the ten galaxies. The data and methods follow exactly like in the fiducial dataset in Figure \ref{fig:lc_constraints_all_in_one}, but the maximum signal-to-noise ratio is reduced from 60 to 20.
}
\label{fig:lc_constraints_all_in_one_snr_20}
\end{figure*}

%\restartappendixfloats

\clearpage

\section{Tables of Results}

\begin{deluxetable*}{cccccccc}[ht!]
\floattable
\tabletypesize{\footnotesize}
\tablecaption{\label{tab:galaxy_model_om_merged}
Galaxy Modeling Results for Osipkov--Merritt $\beta$ Model}
\tablecolumns{8}
\tablewidth{\textwidth}

\tablehead{
\multicolumn{2}{c}{} &
\multicolumn{3}{c}{Model Parameters} &
\multicolumn{3}{c}{Model Parameters} \\
\cline{3-5}\cline{6-8}
\colhead{Galaxy ID} &
\colhead{$r_s^{\mathrm{true}}/R_{\rm eff}$} &
\colhead{$r_s^{\mathrm{model}}/R_{\rm eff}$} &
\colhead{$r_{\rm ani}^{\mathrm{model}}/R_{\rm eff}$} &
\colhead{$\chi^2_{\nu}$} &
\colhead{$r_s^{\mathrm{model}}/R_{\rm eff}$} &
\colhead{$r_{\rm ani}^{\mathrm{model}}/R_{\rm eff}$} &
\colhead{$\chi^2_{\nu}$} \\
\cline{3-5}\cline{6-8}
& &
\multicolumn{3}{c}{(Kin Only)} &
\multicolumn{3}{c}{(Joint)}
}

\startdata
271153 & 12.64 & 25.55 $^{+6.14}_{-8.78}$ & 7.10 $^{+6.58}_{-2.33}$ & 1.04 & 26.88 $^{+5.29}_{-7.82}$ & 11.48 $^{+6.02}_{-3.92}$ & 1.04 \\
281976 & 4.91  & 2.39 $^{+3.09}_{-1.01}$  & 3.20 $^{+3.27}_{-1.75}$ & 0.99 & 8.42 $^{+3.51}_{-3.47}$  & 1.51 $^{+0.26}_{-0.19}$ & 6.56 \\
205892 & 7.77  & 1.90 $^{+1.45}_{-0.64}$  & 3.92 $^{+3.65}_{-2.24}$ & 0.97 & 8.48 $^{+4.64}_{-3.81}$  & 2.34 $^{+0.59}_{-0.38}$ & 3.32 \\
282638 & 6.04  & 2.75 $^{+2.83}_{-1.12}$  & 3.56 $^{+3.88}_{-2.03}$ & 0.96 & 10.03 $^{+4.45}_{-4.10}$ & 1.79 $^{+0.34}_{-0.24}$ & 3.04 \\
266761 & 7.60  & 12.07 $^{+4.91}_{-4.94}$ & 3.40 $^{+3.67}_{-1.27}$ & 1.03 & 12.29 $^{+4.71}_{-4.83}$ & 2.92 $^{+0.85}_{-0.52}$ & 1.57 \\
159156 & 9.48  & 14.48 $^{+1.52}_{-2.63}$ & 2.14 $^{+0.58}_{-0.40}$ & 1.14 & 14.04 $^{+1.77}_{-2.87}$ & 3.60 $^{+1.04}_{-0.64}$ & 2.51 \\
252359 & 7.30  & 8.59 $^{+7.33}_{-5.05}$  & 2.96 $^{+3.92}_{-1.25}$ & 1.18 & 12.46 $^{+4.89}_{-4.75}$ & 3.06 $^{+1.09}_{-0.65}$ & 1.20 \\
241035 & 7.29  & 5.30 $^{+6.24}_{-2.68}$  & 2.28 $^{+3.74}_{-0.91}$ & 1.06 & 11.83 $^{+4.54}_{-4.78}$ & 2.75 $^{+0.71}_{-0.47}$ & 2.12 \\
189099 & 6.55  & 9.54 $^{+0.81}_{-1.41}$  & 1.05 $^{+0.23}_{-0.16}$ & 1.20 & 6.15 $^{+2.85}_{-2.46}$  & 1.47 $^{+0.18}_{-0.15}$ & 6.85 \\
232602 & 8.39  & 1.39 $^{+0.64}_{-0.21}$  & 4.46 $^{+4.10}_{-3.01}$ & 0.97 & 7.56 $^{+5.88}_{-3.78}$  & 2.32 $^{+0.64}_{-0.39}$ & 3.48 \\
\enddata

\tablenotetext{}{\textbf{Note.}
The first column indicates the ID of the galaxy being modeled.
The second column gives the true dark-matter scale radius $r_s^{\mathrm{true}}$ in units of $R_{\rm eff}$.
Columns 3-5 report marginalized parameter constraints from kinematics-only modeling,
while columns 6-8 report results from joint modeling using kinematics and lensing.
Uncertainties correspond to $68\%$ credible intervals.
The reduced $\chi^2_{\nu}$ values quantify the goodness of fit for each modeling case.}
\label{table: om_new}
\end{deluxetable*}
\vspace{-0.6cm}

\begin{deluxetable*}{cccccccc}[h]
\floattable
\tabletypesize{\footnotesize}
\tablecaption{\label{tab:galaxy_model_ml_merged}
Galaxy Modeling Results for Mamon--\L okas $\beta$ Model}
\tablecolumns{8}
\tablewidth{\textwidth}

\tablehead{
\multicolumn{2}{c}{} &
\multicolumn{3}{c}{Model Parameters} &
\multicolumn{3}{c}{Model Parameters} \\
\cline{3-5}\cline{6-8}
\colhead{Galaxy ID} &
\colhead{$r_s^{\mathrm{true}}/R_{\rm eff}$} &
\colhead{$r_s^{\mathrm{model}}/R_{\rm eff}$} &
\colhead{$r_{\rm ani}^{\mathrm{model}}/R_{\rm eff}$} &
\colhead{$\chi^2_{\nu}$} &
\colhead{$r_s^{\mathrm{model}}/R_{\rm eff}$} &
\colhead{$r_{\rm ani}^{\mathrm{model}}/R_{\rm eff}$} &
\colhead{$\chi^2_{\nu}$} \\
\cline{3-5}\cline{6-8}
& &
\multicolumn{3}{c}{(Kin Only)} &
\multicolumn{3}{c}{(Joint)}
}

\startdata
271153 & 12.64 & 21.09 $^{+8.57}_{-8.61}$ & 0.27 $^{+0.05}_{-0.04}$ & 0.95 & 22.27 $^{+8.14}_{-8.64}$ & 0.27 $^{+0.05}_{-0.04}$ & 0.95 \\
281976 & 4.91  & 4.76 $^{+4.04}_{-2.12}$  & 0.03 $^{+0.01}_{-0.01}$ & 2.01 & 7.05 $^{+3.91}_{-3.29}$  & 0.04 $^{+0.01}_{-0.01}$ & 2.43 \\
205892 & 7.77  & 2.08 $^{+1.63}_{-0.70}$  & 4.21 $^{+3.71}_{-2.27}$ & 0.97 & 10.74 $^{+3.46}_{-3.81}$ & 0.16 $^{+0.09}_{-0.04}$ & 1.97 \\
282638 & 6.04  & 3.04 $^{+2.94}_{-1.20}$  & 3.22 $^{+3.93}_{-1.82}$ & 0.96 & 8.01 $^{+5.05}_{-3.62}$  & 0.07 $^{+0.03}_{-0.02}$ & 1.62 \\
266761 & 7.60  & 5.81 $^{+5.90}_{-2.90}$  & 0.18 $^{+0.04}_{-0.04}$ & 0.96 & 4.06 $^{+3.26}_{-1.48}$  & 0.04 $^{+0.01}_{-0.00}$ & 1.09 \\
159156 & 9.48  & 15.43 $^{+0.80}_{-1.58}$ & 0.03 $^{+0.00}_{-0.00}$ & 1.97 & 13.19 $^{+2.37}_{-3.56}$ & 0.04 $^{+0.01}_{-0.00}$ & 1.61 \\
252359 & 7.30  & 1.56 $^{+0.57}_{-0.22}$  & 0.04 $^{+0.01}_{-0.00}$ & 0.95 & 10.07 $^{+5.58}_{-3.84}$ & 0.04 $^{+0.00}_{-0.00}$ & 1.00 \\
241035 & 7.29  & 1.66 $^{+0.72}_{-0.36}$  & 0.10 $^{+0.02}_{-0.02}$ & 0.95 & 9.75 $^{+5.45}_{-4.21}$  & 0.04 $^{+0.01}_{-0.00}$ & 1.50 \\
189099 & 6.55  & 10.30 $^{+0.25}_{-0.53}$ & 0.02 $^{+0.00}_{-0.00}$ & 1.35 & 3.03 $^{+2.10}_{-1.09}$  & 0.05 $^{+0.01}_{-0.01}$ & 4.01 \\
232602 & 8.39  & 2.69 $^{+1.75}_{-0.88}$  & 0.05 $^{+0.02}_{-0.01}$ & 1.60 & 2.56 $^{+2.55}_{-0.96}$  & 0.08 $^{+0.03}_{-0.02}$ & 1.27 \\
\enddata

\tablenotetext{}{\textbf{Note.} The first column indicates the ID of the galaxy being modeled.
The second column gives the true dark-matter scale radius $r_s^{\mathrm{true}}$ in units of $R_{\rm eff}$.
Columns 3-5 report marginalized constraints from kinematics-only modeling, while columns 6-8 report results from joint modeling using kinematics and lensing.
Uncertainties correspond to $68\%$ credible intervals. The reduced $\chi^2_{\nu}$ values quantify the goodness of fit for each modeling case.}
\label{table: ml_new}
\end{deluxetable*}

\begin{deluxetable*}{cccccccc}[ht!]
\floattable
\tabletypesize{\small}
\tablecaption{\label{tab:galaxy_model_constant_merged}
Galaxy Modeling Results for Constant $\beta$ Model}
\tablecolumns{8}
\tablewidth{\textwidth}

\tablehead{
\multicolumn{2}{c}{} &
\multicolumn{3}{c}{Model Parameters} &
\multicolumn{3}{c}{Model Parameters} \\
\cline{3-5}\cline{6-8}
\colhead{Galaxy ID} &
\colhead{$r_s^{\mathrm{true}}/R_{\rm eff}$} &
\colhead{$r_s^{\mathrm{model}}/R_{\rm eff}$} &
\colhead{$\beta$} &
\colhead{$\chi^2_{\nu}$} &
\colhead{$r_s^{\mathrm{model}}/R_{\rm eff}$} &
\colhead{$\beta$} &
\colhead{$\chi^2_{\nu}$} \\
\cline{3-5}\cline{6-8}
& &
\multicolumn{3}{c}{(Kin Only)} &
\multicolumn{3}{c}{(Joint)}
}

\startdata
271153 & 12.64 & 22.30 $^{+8.20}_{-8.85}$ & 0.24 $^{+0.02}_{-0.02}$ & 0.95 & 26.92 $^{+5.10}_{-7.52}$ & 0.24 $^{+0.02}_{-0.02}$ & 0.95 \\
281976 & 4.91  & 3.78 $^{+4.03}_{-1.71}$  & 0.33 $^{+0.05}_{-0.05}$ & 0.94 & 3.35 $^{+1.91}_{-1.05}$  & 0.35 $^{+0.03}_{-0.03}$ & 0.94 \\
205892 & 7.77  & 7.67 $^{+4.75}_{-3.48}$  & 0.15 $^{+0.03}_{-0.03}$ & 0.94 & 5.00 $^{+2.27}_{-1.44}$  & 0.16 $^{+0.03}_{-0.03}$ & 0.94 \\
282638 & 6.04  & 5.21 $^{+5.15}_{-2.32}$  & 0.27 $^{+0.05}_{-0.06}$ & 0.95 & 3.88 $^{+1.79}_{-1.13}$  & 0.29 $^{+0.03}_{-0.03}$ & 0.95 \\
266761 & 7.60  & 2.45 $^{+1.76}_{-0.84}$  & 0.45 $^{+0.05}_{-0.05}$ & 0.94 & 2.24 $^{+0.98}_{-0.59}$  & 0.46 $^{+0.03}_{-0.03}$ & 0.94 \\
159156 & 9.48  & 11.54 $^{+3.38}_{-3.83}$ & 0.39 $^{+0.02}_{-0.03}$ & 0.94 & 13.69 $^{+1.99}_{-2.82}$ & 0.39 $^{+0.02}_{-0.02}$ & 0.94 \\
252359 & 7.30  & 2.94 $^{+1.64}_{-0.86}$  & 0.67 $^{+0.05}_{-0.05}$ & 0.93 & 2.98 $^{+1.11}_{-0.72}$  & 0.65 $^{+0.03}_{-0.03}$ & 0.94 \\
241035 & 7.29  & 7.61 $^{+6.12}_{-3.52}$  & 0.42 $^{+0.04}_{-0.04}$ & 0.94 & 7.59 $^{+4.90}_{-2.81}$  & 0.42 $^{+0.02}_{-0.02}$ & 0.94 \\
189099 & 6.55  & 4.20 $^{+3.13}_{-1.68}$  & 0.31 $^{+0.03}_{-0.02}$ & 0.94 & 2.74 $^{+0.88}_{-0.64}$  & 0.32 $^{+0.02}_{-0.02}$ & 0.95 \\
232602 & 8.39  & 2.43 $^{+2.15}_{-0.91}$  & 0.24 $^{+0.04}_{-0.04}$ & 0.94 & 2.10 $^{+0.76}_{-0.53}$  & 0.26 $^{+0.03}_{-0.03}$ & 0.94 \\
\enddata

\tablenotetext{}{\textbf{Note.} The first column indicates the ID of the galaxy being modeled.
The second column gives the true dark-matter scale radius $r_s^{\mathrm{true}}$ in units of $R_{\rm eff}$.
Columns 3-5 report marginalized constraints from kinematics-only modeling, while columns 6-8 report results from joint modeling using kinematics and lensing.
Uncertainties correspond to $68\%$ credible intervals. The reduced $\chi^2_{\nu}$ values quantify the goodness of fit for each modeling case.}
\label{table: constant_new}
\end{deluxetable*}

\setlength{\tabcolsep}{3pt}
\begin{deluxetable*}{cccccccccc}[ht!]
\floattable
\tabletypesize{\small}
\tablecaption{\label{tab:galaxy_model_gom_merged}
Galaxy Modeling Results for Generalized Osipkov--Merritt $\beta$ Model}
\tablecolumns{10}
\tablewidth{0.99\textwidth}

\tablehead{
\multicolumn{2}{c}{} &
\multicolumn{4}{c}{Model Parameters} &
\multicolumn{4}{c}{Model Parameters} \\
\cline{3-6}\cline{7-10}
\colhead{Galaxy ID} &
\colhead{$r_s^{\mathrm{true}}/R_{\rm eff}$} &
\colhead{$r_s^{\mathrm{model}}/R_{\rm eff}$} &
\colhead{$r_{\rm ani}^{\mathrm{model}}/R_{\rm eff}$} &
\colhead{$\beta_\infty^{\mathrm{model}}$} &
\colhead{$\chi^2_{\nu}$} &
\colhead{$r_s^{\mathrm{model}}/R_{\rm eff}$} &
\colhead{$r_{\rm ani}^{\mathrm{model}}/R_{\rm eff}$} &
\colhead{$\beta_\infty^{\mathrm{model}}$} &
\colhead{$\chi^2_{\nu}$} \\
\cline{3-6}\cline{7-10}
& &
\multicolumn{4}{c}{(Kin Only)} &
\multicolumn{4}{c}{(Joint)}
}

\startdata
271153 & 12.64 & 19.20 $^{+9.88}_{-8.27}$ & 0.23 $^{+0.07}_{-0.09}$ & 0.44 $^{+0.20}_{-0.13}$ & 0.95 & 21.90 $^{+8.52}_{-9.25}$ & 0.18 $^{+0.07}_{-0.07}$ & 0.37 $^{+0.12}_{-0.08}$ & 0.95 \\
281976 & 4.91  & 2.39 $^{+2.34}_{-0.83}$ & 0.03 $^{+0.01}_{-0.01}$ & 0.44 $^{+0.09}_{-0.08}$ & 0.94 & 3.62 $^{+2.71}_{-1.37}$ & 0.03 $^{+0.01}_{-0.00}$ & 0.37 $^{+0.03}_{-0.03}$ & 0.94 \\
205892 & 7.77  & 8.19 $^{+4.66}_{-3.54}$ & 0.10 $^{+0.03}_{-0.04}$ & 0.34 $^{+0.25}_{-0.11}$ & 0.94 & 6.28 $^{+4.85}_{-2.71}$ & 0.09 $^{+0.02}_{-0.03}$ & 0.33 $^{+0.10}_{-0.08}$ & 0.95 \\
282638 & 6.04  & 4.10 $^{+3.37}_{-1.70}$ & 0.05 $^{+0.02}_{-0.01}$ & 0.35 $^{+0.13}_{-0.08}$ & 0.96 & 3.83 $^{+2.75}_{-1.35}$ & 0.04 $^{+0.01}_{-0.01}$ & 0.34 $^{+0.05}_{-0.05}$ & 0.95 \\
266761 & 7.60  & 1.69 $^{+0.84}_{-0.33}$ & 0.05 $^{+0.01}_{-0.01}$ & 0.56 $^{+0.11}_{-0.07}$ & 0.94 & 2.05 $^{+1.32}_{-0.59}$ & 0.04 $^{+0.01}_{-0.00}$ & 0.50 $^{+0.06}_{-0.05}$ & 0.94 \\
159156 & 9.48  & 10.79 $^{+3.83}_{-4.31}$ & 0.04 $^{+0.01}_{-0.01}$ & 0.44 $^{+0.04}_{-0.03}$ & 0.94 & 14.37 $^{+1.57}_{-2.46}$ & 0.04 $^{+0.01}_{-0.00}$ & 0.40 $^{+0.03}_{-0.03}$ & 0.94 \\
252359 & 7.30  & 2.28 $^{+0.77}_{-0.51}$ & 0.05 $^{+0.01}_{-0.00}$ & 0.81 $^{+0.09}_{-0.08}$ & 0.94 & 2.52 $^{+1.71}_{-0.75}$ & 0.04 $^{+0.00}_{-0.00}$ & 0.67 $^{+0.05}_{-0.05}$ & 0.94 \\
241035 & 7.29  & 5.81 $^{+5.31}_{-2.54}$ & 0.05 $^{+0.01}_{-0.01}$ & 0.50 $^{+0.07}_{-0.05}$ & 0.94 & 7.77 $^{+5.86}_{-3.57}$ & 0.04 $^{+0.01}_{-0.00}$ & 0.46 $^{+0.04}_{-0.04}$ & 0.94 \\
189099 & 6.55  & 3.57 $^{+2.81}_{-1.39}$ & 0.03 $^{+0.01}_{-0.00}$ & 0.34 $^{+0.04}_{-0.03}$ & 0.95 & 2.46 $^{+1.25}_{-0.73}$ & 0.03 $^{+0.01}_{-0.01}$ & 0.39 $^{+0.05}_{-0.04}$ & 0.95 \\
232602 & 8.39  & 2.02 $^{+1.59}_{-0.60}$ & 0.05 $^{+0.02}_{-0.01}$ & 0.31 $^{+0.07}_{-0.06}$ & 0.94 & 1.88 $^{+0.96}_{-0.51}$ & 0.05 $^{+0.02}_{-0.01}$ & 0.31 $^{+0.06}_{-0.05}$ & 0.94 \\
\enddata

\tablenotetext{}{\textbf{Note.}
The first column indicates the ID of the galaxy being modeled.
The second column gives the true dark-matter scale radius $r_s^{\mathrm{true}}$ in units of $R_{\rm eff}$.
Columns 3-6 report marginalized constraints from kinematics-only modeling, while columns 7-10 report results from joint modeling using kinematics and lensing.
Uncertainties correspond to $68\%$ credible intervals.
This model includes the additional degree of freedom $\beta_\infty$.
The reduced $\chi^2_{\nu}$ values quantify the goodness of fit.}
\label{table: gom_new}
\end{deluxetable*}

%%%%%%%%%%%%%%%%%%%%%%%%%%%%%%%%%%%%%%%%%%%%%%%%%%%%%%%%
\clearpage
\bibliography{references_try}{}
\bibliographystyle{aasjournalv7}

\end{document}